\documentclass[useAMS,usenatbib]{lib/mnras}

\usepackage{natbib}
\usepackage{epsfig}
\usepackage{scrtime}
\usepackage{graphicx}	
\usepackage{amsmath}	
\usepackage{amssymb}	
\usepackage{float}
\usepackage[caption = false]{subfig}
\usepackage{multirow}
\usepackage{rotating}
\usepackage{ulem}
\usepackage{booktabs}
\usepackage{xcolor}

\newcommand{\ha}{\mathrm{H}\alpha}
\newcommand{\hb}{\mathrm{H}\beta}
\newcommand{\Oiii}{\mathrm{O\ III}}

\newcommand{\mincir}{\raise-3.truept\hbox{\rlap{\hbox{$\sim$}}\raise4.truept\hbox{$<$}\ }}

\title[Classifying low-z SEDs in the PAU survey]{The PAU Survey: Classifying low-z SEDs using Machine Learning clustering}
\author[Ana Luisa González-Morán et al.] {A.L. González-Morán$^{1,2,3}$\thanks{Contact e-mail: anagonzalez@uas.edu.mx}, P. Arrabal Haro$^{4}$, C. {Mu{\~n}oz-Tu{\~n}{\'o}n}$^{1,2}$,
\newauthor J.M. Rodríguez-Espinosa$^{1,2,5}$, J. Sánchez Almeida$^{1,2}$, J. Calhau$^{1,2}$, E. Gazta{\~n}aga$^{6,7}$,
\newauthor F.J. Castander$^{6,7}$, P. Renard$^8$, L. Cabayol$^9$, E. Fernandez$^1$, C. Padilla$^1$,
\newauthor J. Garcia-Bellido$^{11}$, R. Miquel$^{10,12}$, J. De Vicente$^{13}$, E. Sanchez$^{13}$,
\newauthor I. Sevilla-Noarbe$^{13}$ and D. Navarro-Gironés$^{6,7}$.
\\ \\
$^{1}$ Instituto de Astrof\'isica de Canarias, La Laguna, Tenerife, E-38200, Spain \\
$^{2}$ Departamento de Astrof\'isica, Universidad de La Laguna, Spain\\
$^3$ Facultad de Ciencias de la Tierra y el Espacio, Universidad Aut\'onoma de Sinaloa, Cd Universitaria, 80040, Culiac\'an, Sinaloa, M\'exico\\
$^4$ NSF's National Optical-Infrared Astronomy Research Laboratory, 950 N. Cherry Ave., Tucson, AZ 85719, USA\\
$^5$ Instituto de Astrof\'isica de Andaluc\'ia\\
$^6$ Institute of Space Sciences (ICE, CSIC), Campus UAB, Carrer de Can Magrans, s/n, 08193 Barcelona, Spain\\
$^7$ Institut d'Estudis Espacials de Catalunya (IEEC), E-08034 Barcelona, Spain\\
$^8$ Department of Astronomy, Tsinghua University, Beijing 100084, China\\
$^9$ Port d'Informació Científica (PIC-IFAE), Campus UAB, Edifici D $\cdot$ E-08193 Bellaterra, (Cerdanyola del Vallès), Spain\\
$^{10}$ Institut de Física d'Altes Energies (IFAE), The Barcelona Institute of Science and Technology (BIST), Facultat de Ciències,\\
Edifici Cn, Campus UAB, 08193 Bellaterra (Barcelona), Spain\\
$^{11}$ Instituto de Física Teórica (UAM-CSIC), Universidad Autónoma de Madrid, Cantoblanco 28049, Spain\\
$^{12}$ Catalana de Rercerca i Estudis Avançats (ICREA), 08010 Barcelona, Spain\\
$^{13}$ Centro de Investigaciones Energéticas, Medioambientales y Tecnológicas (CIEMAT), Avenida Complutense 40, E-28040 Madrid (Spain)}

\begin{document}

\date{v2 --- Compiled at \thistime\ hrs  on \today\ }

\pagerange{\pageref{firstpage}--\pageref{lastpage}} \pubyear{2016}

\maketitle

\label{firstpage}

\begin{abstract}

We present an application of unsupervised Machine Learning Clustering to the PAU Survey of galaxy spectral energy distribution (SED) within the COSMOS field. The clustering algorithm is implemented and optimized to get the relevant groups in the data SEDs. We find 12 groups from a total number of 5,234 targets in the survey at $0.01 <$ z $< 0.28$. Among the groups, 3,545 galaxies (68\%) show emission lines in the SEDs. These groups also include 1,689 old galaxies with no active star formation. We have fitted the SED to every single galaxy in each group with CIGALE. The mass, age and specific star formation rates (sSFR) of the galaxies range from $0.15 <$ age/Gyr $< 11$; $6 <$ log (M$_{\star}$/M$_{\odot}$) $< 11.26$, and $-14.67 <$ log (sSFR/yr $^{-1}$) $< -8$. The groups are well defined in their properties with galaxies having clear emission lines also having lower
mass, are younger and have higher sSFR than those with elliptical like patterns. The characteristic values of galaxies showing clear emission lines are in agreement with the literature for starburst galaxies in COSMOS and GOODS-N fields at low redshift. The star-forming main sequence, sSFR vs. stellar mass and UVJ diagram show clearly that different groups fall into different regions with some overlap among groups. Our main result is that the joint of low- resolution (R $\sim$ 50) photometric spectra provided by the PAU survey together with the unsupervised classification provides an excellent way to classify galaxies. Moreover, it helps to find and extend the analysis of extreme ELGs to lower masses and lower SFRs in the local Universe.
\end{abstract}
\begin{keywords}
galaxies: star formation – photometry – fundamental parameters – stellar content – starburst
\end{keywords}

\section{Introduction}

The classification of galaxies into different types is as old as the notion of “extragalactic nebulae” \citep{Hubble1926}. Galaxies in the local Universe display a variety of shapes and structural properties. The main classification system still in use is Jeans-Hubble tuning fork diagram \citep{Jeans1928, Hubble1936}, with all the refinements introduced by \cite{Sandage1961} and \cite{deVaucouleurs1959}, based on the morphological properties of galaxies. The basic Hubble classification of galaxies into “early” and “late” types (and their subtypes) has survived because, among other reasons, these types correlate well with other properties of galaxies, such as colours, stellar content or neutral hydrogen, among others \citep{Kennicutt1992, Roberts_Haynes1994, Buta1994, Strateva2001, Sanchez-Almeida2011, Aguerri2012, Moutard2016a}. This is based mostly on  traditional ways of galaxy classification, typically used in the past, which are based on broad band colours.

Classifications based on spectroscopic surveys provide enough spectral resolution to clearly distinguish absorption and emission lines as well as other spectral features from their spectral continuum, which provide information about different physical processes. Emission lines inform about the ionised interstellar medium \citep[ISM; see][for a review on the topic]{Kewley2019}, while absorption lines inform of the properties of the stellar population \citep{Maraston2009}. However, spectroscopic observations require  large integration times whereas broad-band photometric surveys, with filters with full width at half maximum (FWHM) $\sim$ 1000 \AA\, allow to obtain a high signal-to-noise ratio (SNR) with relatively low integration times but the detection of spectral features other than the continuum, given the low spectral resolution, is limited to detecting the features with the highest equivalent widths.

A good compromise comes from narrow-band photometric surveys (FWHM $\sim$ 100 \AA) such as SHARDS \citep{PerezGonzalez2013}, COSMOS SC4K \citep{Sobral2018} or PAU \citep[Physics of Accelerated Universe;][]{Eriksen2020, 2022arXiv220614022S}.
The sets of consecutive narrow band filters employed in this kind of surveys over a wide wavelength range allow the observation of specific spectral features besides the spectral continuum, which are essential to determine galaxy properties. In this way, spectrophotometric surveys achieve the required spectral resolution for robust redshift determination as well as the characterization of several spectral signatures, which are difficult to analyze in their broad-band counterparts \citep[see, e.g.,][]{Cava2015, HernanCaballero2017, ArrabalHaro2018, Sobral2018, Lumbreras-Calle2019, Barro2019}.

PAU surveys a large Northern area of the sky while simultaneously achieving a high number density of galaxies with sub-percent photometric redshift accuracy. This is possible thanks to the photometric camera, PAUCam: a unique combination of a large field-of-view, with 40 narrow-band (NB) filters (13.5nm FWHM) spanning a wavelength range from 450nm to 850nm. PAUCam was commissioned in June 2015 on the William Herschel Telescope (WHT), atop of the Roque de los Muchachos Observatory. This wavelength sampling results in photometric redshifts, with a precision more than an order of magnitude more precise that conventional broad-band surveys, while being able to cover large areas of sky.

The use of unsupervised Machine Learning (ML) classifications over more traditional grouping methods enables a fast processing of large surveys and, most importantly, the capability of identifying hidden interconnections between different parameters of the sample that classical predefined grouping algorithms could miss. Unsupervised ML algorithms have already been used in the past to perform automatic classifications of  large samples of galaxies based on their spectral energy distributions \cite[][among others]{D'Abrusco2009, Sanchez-Almeida2010, D'Abrusco2012, SanchezAlmeida2013, Baron2017, Siudek2018, Turner2021, Dubois2022, Teimoorinia2022}. In particular, \citet{Sanchez-Almeida2010} used an unsupervised k-means cluster analysis algorithm to classify all spectra in the  Sloan Digital Sky Survey data release 7 (SDSS/DR7). They identified as many as 17 different classes of galaxies. This would have been extremely challenging using classical methods due to the huge number of spectra ($\sim$ 174 k) to be processed.

The goal of this paper is to perform an unsupervised ML clustering of the PAU survey within the COSMOS field, focusing on the search for differences in the shape of normalised rest-frame low-z SEDs and linking them to stellar population properties. This work shows the potential of the PAU survey data in this regard.
 
This paper is organized as follows: \S \ref{sec:Data} describes the data sample. In \S \ref{sec:ML classification} we explain the methodology applied to do an unsupervised ML classification. In \S \ref{sec:SED fitting} we describe the procedure applied to perform the SED fitting. In \S \ref{sec:ResultsDiscussion} we present and discuss the results of possible differences associated with the Stellar Populations (SPs) of the galaxies from each class while the conclusions are given in \S \ref{sec:Conclusions}.

\section{Data sample}
\label{sec:Data}

To perform this study we used the full data from the PAU spectro-photometry catalogue provided by the PAU collaboration. PAU spans a fraction of the COSMOS field \citep{Scoville2007, Lilly2009}. The data were taken with the WHT at the Observatorio del Roque de los Muchachos at La Palma, Canary Islands (Spain). The images were obtained with the especially conceived PAUCam instrument \citep{Padilla2019}. PAUCam is an optical camera equipped with 40 narrow band (NB) filters. The NB filters have $\sim$ 130 \AA\ FWHM and are spaced at intervals of 100 \AA, entirely covering the wavelength range from 4500 \AA\ to 8500 \AA\ \citep{Casas2016}, which results in an effective resolution of R $\sim$ 50. The basic properties of these 40 consecutive PAU NB band filters are showed in table \ref{tab:PAU filters}, where in column 1 are the names, in column 2 the effective wavelength, in column 3 the FWHM and, in column 4 the 5$\sigma$ depth.
The entire photometric catalogue comprises 64,151 galaxies up to $i_{AB}$ $<$ 23. These data have been used in photo-z studies published by the PAU collaboration \citep[e.g.][]{Eriksen2019, CabayolGarcia2020, Cabayol2021}. PAU achieves a photometric redshift precision in the COSMOS field of $\sigma_{68}/(1+z)=0.0037$ to $i_{AB}$ $<$ 22.5 \citep{Eriksen2019}. These results have been further improved with enhancements over the original photo-z code \citep{Alarcon2021}, or ML approaches \citep{Eriksen2020, Soo2021}. For this paper we use the 30-bands photo-z from \cite{Ilbert2009} with $\sigma_{68}/(1+z) = 0.007$ at $i_{AB}$ $<$ 22.5 rather than the PAU photo-z in order to use the entire PAU catalog with 64,151 targets against the PAU photo-z catalog with 44,318 targets.\\

\begin{figure}
\hspace*{-0.5cm}
	\includegraphics[width=1.\columnwidth]{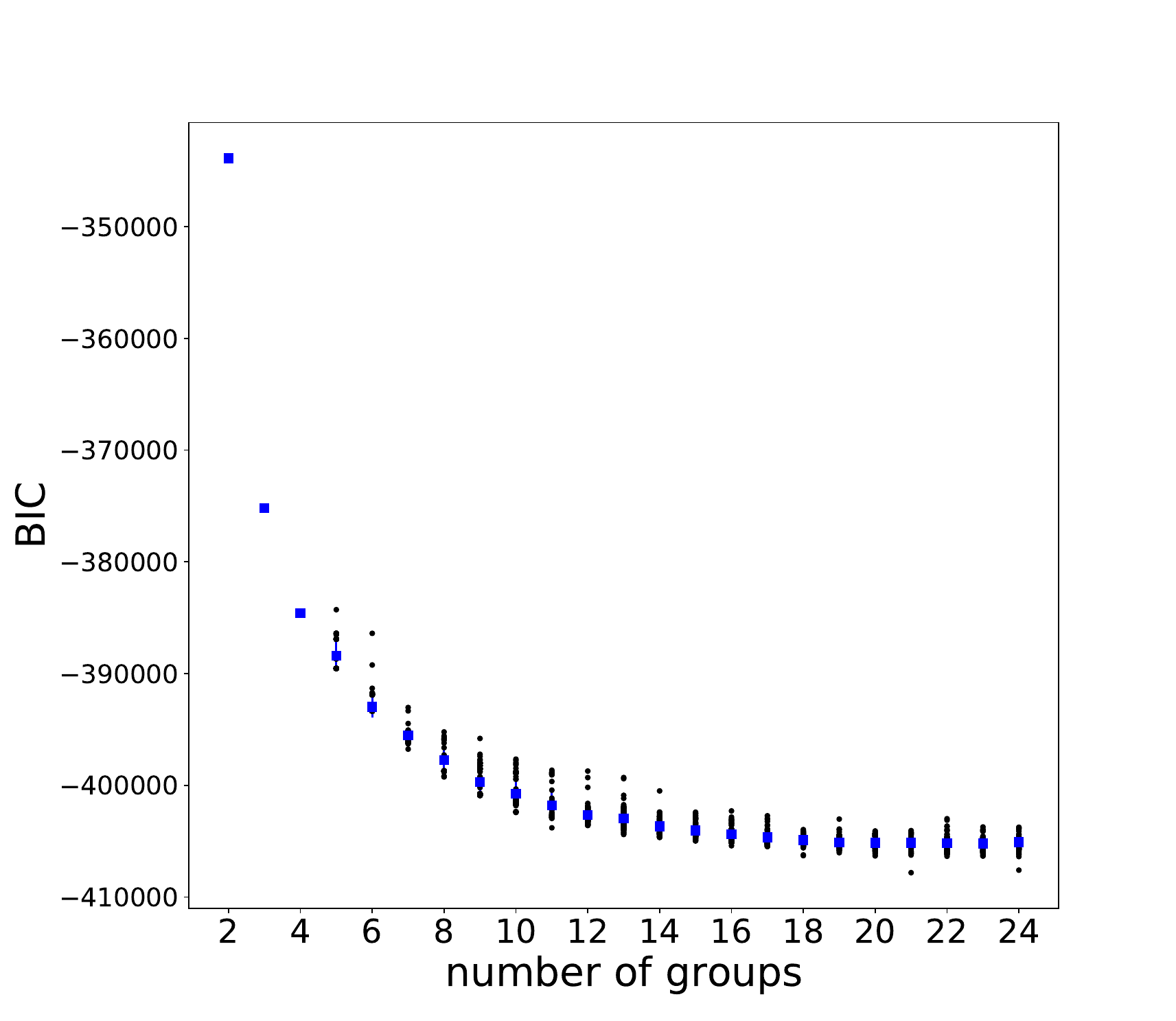}
    \caption{The trend of the BIC parameter with the number of groups. The blue squares represent the BIC mean value among the models resulting in each number of groups.}
    \label{fig:BIC-ncomp}
\end{figure}

\begin{figure}
\hspace*{-0.5cm}
	\includegraphics[width=1.\columnwidth]{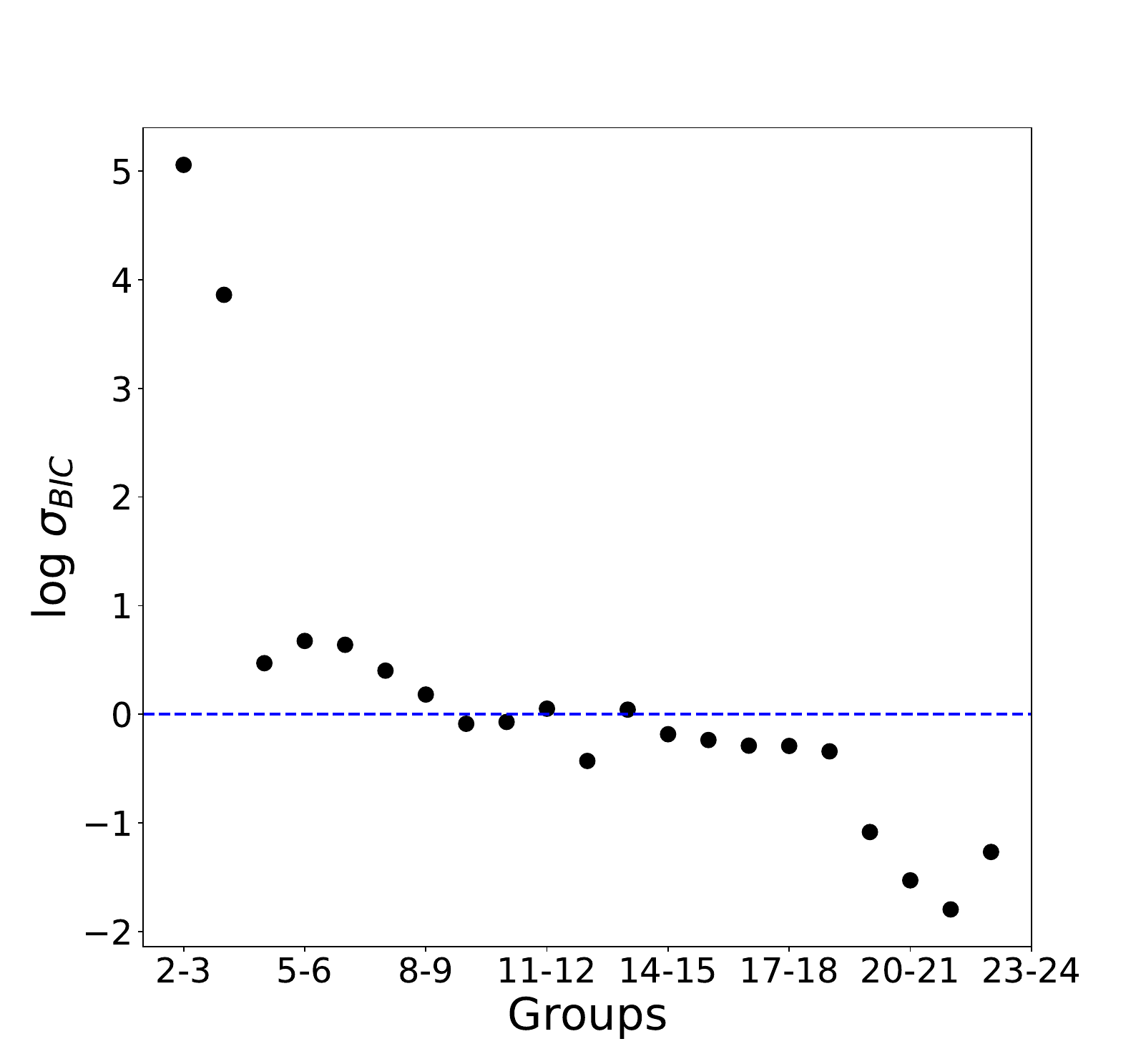}
    \caption{Evaluation of the BIC parameter gradient in terms of the $\sigma_{BIC}$ as a function of the transitions between consecutive numbers of components (see eq.~\ref{eq:sigma_BIC}). This is used to identify up to which number of groups an increase in the amount of groups does not translate into a substantial BIC improvement. The horizontal dashed blue line represents to $\sigma_{BIC}=1$.
    }
    \label{fig:DeltaBIC-ncomp_zoom}
\end{figure}

\begin{figure*}
\begin{center}$
\begin{array}{cccc}
   \includegraphics[width=.25\textwidth]{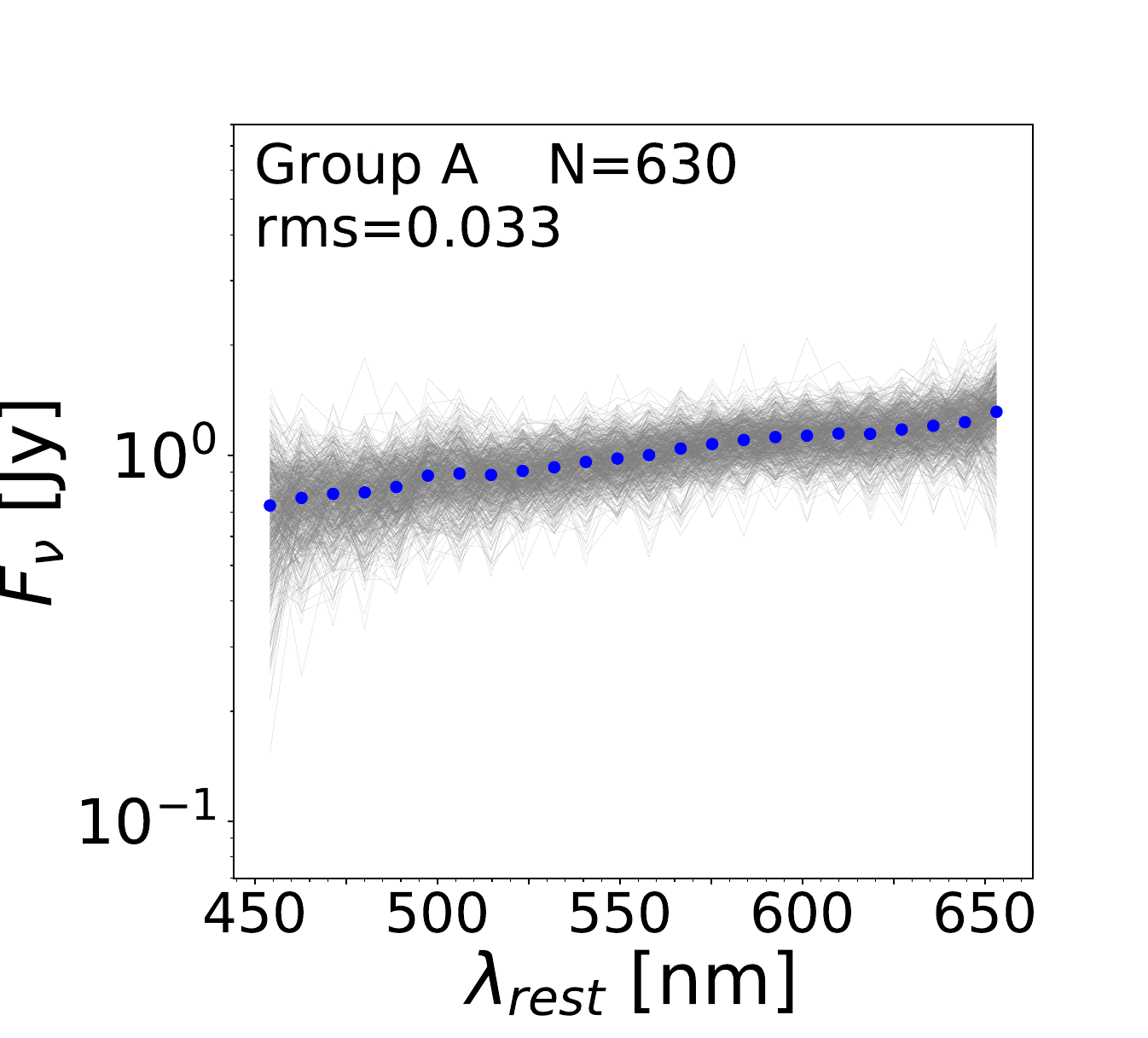} &   
  \hspace{-0.5cm} \includegraphics[width=.25\textwidth]{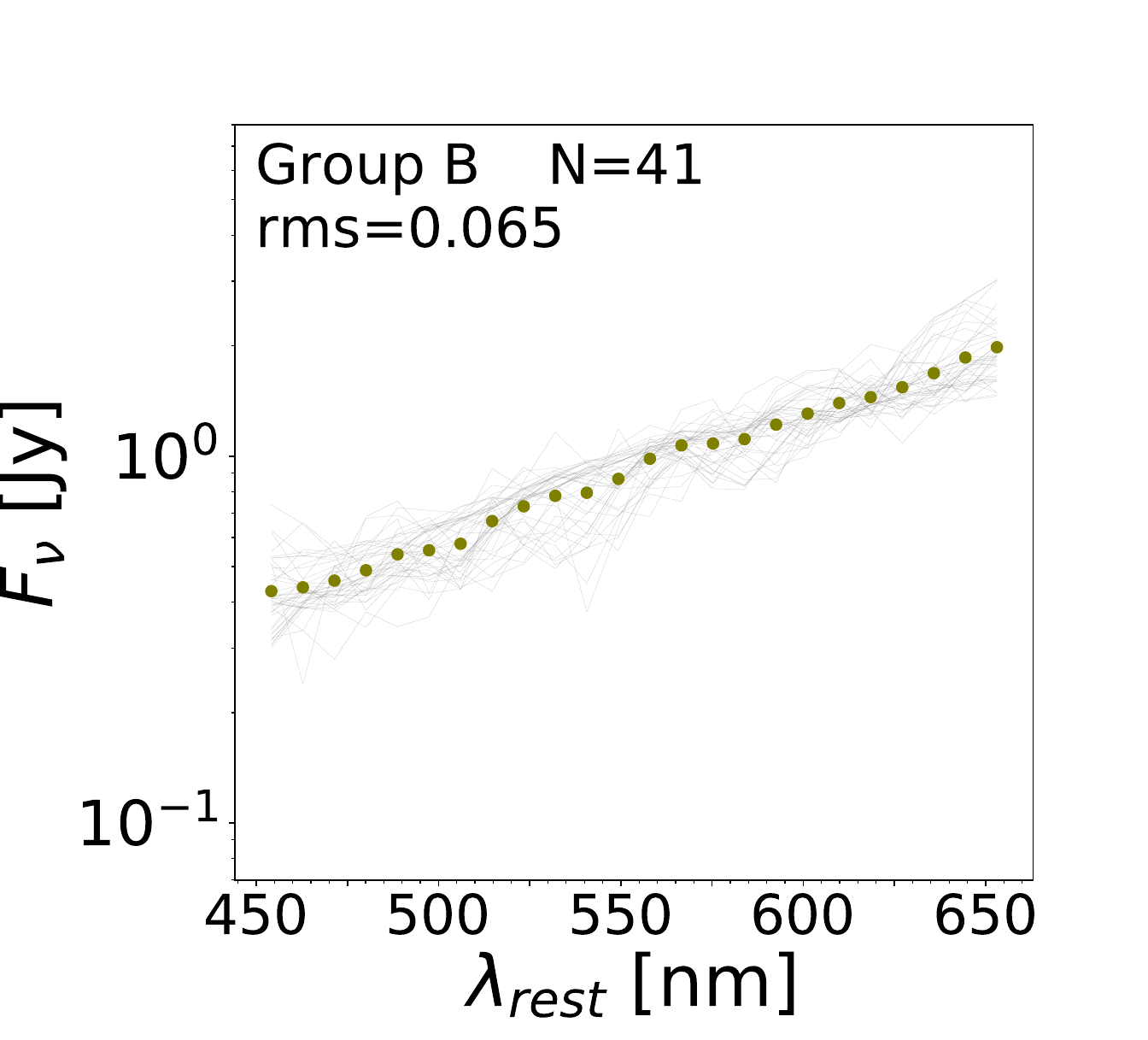}&
 \hspace{-0.5cm}\includegraphics[width=.25\textwidth]{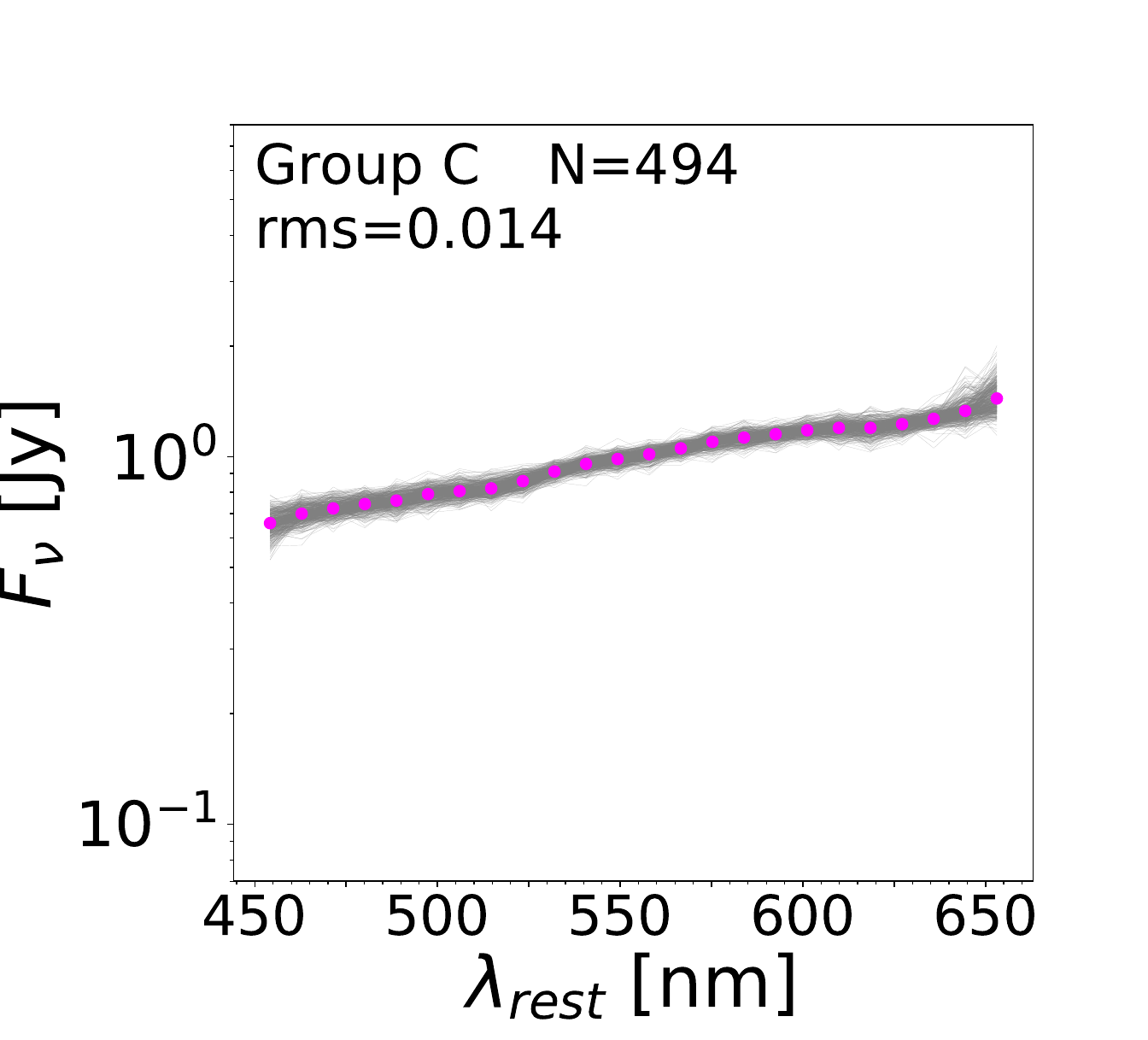} &
 \hspace{-0.5cm}\includegraphics[width=.25\textwidth]{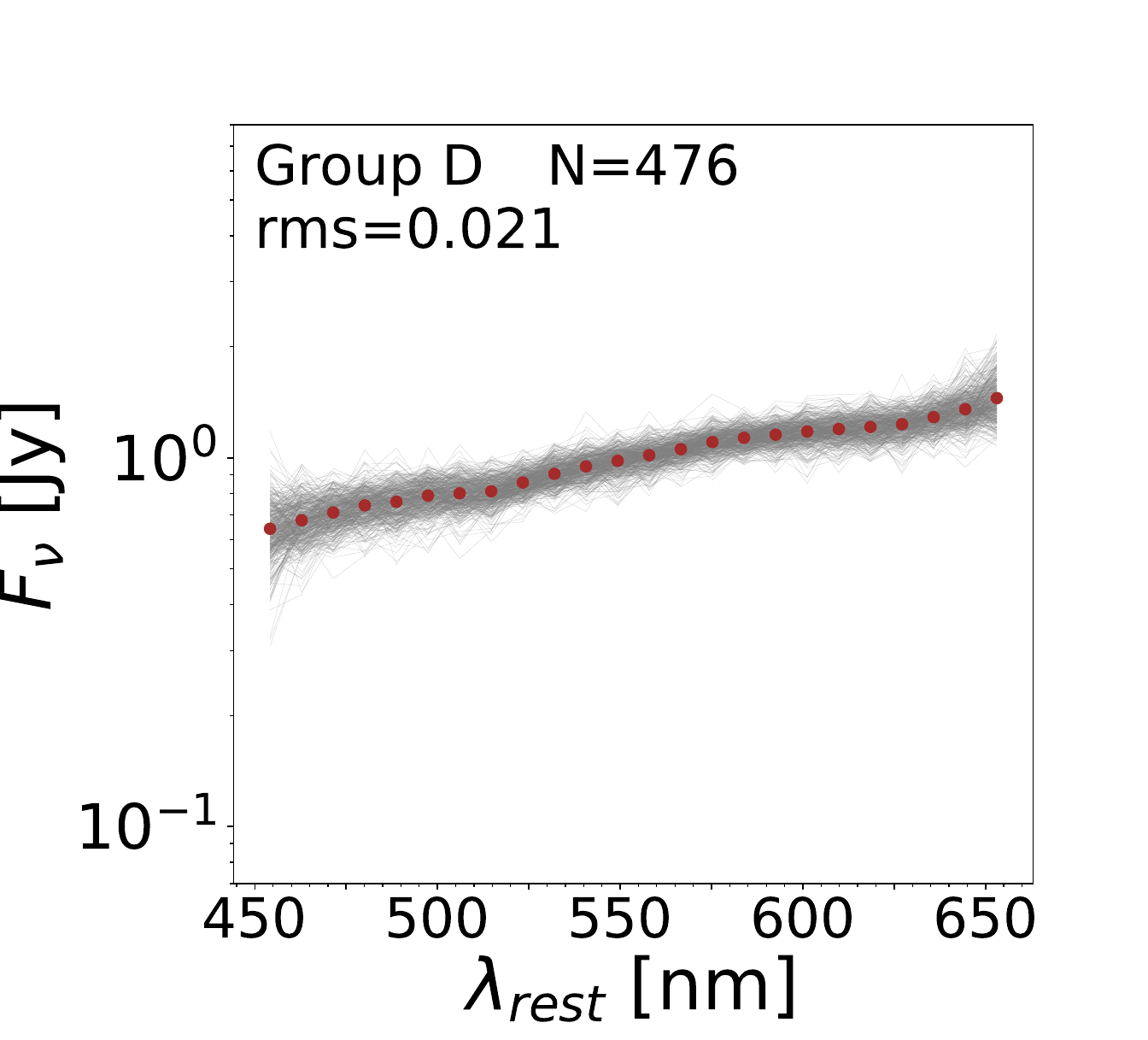}
\end{array}$
\end{center}

\begin{center}$
\begin{array}{cccc}
   \includegraphics[width=.25\textwidth]{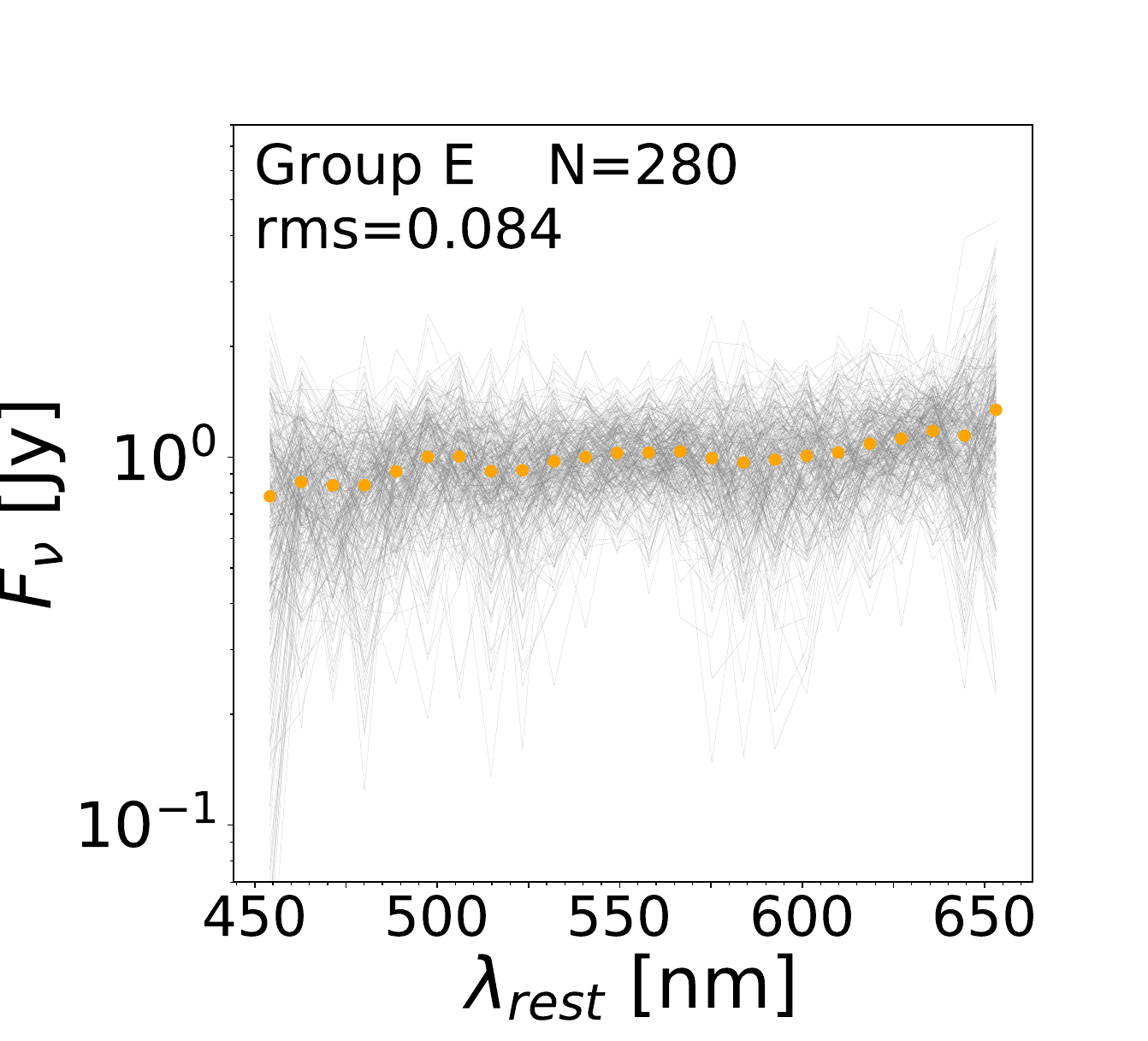} &   
 \hspace{-0.5cm}\includegraphics[width=.25\textwidth]{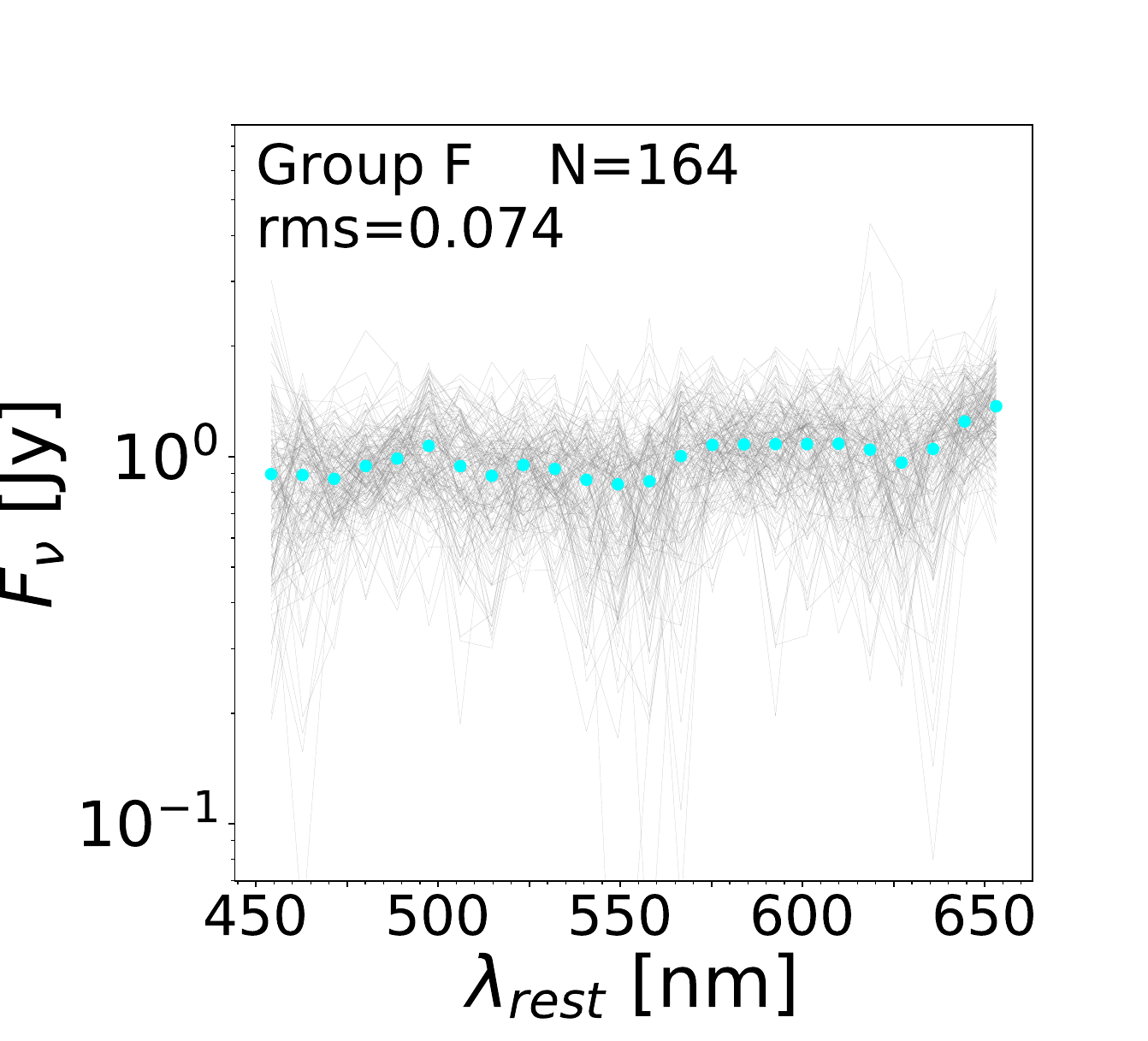}&
 \hspace{-0.5cm}\includegraphics[width=.25\textwidth]{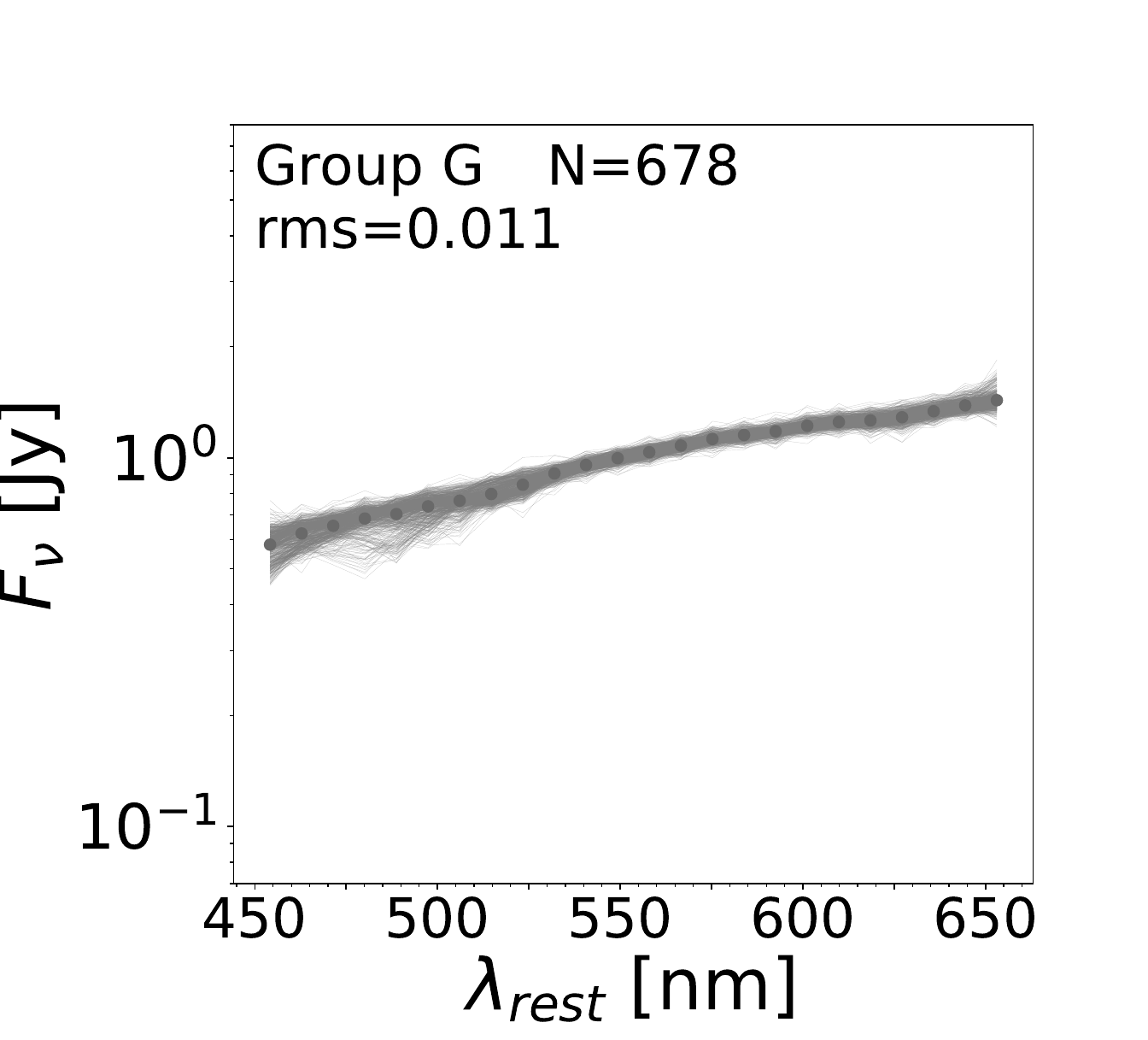} &
 \hspace{-0.5cm}\includegraphics[width=.25\textwidth]{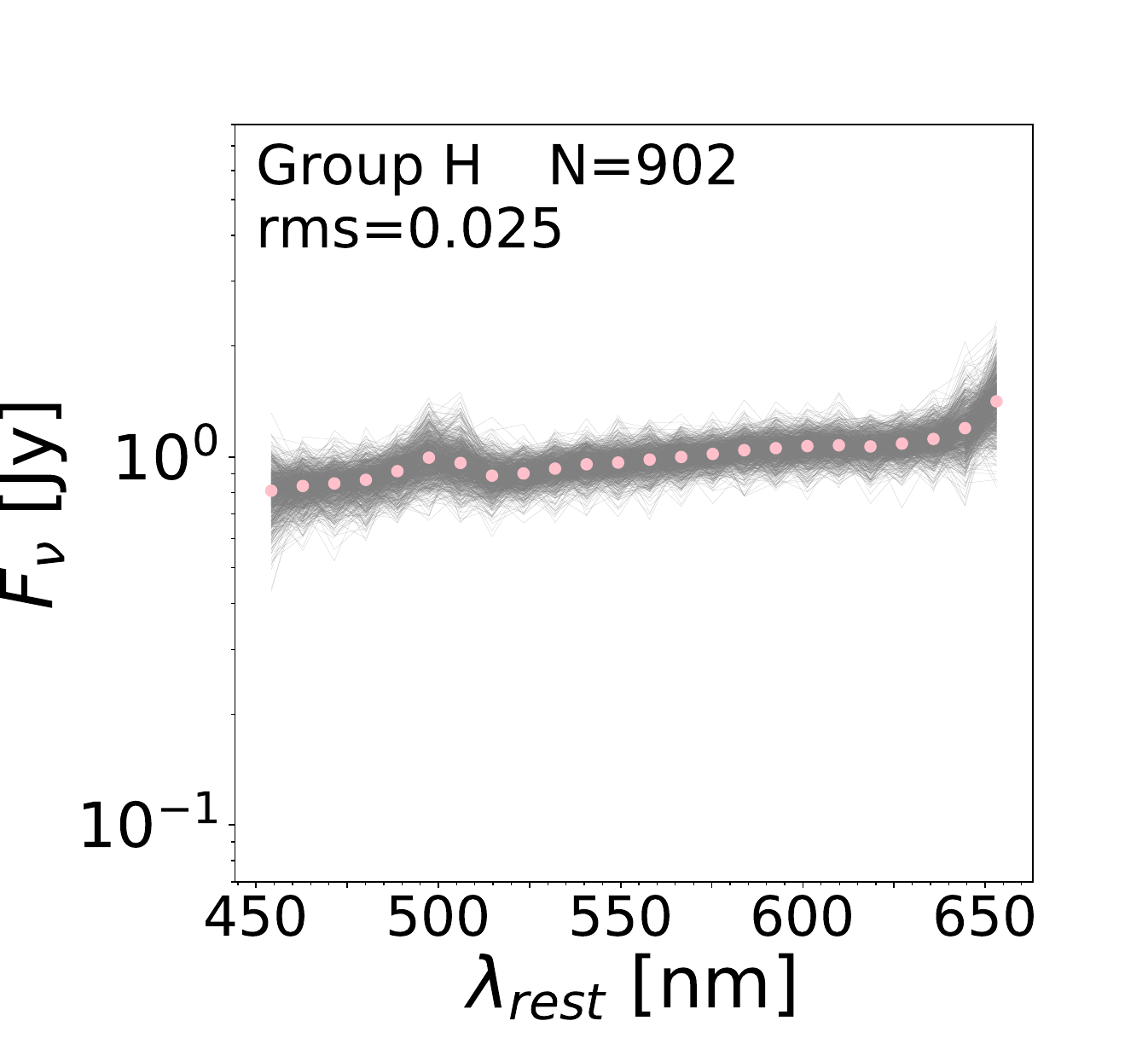}
\end{array}$
\end{center}

\begin{center}$
\begin{array}{cccc}
   \includegraphics[width=.25\textwidth]{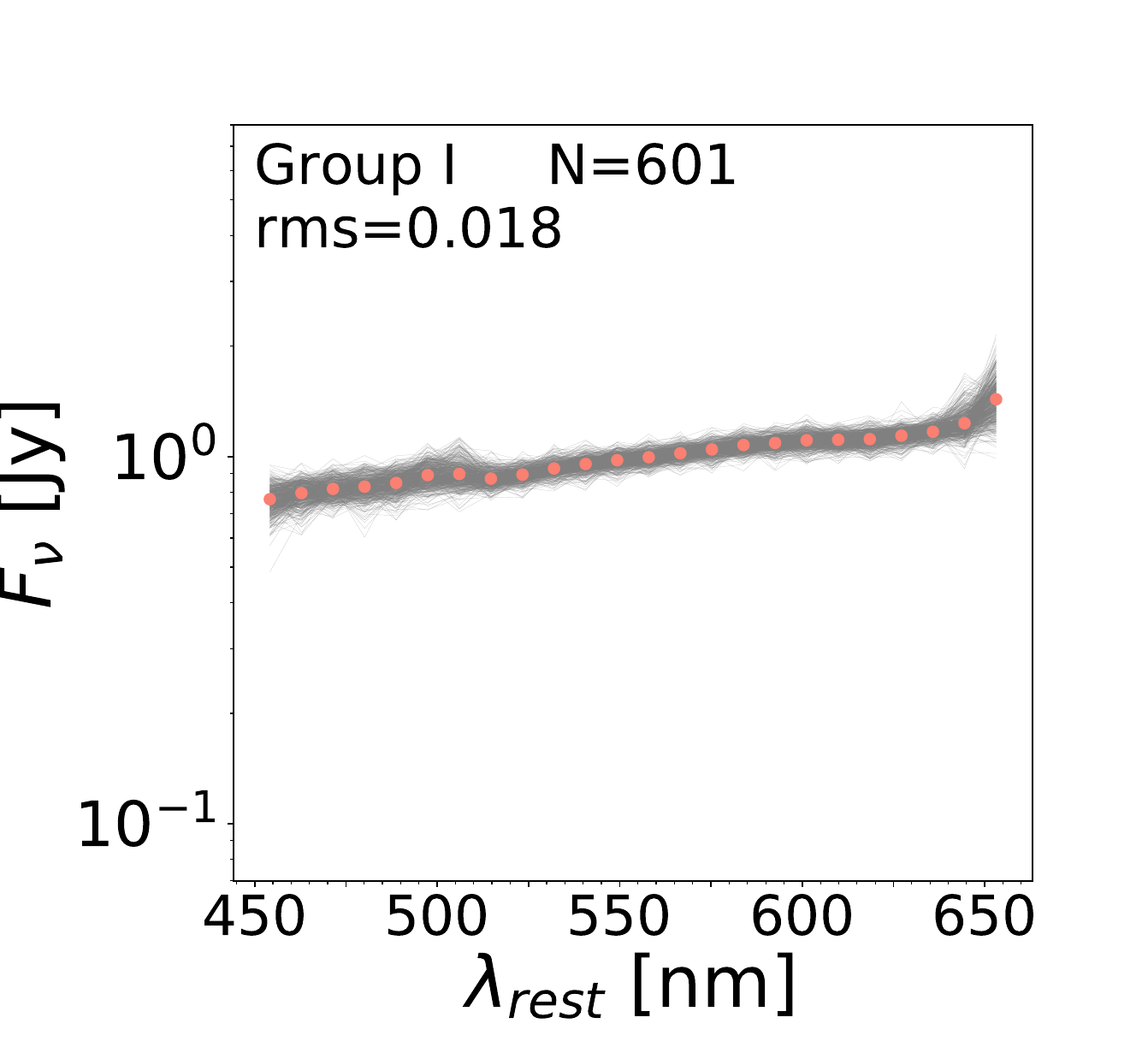} &   
 \hspace{-0.5cm}\includegraphics[width=.25\textwidth]{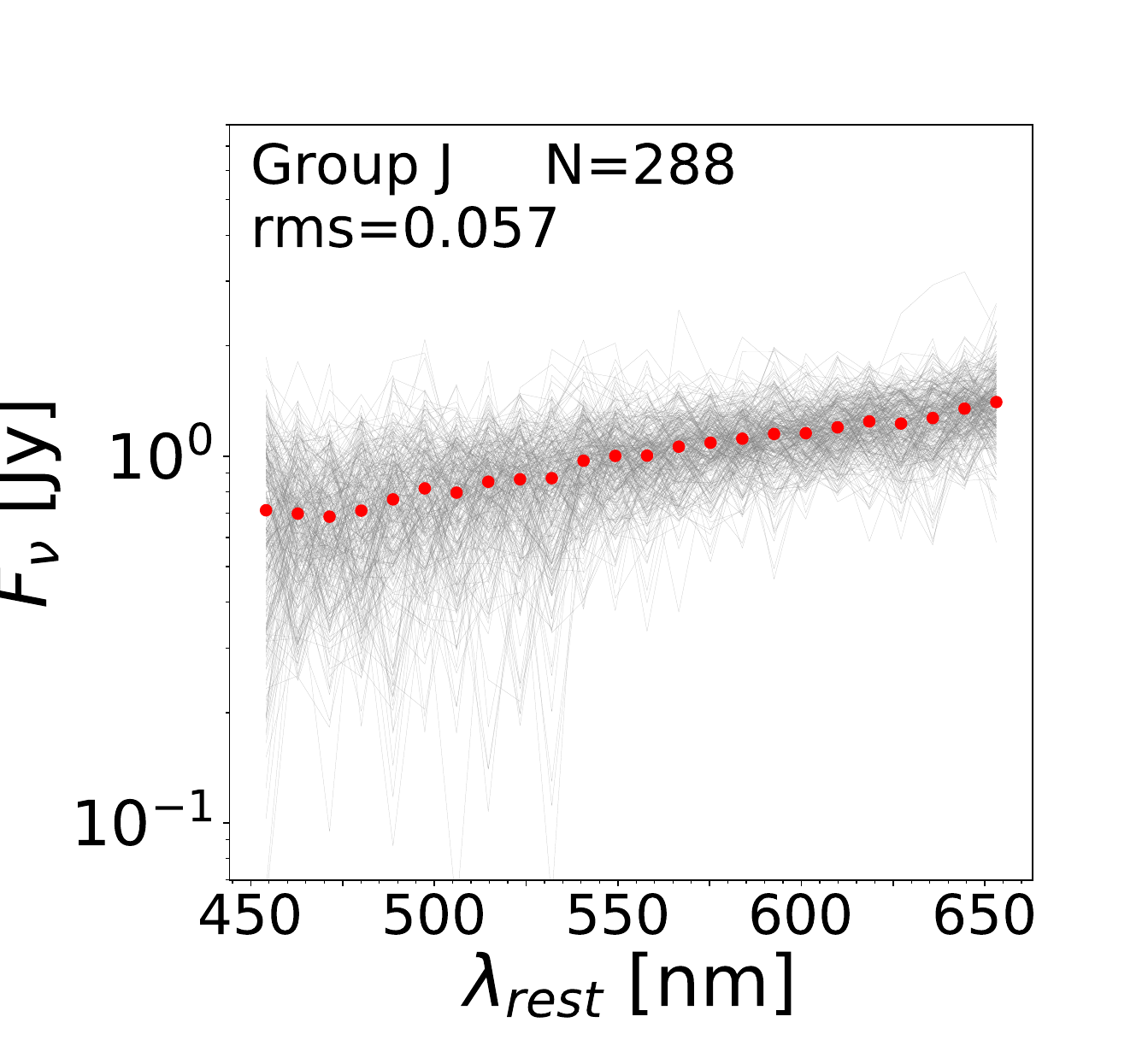}&
 \hspace{-0.5cm}\includegraphics[width=.25\textwidth]{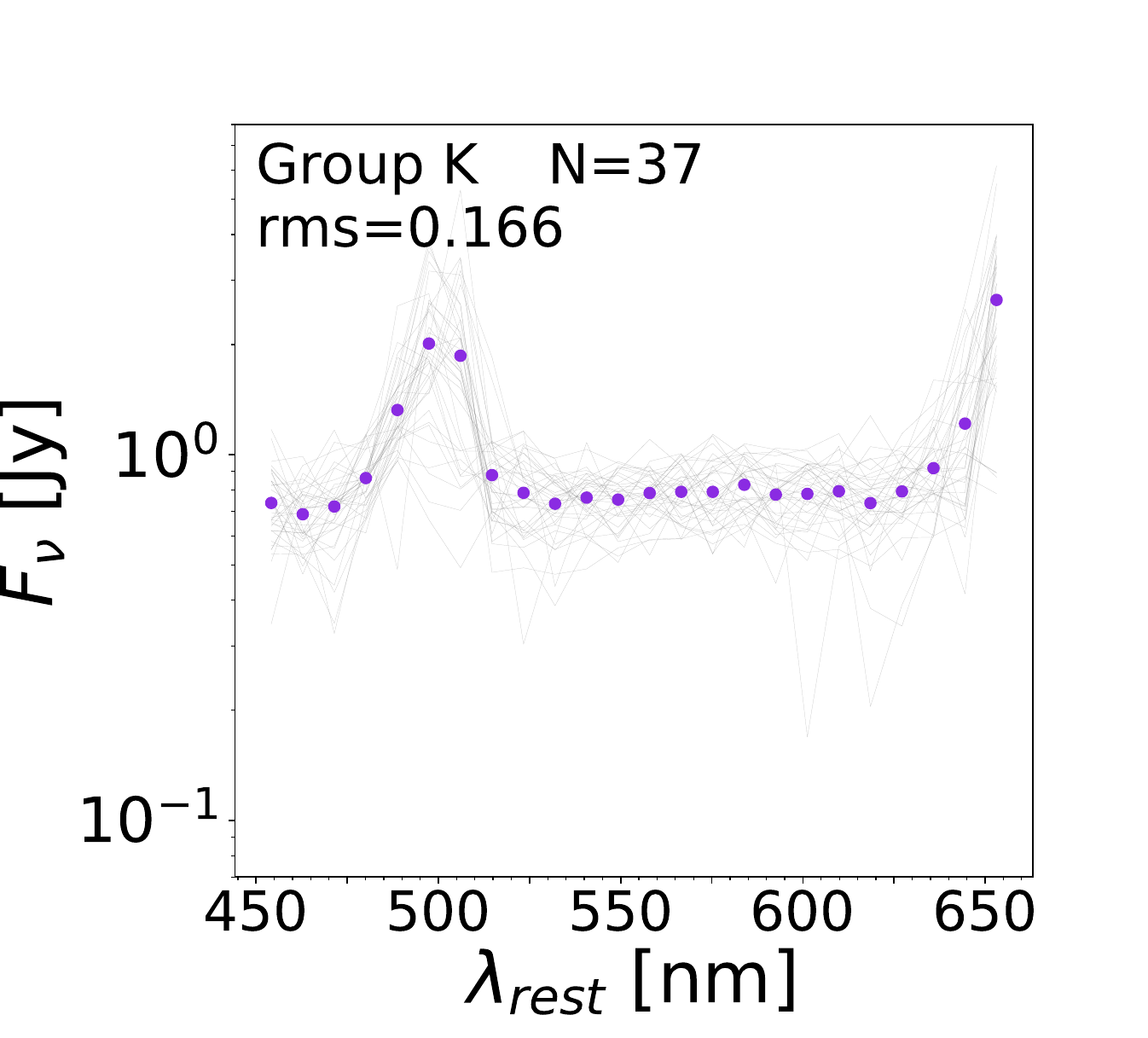} &
 \hspace{-0.5cm}\includegraphics[width=.25\textwidth]{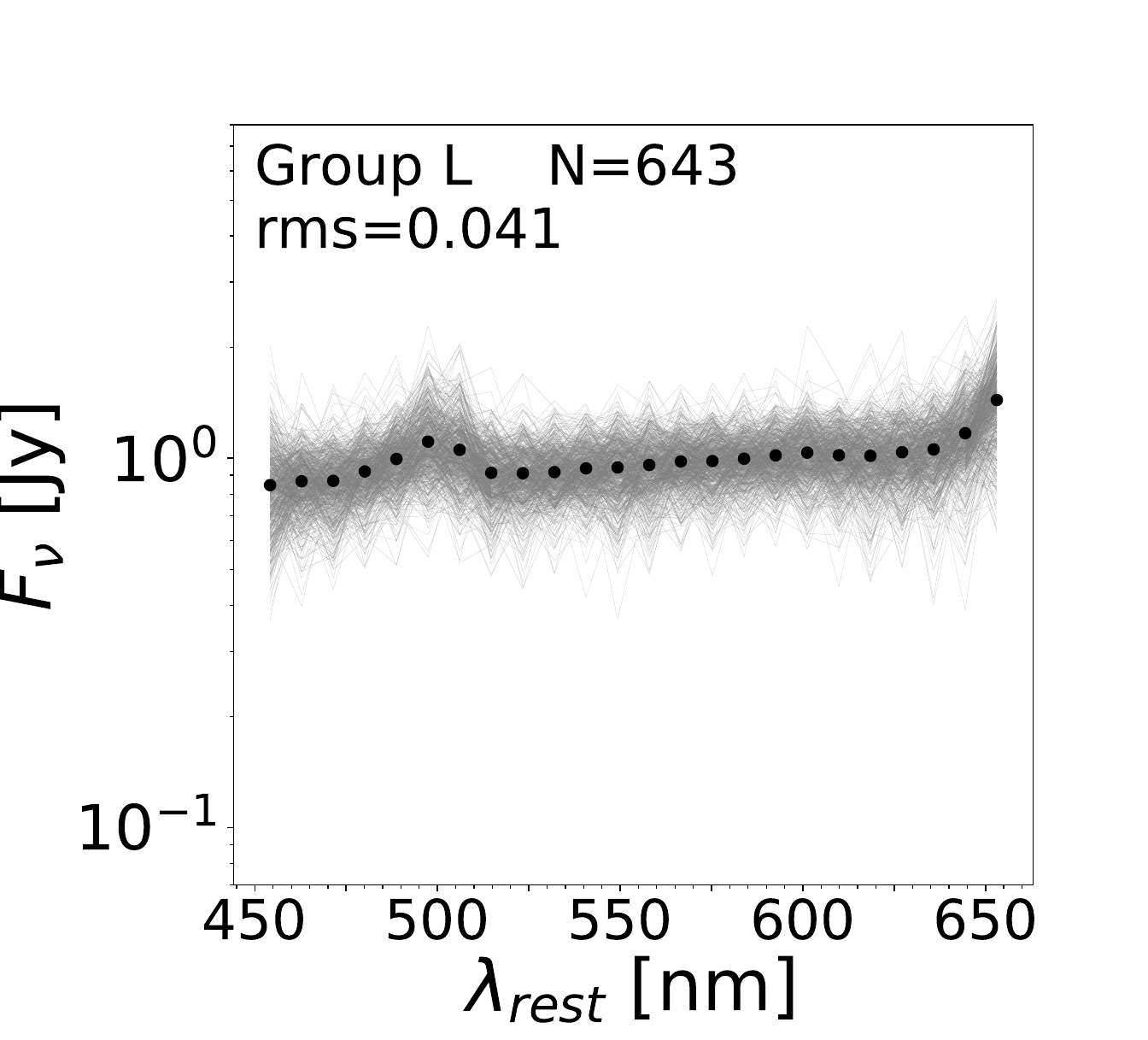}
\end{array}$
\end{center}
\caption{SEDs of the targets in the different groups resulting from the best classification derived via GM clustering. Note that the groups are not equally populated. Also, the continuum level detected in the same wavelength range changes depending on the group. 
Some groups represent the galaxies in a star-forming phase showing clear detection of H$\beta$, [O III]$\lambda$5007\AA\ and H$\alpha$ emission lines. The colour points represent the SEDs means color-coded by different groups.}
\label{fig:ML classes}
\end{figure*}

\begin{figure*}
\begin{center}
\includegraphics[width=2.\columnwidth]{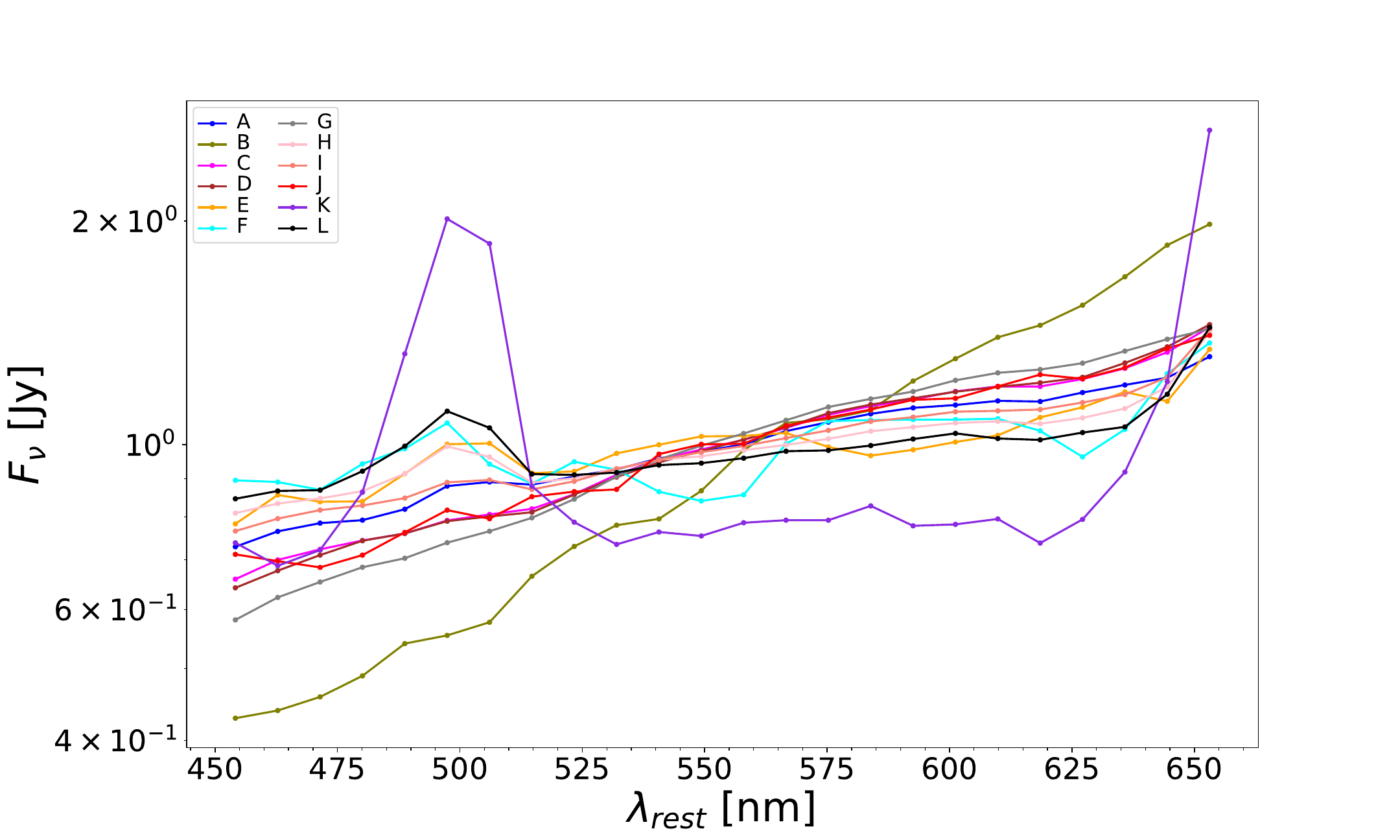}
\end{center}
\caption{Mean SEDs for classes in Fig \ref{fig:ML classes} color-coded for a clearer comparison among the different groups.}
\label{fig:means ML classes}
\end{figure*}


\begin{table}
	\small
	\centering
	\caption{Overview of the PAU filters.}
	\label{tab:PAU filters}
	\begin{tabular}{lccc} 
	\hline
Filter	 & $\lambda_{eff}$ & FWHM & $m_{5\sigma}$ \\
	 & [\AA] & [\AA] & [AB mag] \\
		\hline
NB455	&	4549.76	&	134.79 & 22.81  \\
NB465	&	4644.65	&	134.25  & 22.81  \\
NB475	&	4750.71	&	135.03  & 22.81  \\
NB485	&	4845.87	&	135.03  & 22.81  \\
NB495	&	4948.25	&	130.38  & 22.80  \\
NB505	&	5047.50	&	132.11	& 22.80  \\
NB515	&	5147.12	&	132.89	& 22.79  \\
NB525	&	5249.58	&   132.45	& 22.78  \\
NB535	&	5348.23	&   133.62	& 22.77  \\
NB545	&	5348.23	&   133.62	& 22.79  \\
NB555	&	5552.23	&   133.70	& 22.79  \\
NB565	&	5653.18	&   133.03	& 22.46  \\
NB575	&	5748.19	&   133.08	& 22.71  \\
NB585	&	5847.12	&   132.64	& 22.74  \\
NB595	&	5948.15	&   132.07	& 22.59  \\
NB605	&	6046.08	&   133.18	& 22.59  \\
NB615	&	6145.16	&   133.88	& 22.73  \\
NB625	&	6252.25	&   132.74	& 22.79  \\
NB635	&	6347.43	&   132.23	& 22.23  \\
NB645	&	6443.39	&   133.45	& 22.13  \\
NB655	&   6548.42	&   135.21	& 22.79  \\
NB665	&   6647.45	&   133.76	& 22.80  \\
NB675	&   6748.52	&   134.67	& 22.78  \\
NB685	&   6847.70	&   134.44	& 22.75  \\
NB695	&   6948.68	&   137.05	& 22.65  \\
NB705	&   7050.10	&   134.53	& 22.77  \\
NB715	&   7146.52	&   133.68	& 22.80  \\
NB725	&   7254.03	&   135.80	& 22.69  \\
NB735	&   7354.69	&   136.38	& 22.60  \\
NB745	&   7453.55	&   133.26	& 22.65  \\
NB755	&   7547.60	&   118.09	& 22.52  \\
NB765	&   7657.67	&   118.09	& 22.42  \\
NB775	&   7750.97	&   135.02	& 22.55  \\
NB785	&   7849.11	&   132.19	& 22.42  \\
NB795	&   7949.79	&   134.19	& 22.54  \\
NB805	&   8053.69	&   135.51	& 22.51  \\
NB815	&   8146.03	&   133.54	& 22.49  \\
NB825	&   8259.04	&   132.36	& 22.28  \\
NB835	&   8358.30	&   132.65	& 22.30  \\
NB845	&   8454.65	&   131.47	& 22.39  \\
		\hline
		\multicolumn{4}{l}{Data taken from Mart\'i et al. 2014 and}\\
		\multicolumn{4}{l}{https://pausurvey.org/paucam/filters/}\\
	\end{tabular}
\end{table}


\section{Machine Learning classification}
\label{sec:ML classification}

With the goal of identifying differences in the rest-frame SED of PAU galaxies, we perform an unsupervised clustering using Gaussian Mixture \citep[GM;][]{Duda1973} models. This algorithm presents some advantages over the more classical $k$-means, in which each object can exclusively belong to a single group, depending on the distance to the center of the said group. On the contrary, the GM method assigns probabilities of belonging to different groups. In this way, objects in inter-group regions can belong to one or more groups with similar probabilities, defining transitions between those. Finally, the GM method is also more flexible at the time of defining the covariance of the groups, while that is constrained to be spherical in the $k$-means models. That means that the variance of the different components defining a group must be the same in the $k$-means models, while the components of the GM models are allowed to have different variances, which in turn translates into a more accurate definition of the groups.

\subsection{Input SED sample}

To automatically search for differences in the shape of the SEDs, which trace physical properties between groups, we have to take into account some considerations before performing the ML classification.
First, we must work with rest-frame SEDs. Otherwise, the differences introduced in the observed SEDs by the variety of redshifts present in the sample would interfere in the clustering process, making the redshift become one of the factors driving our classification instead of inner physical properties of the galaxies.

The rest-frame SEDs are then interpolated to a common wavelength grid (24 steps in a range of $454.15 < \lambda_{rest}$/nm $< 653.11$) to be used as the homogeneous input information for our GM model. Once in a common wavelength frame, the SEDs are also normalized to their mean fluxes to avoid getting GM classes purely defined by the bolometric luminosity of the galaxies.

The redshift range selected for this work ($0.01 <$ z $< 0.28$) allows us to cover $\ha$, $\hb$ and [$\Oiii$]$\lambda5007$\AA\ in all the SEDs simultaneously within the spectral coverage of our dataset. These lines can be well identified in the SEDs thanks to the good spectral resolution provided by the PAU spectrophotometry. Note that the redshift interval employed cannot be wider or we would lose homogeneity in the rest-frame SEDs sampled by the PAU filters at higher z. In total, 6,061 objects in the redshift range 0.01 $<$ z $<$ 0.28 are detected in the 40 PAU narrow-band filters, from which 801 were discarded as they are labeled as stars in the PAU's catalogue. 26 other objects were discarded in the common wavelength grid creation. This is because in the interpolation process, if the points fall outside the common wavelength grid, an extrapolation was applied. So, a negative value was obtained for the flux corresponding to the first wavelength for these 26 objects. After this, the final number of galaxies at 0.01 $<$ z $<$ 0.28 in our input sample is 5,234.

\subsection{GM clustering description}
\label{sec:GM_description}

The GM algorithm used in this work models the SEDs as a mixture of d-dimensional Gaussian distributions. The corresponding d-variate Gaussian probability density distribution can be expressed as:

\begin{equation}
    \label{eq:Gauss_xd}
    G(X|\mu,\Sigma)=\frac{\exp[-\frac{1}{2}(X-\mu)^{T}\Sigma^{-1}(X-\mu)]}{\sqrt{2\pi|\Sigma|}},
\end{equation}

\noindent where $\mu$ represents a d-dimensional mean vector and $\Sigma$ is the covariance matrix.

Considering $K$ different groups within the sample, we can define $\pi_{k}$ as the mixing coefficient of the $k$-th Gaussian distribution, associated to the probability of observing a data point from the $k$-th Gaussian distribution. The combination of all $K$ distributions results in the total probability density function:

\begin{equation}
    \label{eq:pd_lin_comb}
p(X)=\sum_{k=1}^{K}\pi_{k}G(X|\mu_{k},\Sigma_{k}).
\end{equation}

The means, covariances and mixing coefficients are estimated maximizing the log-likelihood of $p(X)$ making use of the Expectation-Maximization (EM) algorithm \citep{Dempster1977}, an iterative way of finding \mbox{maximum-likelihood} solutions for incomplete data or data with hidden variables. Given that the EM algorithm employed in the GM models can sometimes provide local optima solutions, the clustering process is executed a hundred times for each number of groups desired in the final configuration, employing different initial random seeds each time as well as different kinds of covariances.

\subsection{Best GM model selection}
\label{sec:best_model_select}

The optimal number of groups in the most common unsupervised clustering methods is usually determined by the minimum of the Bayesian Information Criterion \citep[BIC,][]{Schwarz1978} trend with the number of groups, where BIC is derived from Bayesian statistics and penalises the likelihood in terms of $q$, the number of parameters of the model used and m, the number of independent data points available, as:

\begin{equation}
\label{eq:BIC}
BIC=\chi^2 + q\ln(m),
\end{equation}

However, it is possible that the BIC trend with the number of groups does not converge to a clear minimum but instead continuously decreases with the number of groups. Fig.~\ref{fig:BIC-ncomp} shows the just mentioned behavior for the BIC distribution obtained in our SEDs GM clustering. In this situation, the best number of groups can be estimated from the BIC gradient by identifying the number of groups up to which an increase in the amount of groups does not translate into a substantial BIC reduction. In fact, for this purpose, we evaluated this gradient using a BIC significance, $\sigma_{BIC}$, analysis, as:

\begin{equation}
\label{eq:sigma_BIC}
\sigma_{BIC}=\frac{\overline{BIC}_{n_i}-\overline{BIC}_{n_{i+1}}}{\epsilon_{BIC_{i+1}}},
\end{equation}

\noindent where $\overline{BIC}_{n_i}$ and $\overline{BIC}_{n_{i+1}}$ are the mean of the BIC parameters given by the combinations in the $i$-th and ($i+1$)-th numbers of groups, respectively; and $\epsilon_{BIC_{i+1}}$ is the standard deviation of the BIC parameters in the ($i+1$)-th number of groups. Notice here that, as mentioned in \S\ref{sec:GM_description}, a hundred different solutions from different initial random seeds are computed for each number of groups.

The BIC gradient presented in Fig.~\ref{fig:DeltaBIC-ncomp_zoom} shows that most classifications with a number of groups higher than 10 are within 1$\sigma$, although a well defined decreasing-increasing $\sigma_{BIC}$ occurs between the 12- and 13-classes solutions. Increasing the amount of groups beyond that does not result in a significant improvement. Based on this criterion, we select 12 as the best number of groups for our SED clustering.

Among all the different solutions obtained for 12-groups classification with our SED sample, we pick that with the highest silhouette score \citep{Rousseeuw1987}. This coefficient is defined as: 

\begin{equation}
    \label{eq:sil_point}
    s(x_{i})=\begin{cases} \frac{b(x_{i})-a(x_{i})}{\max\{a(x_{i}),b(x_{i})\}} & \mbox{if } |C_{k}|>1\\
    0 & \mbox{if } |C_{k}|=1. \end{cases}
\end{equation}

\noindent where $a(x_{i})$ is the mean distance between a given SED $x_{i}$ belonging to the group $C_{k}$ (with $|C_{k}|$ members) and the rest of SEDs in its same group, while $b(x_{i})$ is the minimum distance between said SED and all the other group. In this way, the higher the silhouette coefficient, the better defined the SED is within its group. This indicator can be extended to a mean silhouette score of each individual group or of the complete sample, providing information of how differentiated the groups are in the overall classification. Following this criterion, the model with the highest silhouette score among those resulting in 12 groups is selected as the best unsupervised classification of the SED sample.

The SEDs of the targets separated in the different groups from the selected clustering model are shown in Fig. \ref{fig:ML classes}, as well as the number of elements in each group. In order to visualize a clearer comparison among the different groups, Fig. \ref{fig:means ML classes} shows the mean SEDs for classes in Fig \ref{fig:ML classes}. A first glance to the SEDs groups reveals obvious differences in size and physical properties such as continuum pattern or presence/absence of emission lines. Four groups (B, C, D, and G) do not show emission lines, and some groups (B, C, D, E, F, and J) present absorption lines such as Magnesium (Mg $\lambda$5175\AA) and Sodium (Na I $\lambda\lambda$5889, 5895 \AA\AA) which are characteristic of elliptical galaxies. Four groups (A, E, F, and J) have a large scatter with coincidence factor (see below) minimal $<$ 43\% and rms $>$ 0.06. Five groups (A, H, I, K, and L) present a clear detection in emission lines such as $\hb$, [$\Oiii$]$\lambda$5007\AA\, and $\ha$. In particular, group K hosts a small amount of objects with intense emission lines.

To further evaluate the robustness of the different classes in our fiducial clustering model, we compute the coincidence factor as defined in \citet{Sanchez-Almeida2010}. This parameter is calculated as the fraction of elements in each group that remain in the same group under a different 12-groups classification (i.e. using a different initial random seed of the GM model). According to this parameter, those groups with higher coincidence factor can be considered as more resolved for the clustering algorithm, while groups with lower coincidence factor could be more subtle and/or poorly defined.

\begin{figure}
\hspace*{-0.5cm}
	\includegraphics[width=1.\columnwidth]{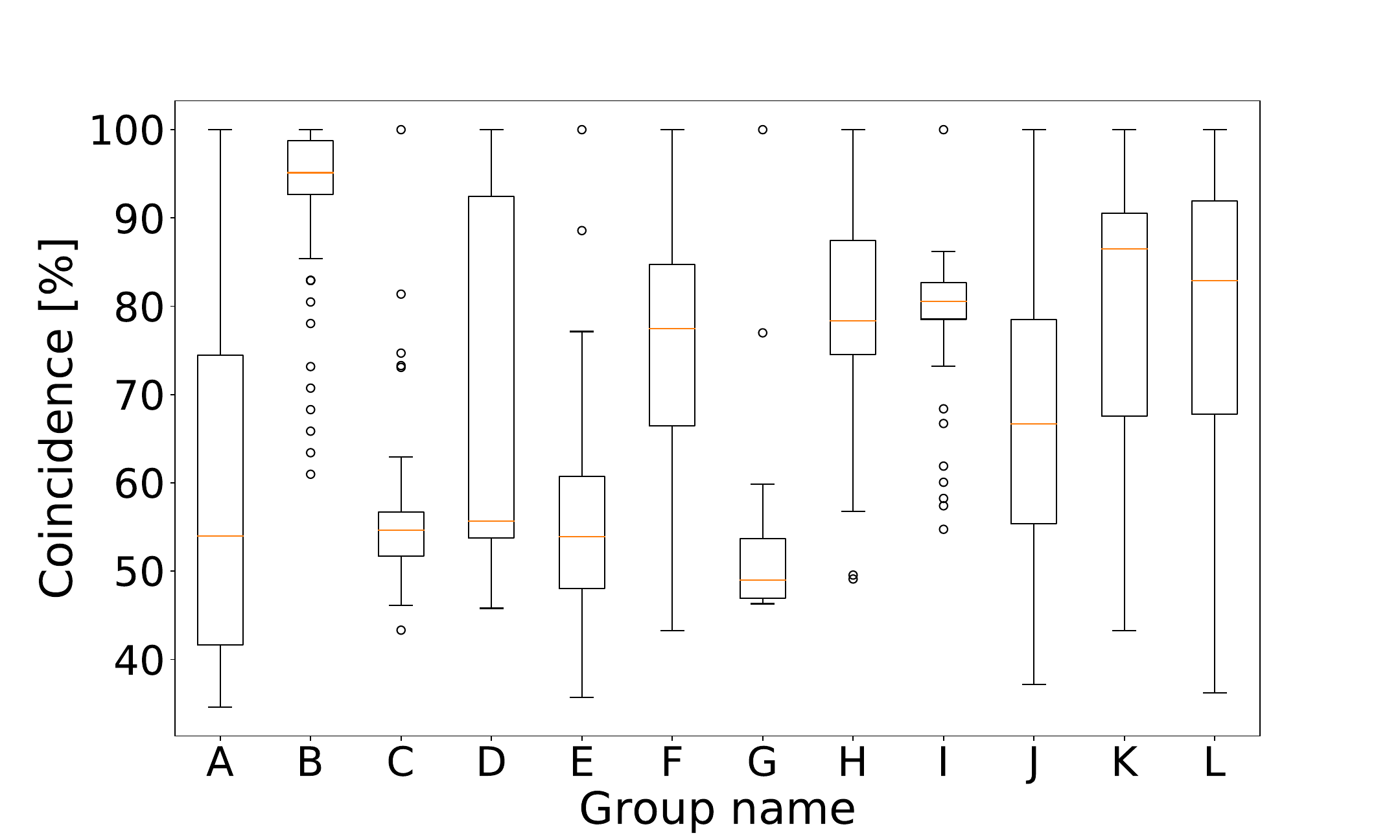}
    \caption{Coincidence factor vs Group. This represents the percentage of the number of targets that fall in the same class in each group for 100 independent GM runs. The horizontal markers within each rectangle indicate the median value of the coincidence factor. The quartiles of the coincidence factor are shown with rectangles, while the complete error bars show the entire range of the coincidence factor. Outliers are determined in terms of the interquartile range and are marked by empty circles.}
    \label{fig:coincidence-Group}
\end{figure}

The coincidence factors for the 12 groups under 100 different GM classifications (see Fig.~\ref{fig:coincidence-Group}) suggest that groups A, E, J and L are the worst classified (if we consider the minimal value of the coincidence factor), which in turn relates with these groups presenting the largest inner SEDs scatter (see Fig.~\ref{fig:ML classes}). The statistical properties of the targets within each group classified using ML are shown in table \ref{tab:properties ML}.

\begin{table}
	\small
	\centering
	\caption{Statistical properties of the targets within each group classified using Machine Learning.}
	\label{tab:properties ML}
		\begin{tabular}{lcccc} 
			\hline
Name & N & rms$^a$ & Coinc. min$^b$ & Coinc. median$^c$ \\
		\hline
Group A	& 630 &	0.033	& 34\% &54\% \\
\\
Group B	& 41 &	0.065	& 61\% & 95\% \\
\\
Group C	& 494 &	0.014	& 43\% & 55\% \\
\\
Group D	&	476	&	0.021	&	46\% & 56\% \\
\\
Group E	&	280 &	0.084	&	35\% & 54\% \\
\\
Group F	&	164	&	0.074	&	43\% & 78\% \\
\\
Group G	&	678	&	0.011	&	46\% & 49\% \\
\\
Group H	&	902 &	0.025	&	49\% & 78\% \\
\\
Group I	&	601	&	0.018	&	55\% & 80\% \\
\\
Group J	&	288	&	0.057	&	37\% & 67\% \\
\\
Group K	&	37	&	0.166	&	43\% & 87\% \\
\\
Group L	&	643	&	0.041	&	36\% & 83\% \\
		\hline
        \multicolumn{5}{l}{$^a$ The rms is derived taking into account the difference}\\
        \multicolumn{5}{l}{between the SED mean and all SED. The standard}\\
        \multicolumn{5}{l}{deviation of this distribution is considered the rms.}\\
		\multicolumn{5}{l}{$^b$ Minimal value of the coincidence factor. This means}\\
		\multicolumn{5}{l}{that at least this percent of targets fall in the same}\\
		\multicolumn{5}{l}{classification when the code is run 100 times.}\\
        \multicolumn{5}{l}{$^c$ Median value of the coincidence factor.}
		\end{tabular}
\end{table}

\section{Analysis of SED fitting}
\label{sec:SED fitting}

\begin{figure*}
    \hspace*{-1cm}
	\includegraphics[width=2.4\columnwidth]{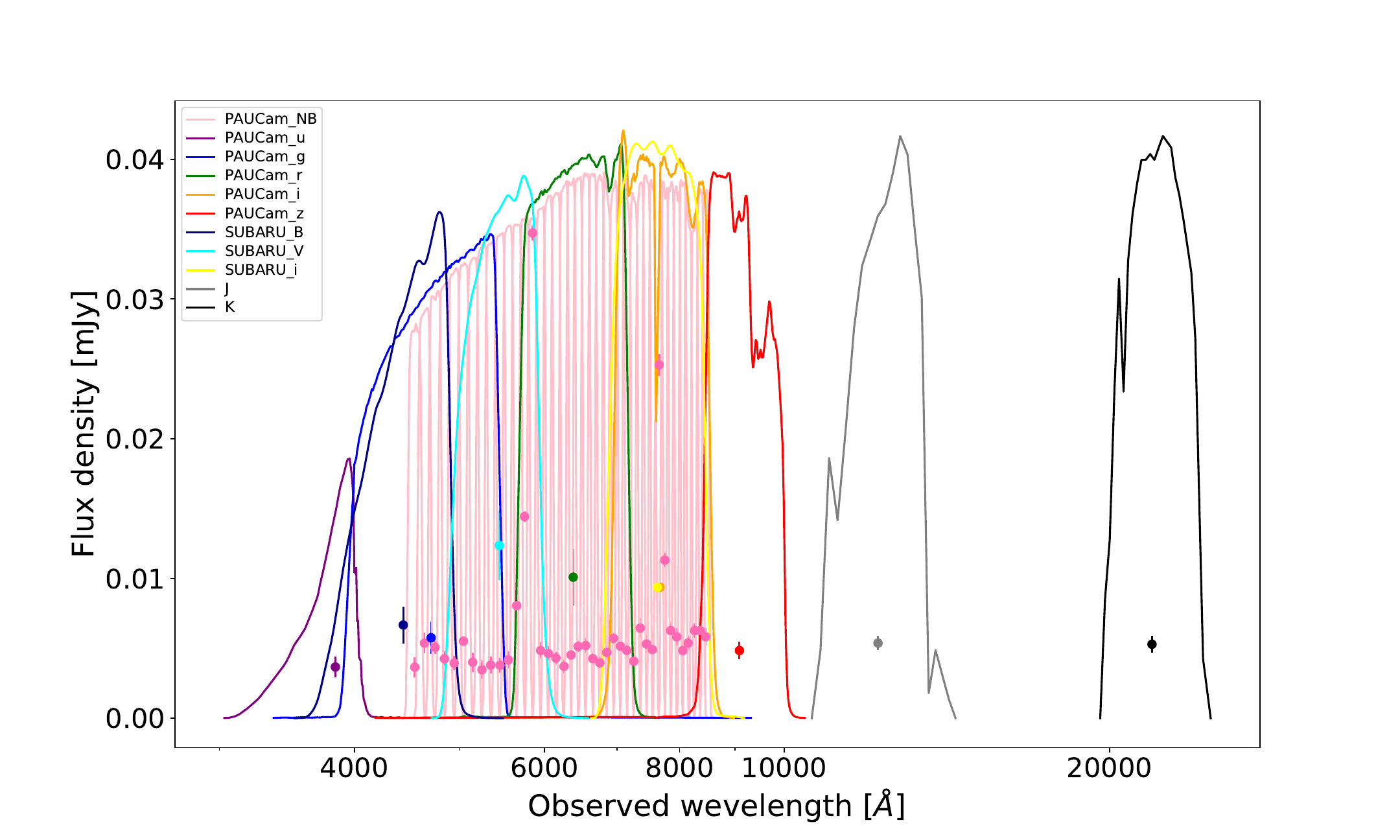}
    \caption{Observed SED of one target from our sample (ID49378). Pink dots are the photometric fluxes obtained from NB filters and the rest dots are the photometric fluxes obtained from BB filters. The transmission profile of filters used to get the fluxes are overplotted.}
    \label{fig:Entire data set}
\end{figure*}

The analysis of stellar populations of the galaxies was made by fitting the SED to each single galaxy. However, modeling the SED of galaxies is not so easy as galaxies with different properties can have broadly similar SEDs. This is particularly the case when the SED wavelength ranges are too short as is the case of the SED obtained with PAU data. To circumvent this, we extend the wavelength coverage adding 10 broad bands (BB) filters to the 40 NB SED. Note that, these BB filters are not considered in the ML classification. Fig. \ref{fig:Entire data set}  shows an example of the complete SED to be used in this section to fit the stellar population models. In the figure, the shape of the filters used for each data point is also drawn. The points correspond to the photometric fluxes for one target, ID49378, belonging to group K (see Fig. \ref{fig:ML classes}).

\begin{table}
	\centering
	\caption{Selected parameters values for model to analyze the SEDs.}
	\label{tab:CIGALE parameters}
	\begin{tabular}{lccc} 
	\hline
Parameter	 & Min & Max & N\\
		\hline
$\tau_{0}^1$ [Myr]	&	50	&	30000 & 11  \\
\\
$f_{burst}^2$	&	0.0	&	0.2  & 9 \\
\\
$t_{0}^3$ [Myr]	&	10	&	12000 & 10 \\
\\
Z$^4$ [Z$_{\odot}$]	&	0.0001	&	0.05	& 6  \\
\\
log U$^5$	&	-2	&	-2	& 1  \\
\\
Av$_{ISM}^6$ [mag] 	&	0.0	& 4.0 & 21  \\
		\hline
		\multicolumn{4}{l}{$^1$ e-folding time of the main stellar population model.}\\
		\multicolumn{4}{l}{$^2$ The burst strength, $f_{burst}$, is defined as the fraction}\\
		\multicolumn{4}{l}{of stars formed in the second burst relative to the}\\
		\multicolumn{4}{l}{total mass of stars ever formed.}\\
		\multicolumn{4}{l}{$^3$ Age of the main stellar population in the galaxy.}\\
        \multicolumn{4}{l}{$^4$ Metallicity.}\\
        \multicolumn{4}{l}{$^5$ Ionization Parameter.}\\
        \multicolumn{4}{l}{$^6$ V-band attenuation in the interstellar medium.}
	\end{tabular}
\end{table}

The stellar population modeling was performed using the Code Investigating GALaxy Emission \citep[CIGALE;][]{Noll2009, Boquien2019}. CIGALE has already been successfully applied to PAU NB data before to derive rest-frame colours and luminosities \citep{Johnston2021} also stellar masses and sSFR \citep{Tortorelli2021} as well as the D4000 spectral break index \citep{Renard2022}.

CIGALE builds grids of models based on stellar spectra from the Star Formation History (SFH) and stellar population models. It includes as well models for the nebular emission (lines and continuum), the attenuation of the stellar and nebular emission assuming an attenuation law, dust emission, and emission of an active nucleus. The resulting grid of models is fitted to the photometric data, and the galaxy properties are estimated analyzing the posterior likelihood distribution, producing a best-fit model, and a Bayesian estimate for each parameter.

In CIGALE it is possible to use a double exponential SFH consisting of a first decaying exponential corresponding to the long-term star formation responsible for the bulk of stellar mass, plus a second exponential that models recent starbursts \citep[e.g.][]{Papovich2001, Perez-Gonzalez2003, Rodriguez-Espinosa2014, Grazian2015, Lumbreras-Calle2019, Arrabal-Haro2020}.

For SED modeling, we use \cite{Bruzual-Charlot2003} stellar population models with a Salpeter Initial Mass Function \citep[IMF;][]{Salpeter1955}, as well as a modified \cite{Charlot_Fall2000} attenuation law for dust extinction. The dust emission templates are from \cite{Dale2014}. Besides, the models were constructed by varying the interstellar medium (ISM) properties of metallicity (Z), the V-band attenuation in the ISM (Av$_{ISM}$), the ionization parameter (log U) for nebular emission, the e-folding time of the main stellar population model ($\tau_0$), the age of the main stellar population in the galaxy ($t_0$), and the burst parameter ($f_{burst}=0$) over a wide parameter space as indicated in Table \ref{tab:CIGALE parameters}. The $f_{burst}=0$ parameter accounts for the relative mass of the young burst with respect to the old population or main population, because it does not necessarily have to be old. When $f_{burst}=0$, a single population is considered.

\begin{figure*}
\begin{center}$
\begin{array}{cc}
   {\includegraphics[width=.5\textwidth]{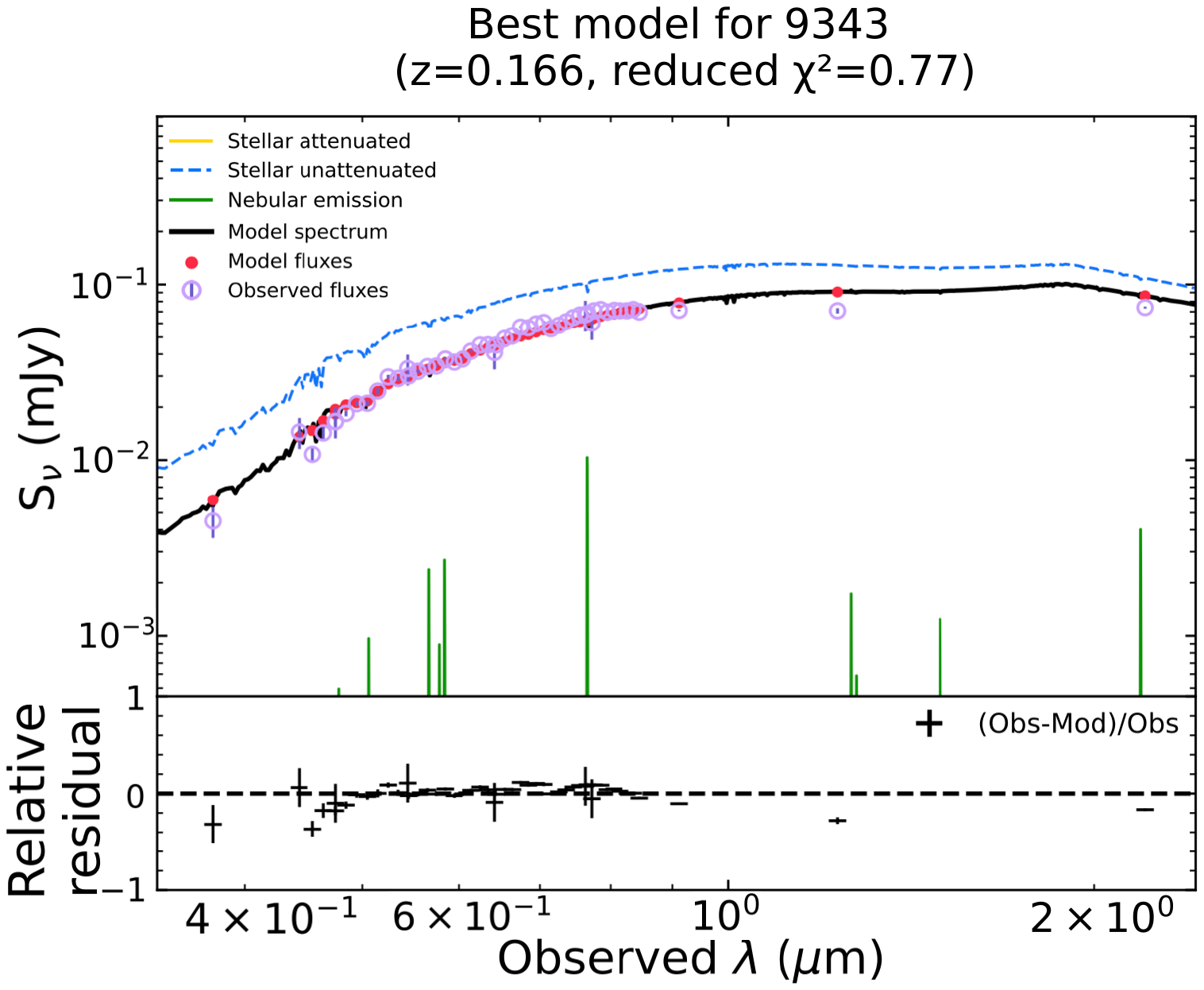}} &
   {\includegraphics[width=.5\textwidth]{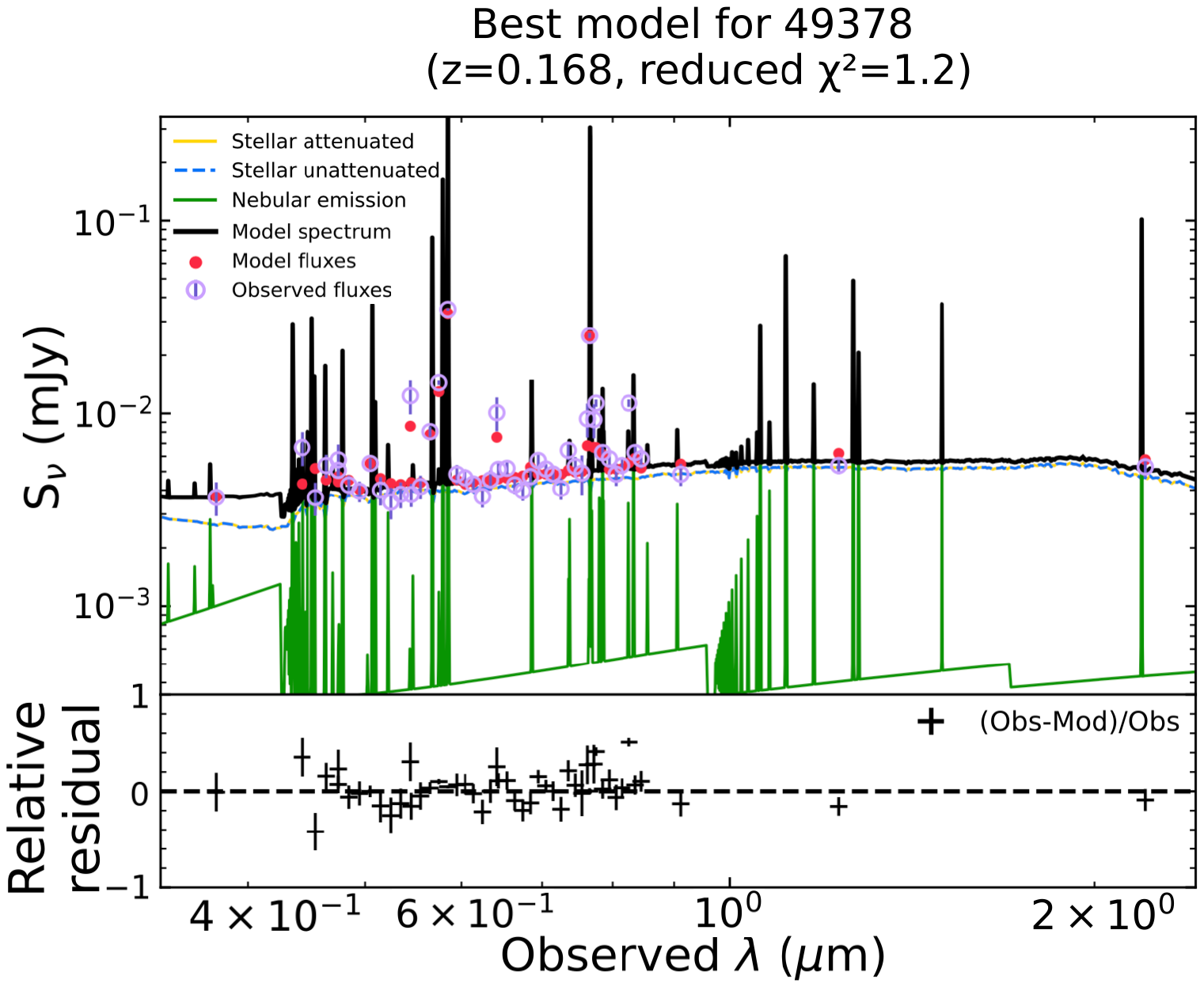}}
\end{array}$
\end{center}
\caption{Example of SED fittings using CIGALE for two targets. The purple circles are the photometric fluxes used in the fit, while the black line represents the best model. The dust free and attenuated stellar emission are also represented by the dashed blue line and yellow line, respectively. The residuals between the observed and model fluxes are shown on the bottom panels.}
\label{fig:sed fitting}
\end{figure*}

An example of the SEDs fitting for two targets belonging to different ML classified groups is shown in Fig. \ref{fig:sed fitting}. The target on the left, ID 9343 belongs to group G, with no strong nebular emission. The panel on the right shows ID 49378, a target belonging to group K, which has a significant nebular emission. The purple circles represent the observed fluxes and the black line represents the best model spectrum. The residuals between the observed and model fluxes are shown on the bottom panel. From the physical properties obtained using CIGALE we estimated the sSFR=SFR/M$_{\star}$ of log(sSFR/yr$^{-1}$)=$-13.94\pm1.90$ and $-8.86\pm0.24$ for ID9343 and ID 49378, respectively. 
Table \ref{tab:Physical properties} shows the mean physical properties for the targets belonging to each group. These were derived fitting CIGALE to each individual object. As examples, for a very well populated class (G group), with no strong emission lines observed, we see that their SEDs fitting give that this group is the most massive (log (M$_{\star}$/M$_{\odot}$) $\sim$ 10.5), oldest (age $\sim$ 10 Gyr), and with the lowest sSFR (log(sSFR/yr$^{-1}$) $\sim$ -12.4) while group K is the least massive (log (M$_{\star}$/M$_{\odot}$) $\sim$ 8), youngest (age $\sim$ 1 Gyr), and with the highest sSFR (log(sSFR/yr$^{-1}$) $\sim$ -8.7). These results are in agreement with the SED shapes observed in each group (see Fig. \ref{fig:ML classes}).
 
\begin{table}
	\small
	\centering
	\caption{Physical properties of the targets within each group classified using Machine Learning.}
	\label{tab:Physical properties}
		\begin{tabular}{lccccc} 
			\hline
Name & Age & log M$_{\star}$ & $f_{burst}^{a}$ &log sSFR & N \\
& [Gyr] & M$_{\odot}$ & [10$^{-4}$]  & yr$^{-1}$ &\\
		\hline
Group A	&	3.4$^{+2.9}_{-2.0}$	&	8.78$^{+0.79}_{-0.34}$	& 5$^{+4}_{-3}$	& $-9.66\pm0.60$	&630	\\
\\
Group B	&	7.5$^{+1.5}_{-1.5}$	&	10.61$^{+0.29}_{-0.29}$	& 7$^{+3}_{-3}$	& $-10.25\pm0.36$	&41	\\
\\
Group C	&	8.5$^{+0.8}_{-3.0}$	&	10.20$^{+0.68}_{-0.23}$	&1$^{+1}_{-0}$	& $-11.19\pm0.75$ &494	\\
\\
Group D	&	8.3$^{+1.0}_{-3.2}$	&	9.70$^{+0.82}_{-0.29}$	& 1$^{+1}_{-2}$	& $-10.78\pm0.77$	&476	\\
\\
Group E	&	1.5$^{+1.9}_{-0.7}$	&	8.20$^{+0.79}_{-0.37}$	& 10$^{+10}_{-5}$	& $-9.12\pm0.43$	&280	\\
\\
Group F	&	1.4$^{+1.6}_{-0.7}$	&	8.20$^{+0.60}_{-0.34}$	& 11$^{+6}_{-4}$	&$-9.02\pm0.37$	 &164	\\
\\
Group G	&	9.5$^{+0.4}_{-0.7}$	&	10.50$^{+0.45}_{-0.26}$	& 0$^{+1}_{-0}$		& $-12.42\pm1.42$	&678	\\
\\
Group H	&	2.6$^{+2.7}_{-1.8}$	&	9.06$^{+0.53}_{-0.29}$	& 6$^{+4}_{-2}$		& $-9.48\pm0.41$	&902	\\
\\
Group I	&	4.2$^{+2.1}_{-2.6}$	&	9.62$^{+0.65}_{-0.30}$	& 4$^{+4}_{-2}$		& $-9.96\pm0.49$ &601	\\
\\
Group J	&	3.4$^{+3.3}_{-1.8}$	&	8.20$^{+1.35}_{-0.39}$	& 8$^{+7}_{-4}$		& $-9.48\pm0.64$	&288	\\
\\
Group K	&	0.9$^{+3.2}_{-0.7}$	&	8.09$^{+1.52}_{-0.39}$	& 35$^{+166}_{-30}$		& $-8.71\pm0.43$	&37	\\
\\
Group L	&	1.3$^{+1.9}_{-0.8}$	&	8.51$^{+0.55}_{-0.33}$	& 9$^{+6}_{-4}$		& $-9.10\pm0.34$	&643	\\
		\hline
		\end{tabular}
\end{table}

\section{Results and Discussion}
\label{sec:ResultsDiscussion}
We have classified the galaxies in the COSMOS field using the data in the PAU survey and an unsupervised ML clustering procedure. The algorithm classified the 5,234 galaxies in the sample into twelve groups. Each group has its particular SED with different slopes and patterns. Four groups (B, C, D, and G) do not show emission lines, and some groups (B, C, D, E, F, and J) present absorption lines such as Mg $\lambda$5175\AA\ and Na I $\lambda\lambda$5889, 5895 \AA\AA\ which are characteristic of elliptical galaxies. Four groups (A, E, F, and J) have a large scatter with coincidence factor minimal $<$ 43\% and rms $>$ 0.06. Five groups (A, H, I, K, and L) present a clear detection in emission lines such as $\hb$, [$\Oiii$]$\lambda$5007\AA\, and $\ha$. The most extreme case in the emission pattern is group K,  which includes galaxies with intense emission lines. In this sense and more specifically, 68\% of the total sample of 5,245 galaxies in the PAU survey at 0.01 $<$ z $<$ 0.28 show emission lines in their spectra. The 12 groups are well populated although the number of galaxies and scatter among galaxies shows a wide range. The details are summarized in \S \ref{sec:best_model_select}, Fig. \ref{fig:ML classes} and table \ref{tab:properties ML}. Specially, group K with very intense emission lines has the smallest number of members (37).

Once the ML classification was done, we extend the wavelength range coverage adding BB photometric data to extend the wavelength coverage and characterize the classes properties using CIGALE SED fittings.
From the analysis of physical properties obtained using CIGALE, we found that the range of age, mass, and sSFR of the galaxies are $0.15 <$ age/Gyr $<11$, $6 <$ log (M$_{\star}$/M$_{\odot}$) $< 11.26$  and $-14.67 <$ log (sSFR/yr $^{-1}$) $< -8$, respectively.\\

The stellar mass, age and sSFR are summarized in \S \ref{sec:SED fitting} and table \ref{tab:Physical properties}. Specifically, group G is the most massive, oldest, and with the lowest sSFR while group K is the least massive, youngest, and with the highest sSFR.\\

\begin{figure*}
\begin{center}$
\begin{array}{ccc}
   {\includegraphics[width=.33\textwidth]{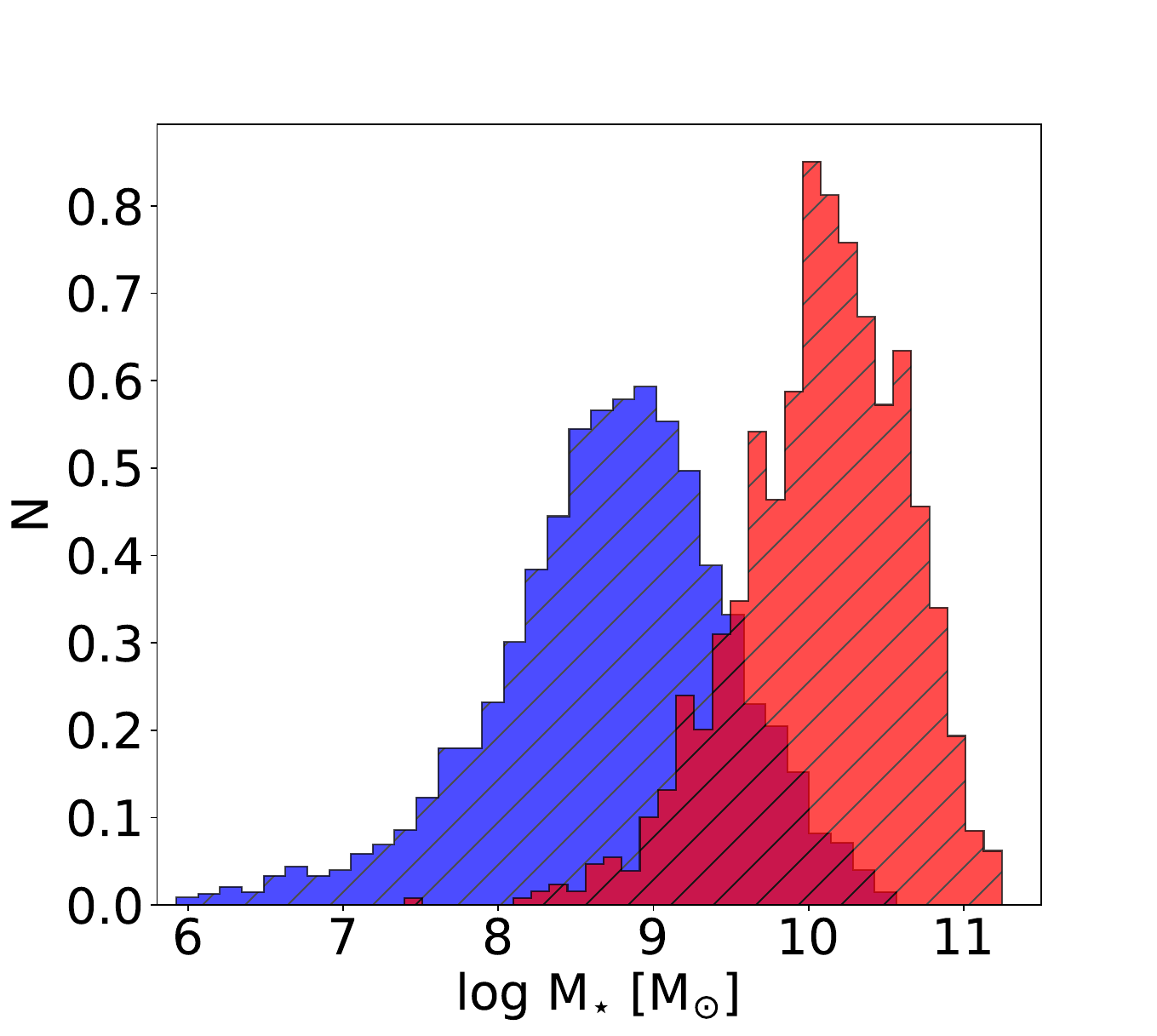}} &   
   {\includegraphics[width=.33\textwidth]{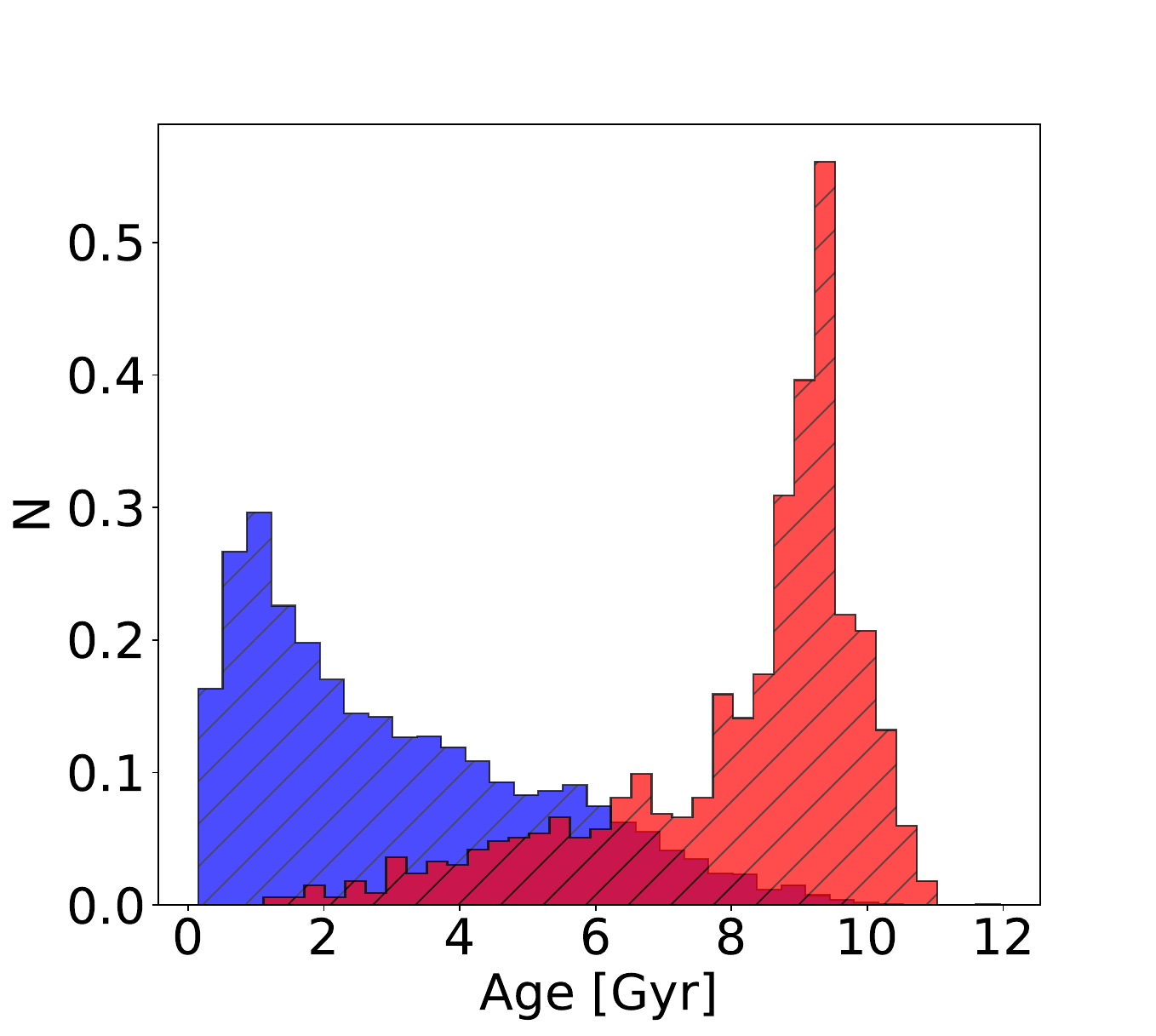}} &
   {\includegraphics[width=.33
   \textwidth]{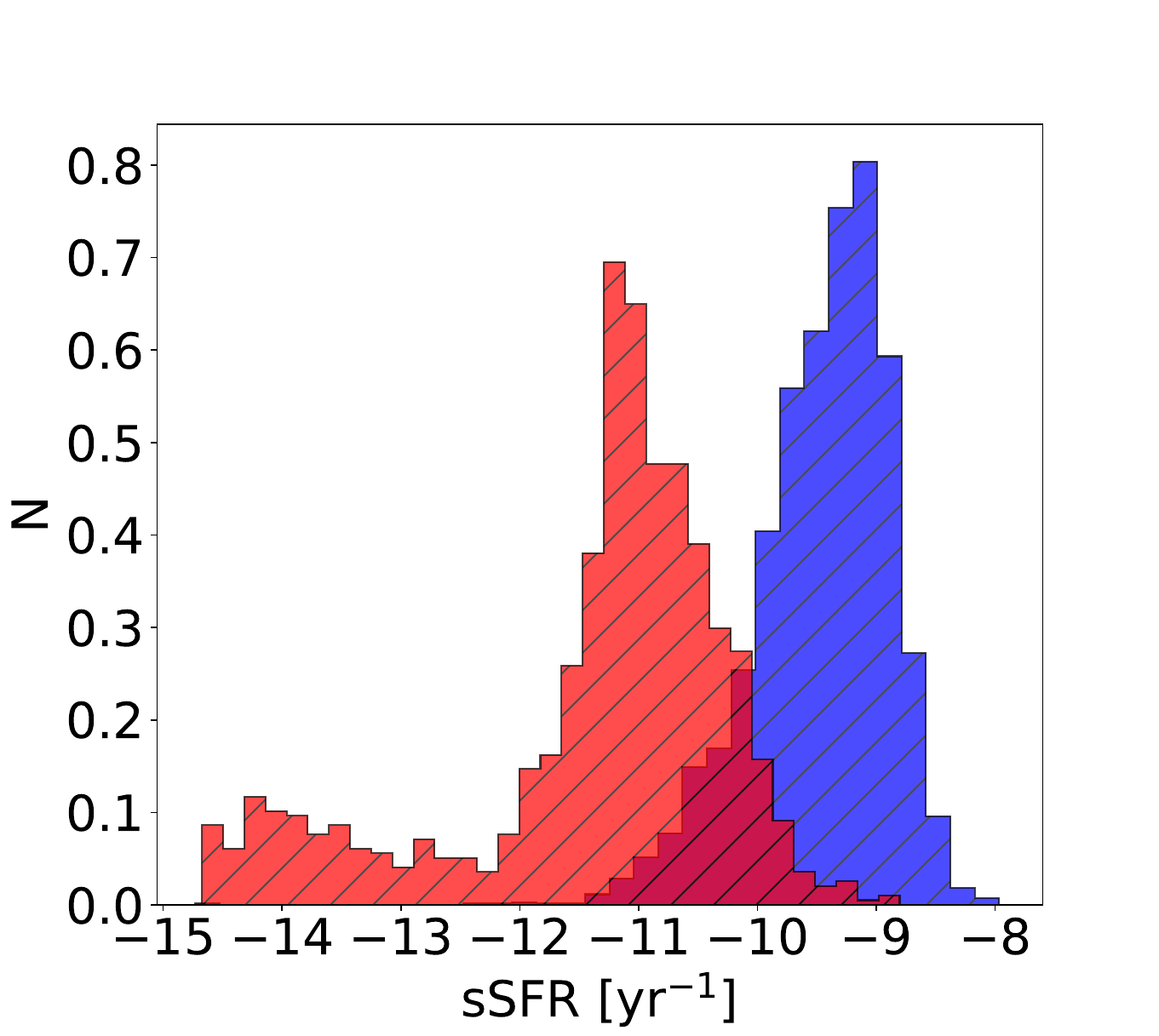}}   
\end{array}$
\end{center}
\caption{Stellar mass, age and sSFR distributions for the galaxies with emission lines (68\%) in the SEDs represented in blue colour and, for galaxies without emission lines (32\%) in the SEDs represented in red.}
 \label{fig:Hist_ELG_and_noELG}
\end{figure*}

We note that the groups that show emission lines have mean values of age $=3.02\pm2.16$ Gyr, log (M$_{\star}$/M$_{\odot}$)$= 8.72 \pm 0.75$ and, log (sSFR/yr $^{-1}$)$= -9.46 \pm 0.57$, respectively. While the groups that do not show emission lines in the SEDs have mean values of age $=8.14\pm1.94$ Gyr, log (M$_{\star}$/M$_{\odot}$)$= 10.08 \pm 0.54$ and, log (sSFR/yr $^{-1}$)$= -11.36 \pm 1.18$. The distributions are shown in Fig. \ref{fig:Hist_ELG_and_noELG}.

Our results for groups that show emission lines are in agreement inside 1$\sigma$ with those of log (M$_{\star}$/M$_{\odot}$) $\sim 8.90$ and log (sSFR/yr $^{-1}$)$=-9.52$ given by \cite{Hinojosa2016} for starburst galaxies in the COSMOS field at $0 <$ z $< 0.5$ as well as with those of log (M$_{\star}$/M$_{\odot}$) $\sim 8.50$ given by \cite{Lumbreras-Calle2019} for star-forming galaxies at z $< 0.36$ in the GOODS-N from the SHARDS survey.

In summary, the groups are well defined in their properties with galaxies with clear emission lines being in the lower mass, younger and higher sSFR regime than those with quiescent-like patterns.\\

\begin{figure}
\hspace*{-0.5cm}
	\includegraphics[width=1.\columnwidth]{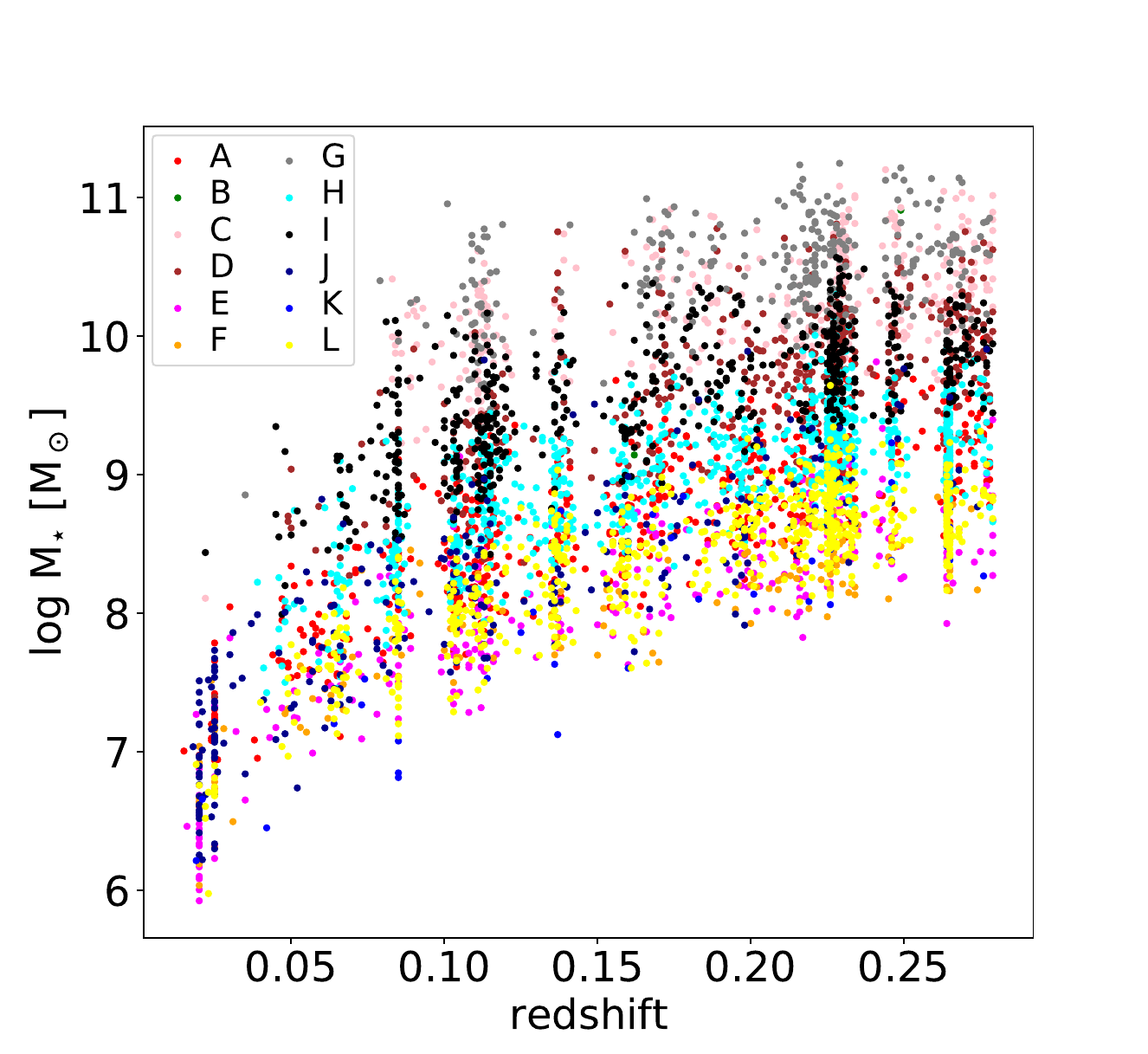}
    \caption{Stellar mass versus redshift for all the targets in each one of the 12 groups obtained by the unsupervised ML algorithm. The symbol colour represents the name of the groups. Note that the targets belonging to each group are spread across the entire redshift range.}
    \label{fig:Mstar_redshift}
\end{figure}

The resulting stellar mass versus redshift for all the targets in each group is shown in Fig. \ref {fig:Mstar_redshift}. Here it can be seen that groups spread across the entire redshift range, with targets with the lowest redshift in the low-mass zone and targets with the highest redshift located in the higher-mass region resulting from a luminosity bias. The figure shows clearly that the separation in groups is not dominated by the redshift with targets at almost all redshifts for all of the groups.

\subsection{SF main sequence}
\label{sec:SF main sequence}
\textcolor{orange}{}
The mass growth of galaxies is mainly through star formation. The more massive galaxies undergo a larger fraction of their star formation at early times whilst less massive ones are still forming stars at a high rate today. Indications of how this process takes place can be revealed through the star-forming main sequence (SFMS), a tight quasi-linear relation between stellar mass, and the star formation rate in log scale \citep{Renzini2015,Duarte-Puertas2017,Belfiore2018,Sanchez2019,Shin2021,Vilella-Rojo2021}.\\

\begin{figure}
\hspace*{-0.5cm}
	\includegraphics[width=1.\columnwidth]{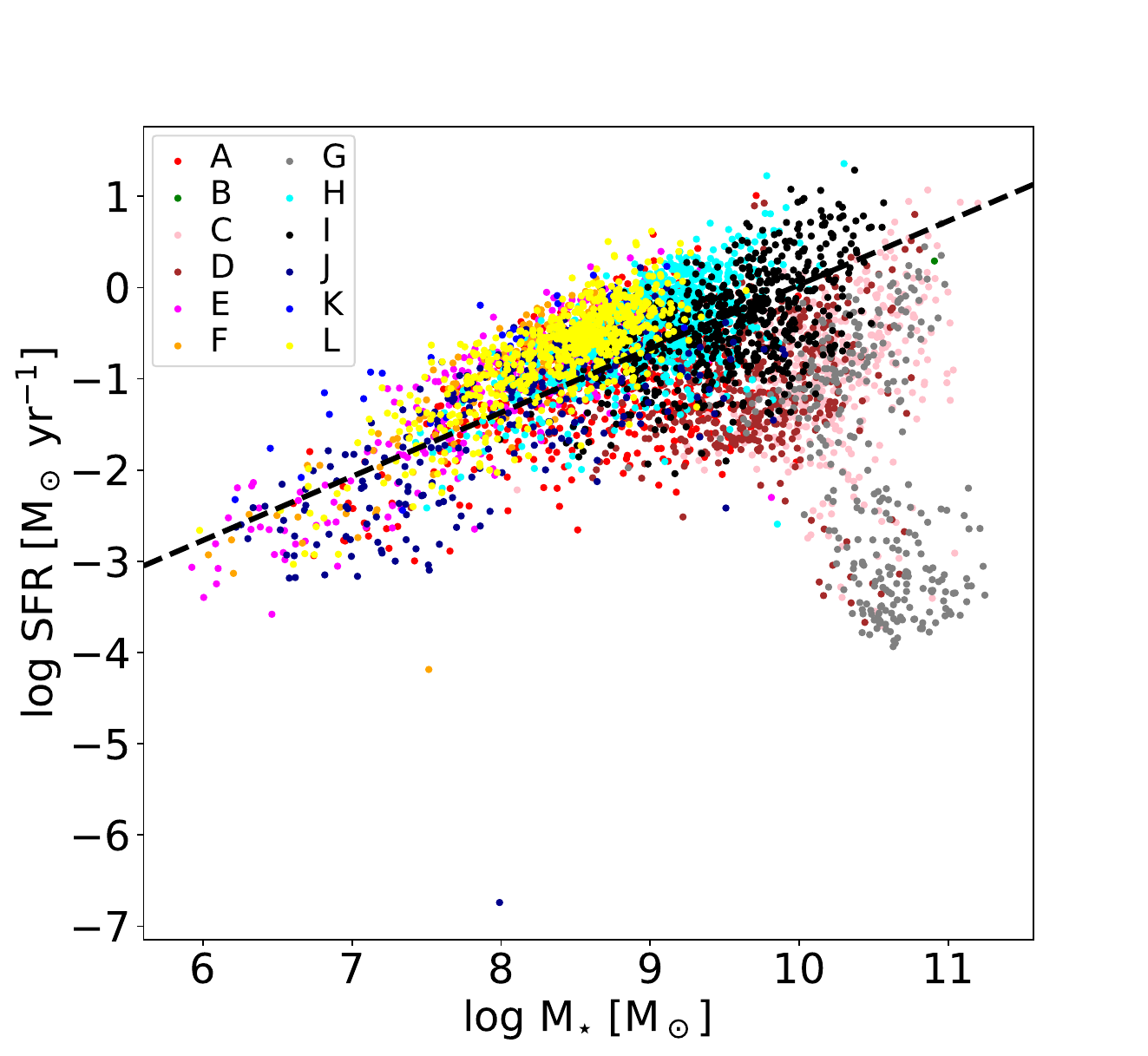}
    \caption{SFMS for all the targets in each one of the 12 groups obtained by the unsupervised ML algorithm. The symbol colour represents the same as in Fig. \ref{fig:Mstar_redshift}. The dashed line corresponds to the SFMS for star-forming galaxies from \citet{Whitaker2012} at z=0. Different groups fall into different regions with some overlap between groups. The points separated from the main trend are the oldest quiescent galaxies.}
    \label{fig:Main sequence}
\end{figure}

\begin{figure}
\hspace*{-0.5cm}
	\includegraphics[width=1.\columnwidth]{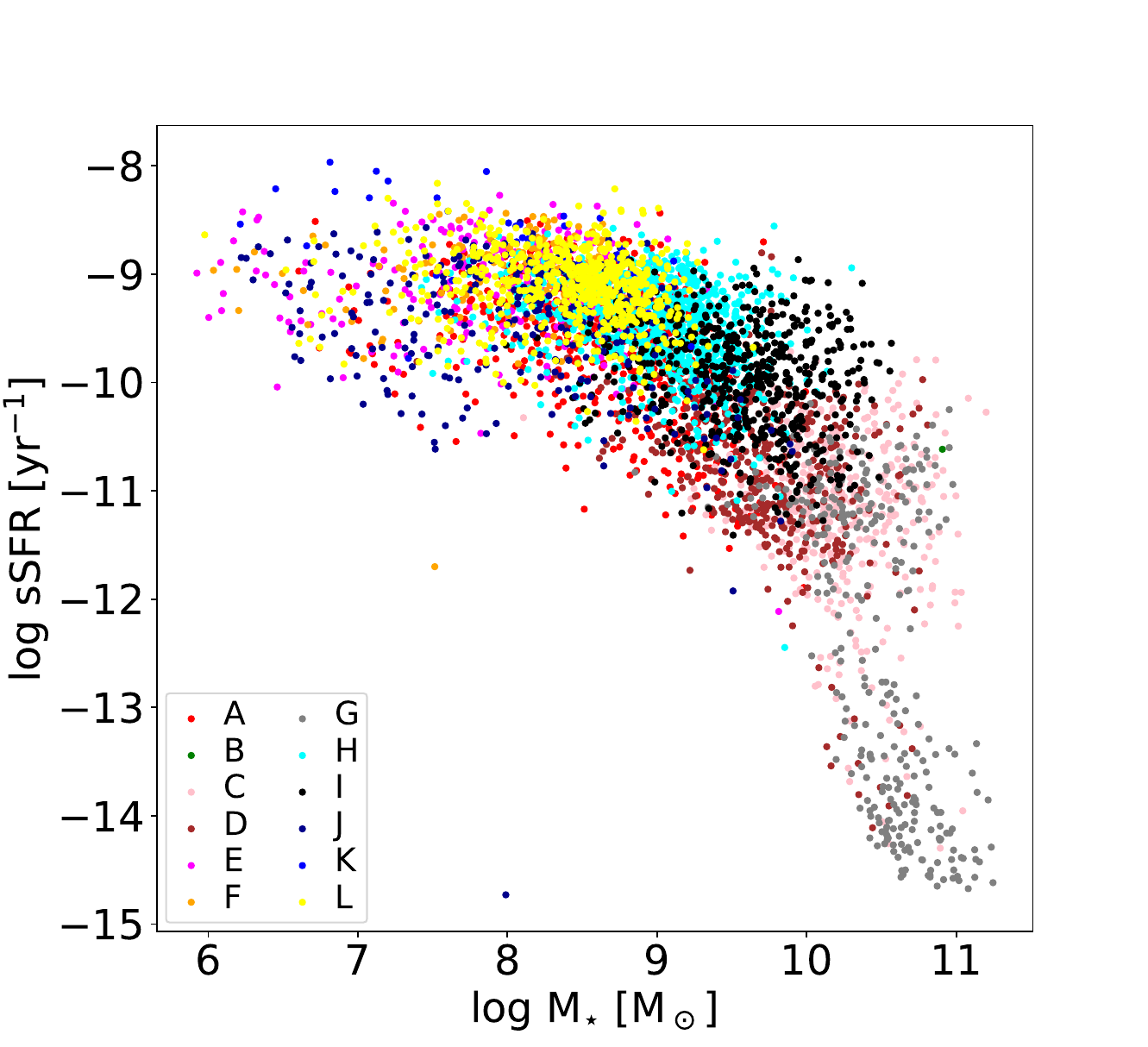}
    \caption{sSFR vs. stellar mass for all the targets in each one of the 12 groups obtained by the unsupervised ML algorithm. The symbol colour represents the same as in Fig. \ref{fig:Mstar_redshift}.}
    \label{fig:sSFR-Mstar}
\end{figure}

The SFMS is presented in Fig. \ref{fig:Main sequence}. From here, we can see that different groups fall into different regions, with some overlap. As discussed in \S \ref{sec:ML classification}, the GM classification allows overlapping between groups.\\ 
To clarify if the behavior observed in the SFMS is dominated by populations with different stellar mass, in Fig. \ref{fig:sSFR-Mstar} we plot the sSFR vs. stellar mass. The behavior is similar to that observed in the SFMS. For example, groups G and C, gray and pink points, have similar stellar mass ($\sim$10$^{10.5}$ M$_{\odot}$) but different ages (9.5 and 8.5 Gyr, respectively).  They fall in coincident regions with a slight offset between them. However, group G (which contains the oldest quiescent galaxies and with the lowest sSFR, see table \ref{tab:Physical properties}) is almost totally outside of the SFMS. Also, group K, which has intense emission lines, (see Fig. \ref{fig:ML classes}) falls in the region with the highest sSFR.\\

The fraction of PAU galaxies in the group K (0.05\%) is similar to the fraction of HII galaxies, HIIG, (0.02\%) from Ch\'avez et al. 2014 selected from the SDSS DR7 spectroscopic catalogue \citep{Abazajian2009} for having the strongest emission lines relative to the continuum at 0.01 $<$ z $<$ 0.2. HIIG are compact low mass systems (M$_{\star}<$ 10$^9$ M$_{\odot}$) with the luminosity almost completely dominated by a young (age $<$ 5 Myr) massive burst of star formation \citep{Terlevich1981, Melnick1988, Bordalo2011, Chavez2014}. By selection, they are the population of extragalactic systems with the strongest narrow emission lines ($\sigma < 90 $ km/s). HIIG fall well above the overall average for star-forming galaxies of log(sSFR) $\sim -10$ yr$^{-1}$ \citep{Guo2015} and their sSFR approach to the largest starburst galaxies such as ULIRGS with log(sSFR) $\sim -8$ yr$^{-1}$ \citep{Doran2013}. However, if we consider the sSFR of the present burst alone, the current starbursts in HIIG are producing new stars at a much higher rate of log(sSFR) $\sim -7$ yr$^{-1}$ \citep{Telles2018}. In fact, the most metal-poor compact starbursts at all redshifts tend to appear as HIIG \citep{Kunth2000, Gil_de_Paz2003, Amorin2012, Izotov2012, Kehrig2016, Amorin2017, Kehrig2018, Wofford2021} besides they can be observed even at large redshifts becoming interesting standard candles \citep{Melnick2000, Plionis2011, Terlevich2015, Chavez2016, Yennapureddy2017, Ruan2019, Gonzalez-Moran2019, Wu2020, Gonzalez-Moran2021, Tsiapi2021, Mehrabi2022}. Therefore, it could be interesting to expand the  analysis of group K by doing a spectroscopic follow-up in order to detect if these targets are HIIG. Although the applied methodology is different, ML-classified photometric methodology could be a fast way to choose HIIG candidates for incoming large surveys.

The results mentioned above pose an extra confirmation on the physical meaning of the classification using ML. Different groups follow patterns associated to different stellar populations and the classification using ML successfully separates an extreme population (G group) from the general SFMS trend.

\begin{figure*}
\begin{center}$
\begin{array}{cc}
   {\includegraphics[width=.49\textwidth]{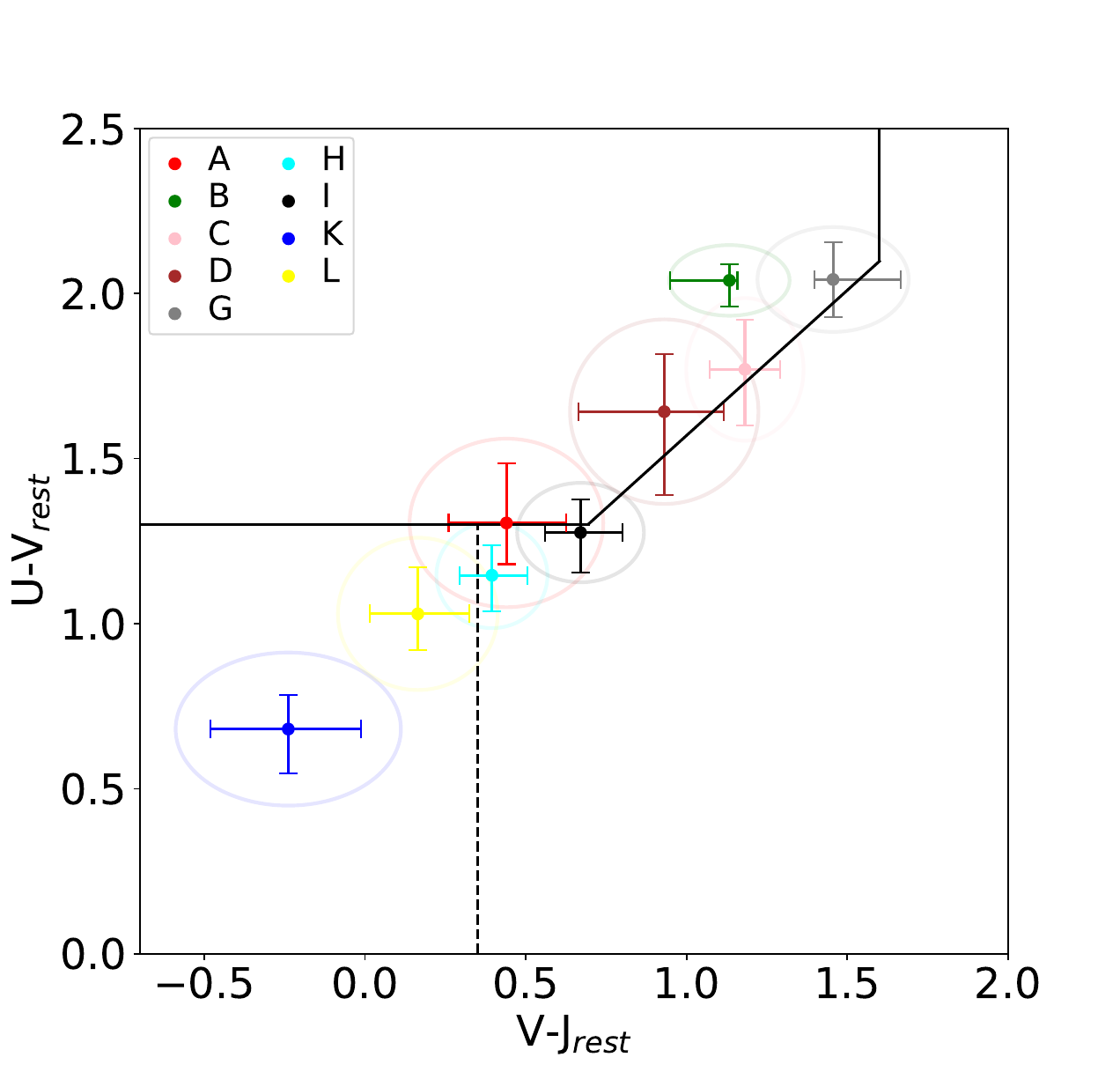}} &   
   {\includegraphics[width=.505\textwidth]{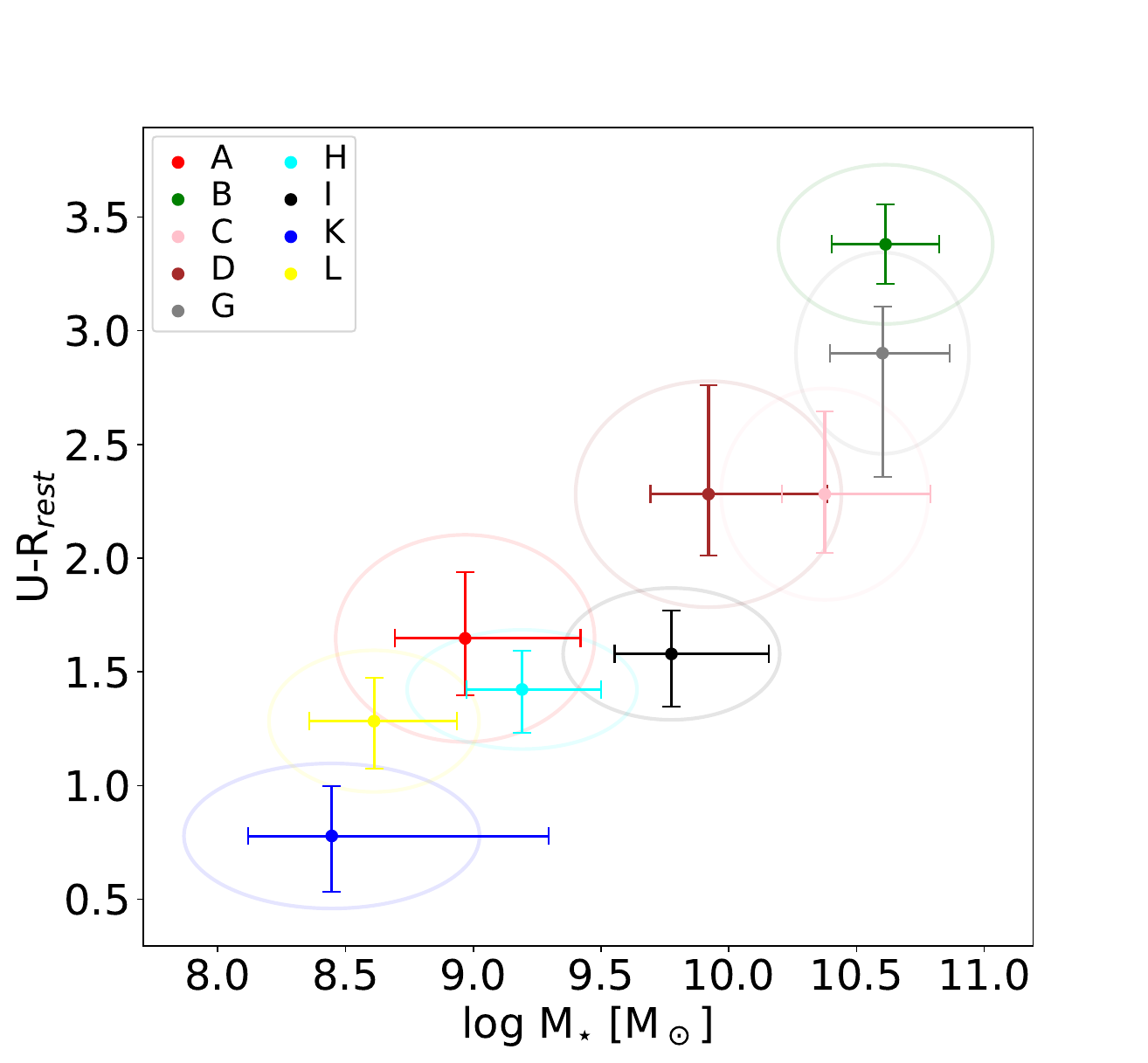}}
\end{array}$
\end{center}

\vspace*{-0.69cm}
\begin{center}$
\begin{array}{cc}
   {\includegraphics[width=.49\textwidth]{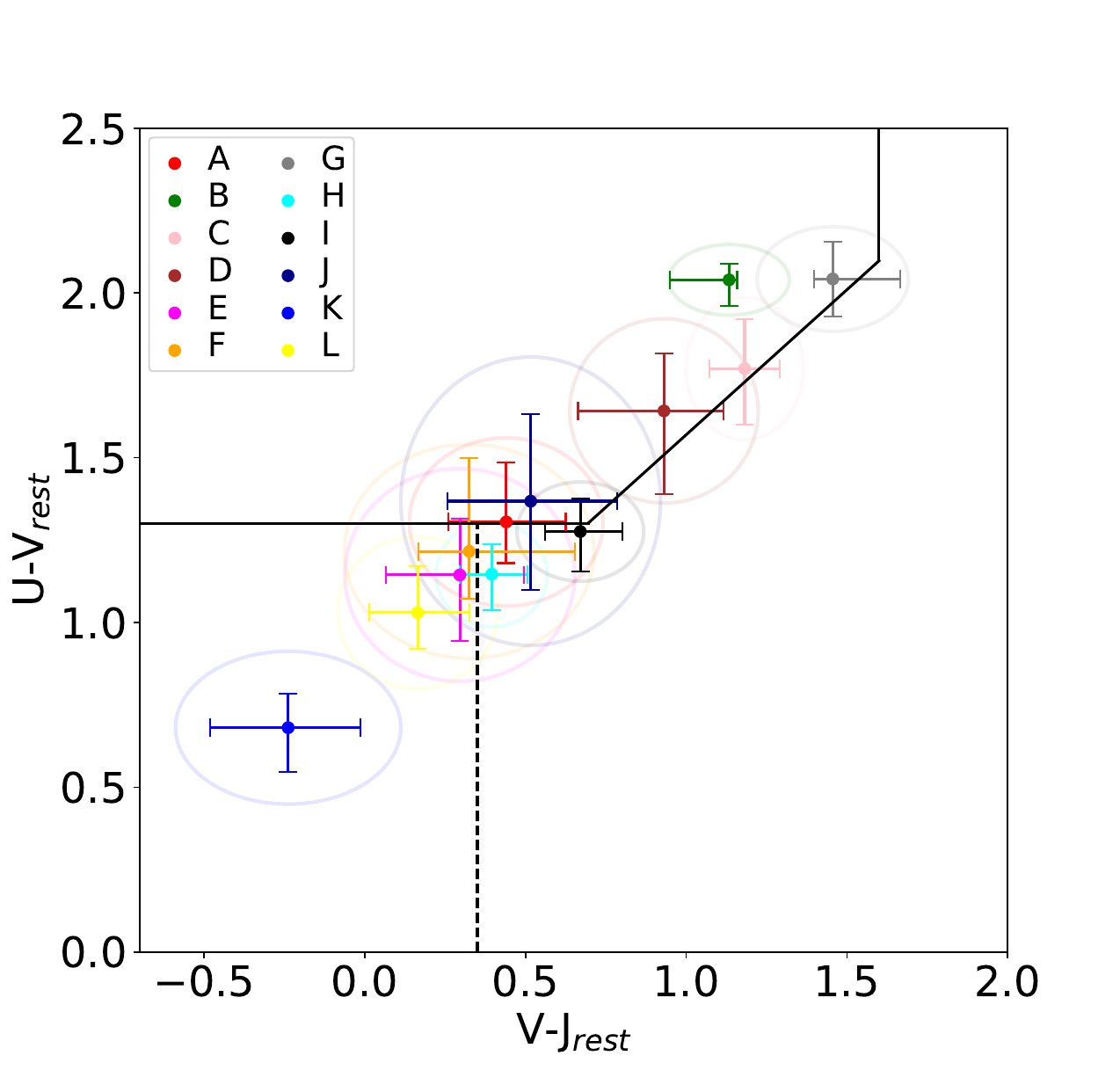}} &   
   {\includegraphics[width=.505\textwidth]{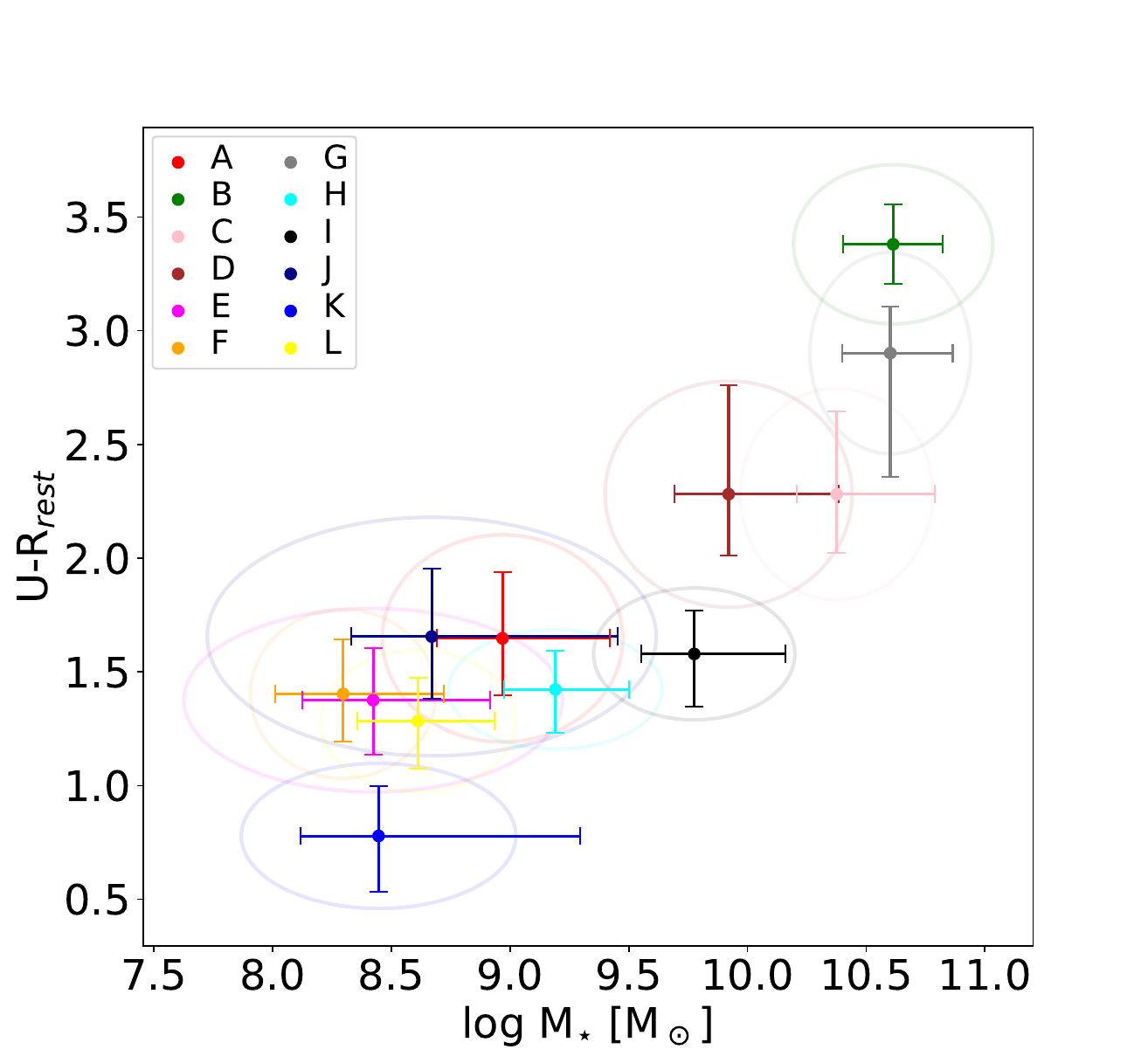}}
   \end{array}$
\end{center}
\caption{Colour-colour and stellar mass-colour diagrams for our sample. The symbol colour represents the class obtained by the unsupervised ML algorithm. On the left panels are the UVJ diagrams whose limits for the quiescent regions (top-left region) are taken from \citet{Whitaker2011} for z $<$ 0.5. On the right panels are the rest-frame U$-$R versus stellar mass relation. The error bars correspond to the first and the third quartile of the distribution of the parameters, while the ellipses are centered in the mean and the axes correspond to the standard deviation of the distribution of the parameters. On the top panels, we have excluded the classes E, F, and J with large scatter in the SEDs (see Fig. \ref{fig:ML classes}) and one of the worst classified by the robustness analysis.}
 \label{fig:color plots}
\end{figure*}

\subsection{The colour of the galaxies}
\label{sec:The colour of the galaxies}
 
The dependence of galaxy colour on morphological type is well established since the pioneer works by \cite{de_Vaucouleurs1961} and colour-colour diagrams have also been used to separate star-forming from quiescent galaxies \citep[see e.g.,][]{Madau1996, Fioc_Rocca-Volmerange1999, Ferreras1999}. Among the possible diagrams, the most popular is the UVJ diagram, i.e., rest-frame U$-$V versus V$-$J \citep{Labbe2005, Wuyts2007, Williams2009}. The effectiveness of the UVJ diagram comes from the fact that the combination of these two colors can break the degeneracy between age and dust reddening.  The UVJ diagram has been used for more than merely grouping galaxies into two categories. For example, UVJ colours have been used to infer the star formation rate and dust attenuation for star-forming systems \citep{Fang2018}, and the stellar ages for quiescent systems \citep{Belli2019}.

We derive U$-$V and V$-$J colours from the rest-frame fluxes, with the potential advantage that these come from directly observed photometry \citep[e.g.][]{Taylor2009}.  Fig. \ref{fig:color plots} presents the mean of the rest-frame colour-colour and colour-mass relations of the PAU galaxies, classified into 12 classes with the unsupervised ML algorithm. On the left panels are the UVJ diagrams.
The separation of the groups found in Fig \ref{fig:color plots} is clear. Galaxies in each ML classified group have similar properties and span only a small region of the available parameter space.

Comparing our results with those predicted using the age$-$colour relation given by \cite{Belli2019}, these are in agreement inside 1$\sigma$. This is a relation between the stellar age and the rest-frame U$-$V and V$-$J colours, which can be used to estimate the age of quiescent galaxies, given their colour. For example, for group G, we obtained using CIGALE an age of $9.5 + 0.4 -0.7$ Gyr and using the mentioned age$-$colour relation an age of 9.77 Gyr.

Following \cite{Lumbreras-Calle2019}, we applied the V$-$J colour threshold criterion to separate Emission Line Galaxies (ELGs) from non-ELGs in blue galaxy samples (V$-$J $<$ 0.35). We found that 20\% (groups K and L) of the galaxies with emission lines in the SEDs (68\% of the total sample used in this work) fall in the ELGs region for blue galaxies on the UVJ diagram. Besides, from the analysis of the main integrated properties of their stellar populations via SED fitting, these galaxies have the lowest stellar masses and highest sSFR. This suggests that the new V$-$J colour criterion to separate ELGs from non-ELGs in blue galaxies by \cite{Lumbreras-Calle2019} could be used as well to select the extreme ELGs in blue galaxy samples.

To highlight the excellent spectral coverage of the PAU narrow bands, we compare our results with previous applications of unsupervised clustering over spectroscopic surveys, in particular, \cite{Siudek2018} also used an unsupervised machine-learning algorithm to classify the VIMOS Public Extragalactic Redshift Survey (VIPERS) into 12 groups. The number of groups, colours, and the SFMS trend show a similar classification to ours, although the redshift range is different ($0.4 <$ z $< 1.3$) so the results can not be compared directly to ours ($0.01 <$ z $< 0.28$). However, interestingly they found 3 classes of red passive galaxies with similar properties as us in the B, C, and G groups (they reported log (M$_{\star}$/M$_{\odot}$) $\sim 10.8$ and log (sSFR/yr$^{-1}$) $\sim -12$, while our values are log (M$_{\star}$/M$_{\odot}$) $\sim 10.5$ and log (sSFR/yr$^{-1}$) $\sim -11.5$). Besides, if we compare the SFMS, the trend is very similar, but the SFR is different perhaps due to a redshift evolution. These same photometric ML classified groups (B, C, and G) show a log (SFR/M$_{\odot}$ yr$^{-1}$) $\sim -3$, while the spectroscopic ML classified classes have log (SFR/M$_{\odot}$ yr$^{-1}$) $\sim -1$. Regarding the colours in the UVJ diagram, both trends are similar, but the \cite{Siudek2018} U$-$V colors are bluer than ours. Besides, in general, galaxies reported by them are more massive than ours. At the same time, group K from our classification falls in a region in the V$-$J colour that does not appear in the \cite{Siudek2018} classification. It is important to highlight that this group has a log (M$_{\star}$/M$_{\odot}$) $\sim 8$, which is an order of magnitude lower than the class with the lowest stellar mass reported in \cite{Siudek2018} with a log (M$_{\star}$/M$_{\odot}$) $\sim 9$, however, both groups have a log (sSFR/yr$^{-1}$) $\sim 8.6$. Although our photometric ML classified work has some similarities with the spectroscopic ML classified work from \cite{Siudek2018}, we emphasize that the \cite{Siudek2018} sample belongs to a higher redshift than ours, so both results can not be compared directly.

\section{Conclusions.}
\label{sec:Conclusions}

We have used an unsupervised ML classification associated with the shape of 5,234 low-redshift SEDs from the PAU survey in the COSMOS field. From the analysis of the SEDs obtained from these data, we have found the following:\\

1.- The GM clustering algorithm is implemented and optimized to get relevant classes. We have chosen 12 as the optimal number of classes based on the analysis of the BIC parameter gradient and the silhouette score.\\

2.- The number of targets belonging to each ML classified group is different, so the groups are not equally populated. Also, the continuum pattern of the different groups is different; 
four groups (B, C, D, and G) do not show emission lines, and some groups (B, C, D, E, F, and J) present absorption lines such as Mg $\lambda$5175\AA\ and Na I $\lambda\lambda$5889, 5895 \AA\AA. Four groups (A, E, F, and J) have a large scatter with coincidence factor minimal $<$ 43\% and rms $>$ 0.06. Five groups (A, H, I, K, and L) present a clear detection in emission lines such as $\hb$, [$\Oiii$]$\lambda$5007\AA\, and $\ha$ emission lines. In particular, group K shows intense emission lines in their SEDs. In summary, 68\% of the total sample of 5,245 galaxies in the PAU survey at $0.01 <$ z $< 0.28$ are in a star-forming phase. The groups are not biased by redshift with targets in all the redshifts values.\\

3.- The differences in the galaxy population among the different classes have been studied. The stellar population and other physical properties have been explored using the CIGALE code. The mass, age and sSFR of the galaxies range from $0.15<$ age/Gyr $<11$, $6 <$ log (M$_{\star}$/M$_{\odot}$) $< 11.26$ and $-14.67 <$ log (sSFR/yr$^{-1}$) $< -8$. The ML classified groups are well-defined in their properties. Galaxies showing clear emission lines typically fall in the lower mass, younger and higher sSFR regime (mean values of log (M$_{\star}$/M$_{\odot}$) $= 8.72 \pm 0.75$, $3.02\pm 2.16$ Gyr, and, log (sSFR/yr$^{-1}$) $= -9.46 \pm 0.57$) than galaxies in the groups that do not show emission lines in their SEDs (mean values of log (M$_{\star}$/M$_{\odot}$) = $10.08 \pm 0.54$, $8.14\pm1.94$ Gyr, and, log (sSFR/yr$^{-1}$) $= -11.36 \pm 1.18$). The SFMS and sSFR vs. stellar mass plots show that different groups fall into different regions with some overlap among groups.\\

4.- We applied the new V$-$J colour criterion to separate ELGs from non-ELGs in blue galaxy samples (V$-$J $< 0.35$), as suggested in \cite{Lumbreras-Calle2019}. We found that 20\% of the galaxies with emission lines in the SEDs fall in the ELGs region for blue galaxies on the UVJ diagram. Besides, these galaxies have the lowest stellar masses and highest sSFR of the entire sample suggesting that the V$-$J colour criterion applied could be used to select the extreme ELGs.\\

5- The fraction of galaxies at low- z in the PAU Survey with emission lines is 68\% and their characteristic values of mass, age, and sSFR are consistent with those reported by other medium-band works in the COSMOS \citep{Hinojosa2016} and GOODS-N \citep{Lumbreras-Calle2019} fields.\\

6.- We have demonstrated that the joint of low- resolution (R $\sim$ 50) photometric spectra provided by the PAU survey and unsupervised clustering represents an excellent opportunity to classify galaxies. Moreover, it helps to find and extend the analysis of extreme ELGs to lower masses and lower SFRs in the local Universe.

\section*{DATA AVAILABILITY}
The data underlying this article are available from the PAU spectro-photometry catalogue provided by the PAU collaboration. Other datasets were derived from sources in the public domain available at the PAU webpage https://pausurvey.org.

\section*{Acknowledgements}
This work has been supported by the Ministry of Science and Innovation of Spain, project PID2019-107408GB-C43 (ESTALLIDOS), and the Government of the Canary Islands through EU FEDER funding, projects PID2020010050 and PID2021010077. This article is based on observations made in the Observatorios de Canarias of the Instituto de Astrofísica de Canarias (IAC) with the WHT operated on the island of La Palma by the Isaac Newton Group of Telescopes (ING) in the Observatorio del Roque de los Muchachos. The PAU Survey is partially supported by MINECO under grants CSD2007-00060, AYA2015-71825, ESP2017-89838, PGC2018-094773, PGC2018-102021, PID2019-111317GB, SEV-2016-0588, SEV-2016-0597, MDM-2015-0509 and Juan de la Cierva fellowship and LACEGAL and EWC Marie Sklodowska-Curie grant No 734374 and no.776247 with ERDF funds from the EU Horizon 2020 Programme, some of which include ERDF funds from the European Union. IEEC and IFAE are partially funded by the CERCA and Beatriu de Pinos program of the Generalitat de Catalunya. Funding for PAUS has also been provided by Durham University (via the ERC StG DEGAS-259586), ETH Zurich, Leiden University (via ERC StG ADULT-279396 and Netherlands Organisation for Scientific Research (NWO) Vici grant 639.043.512), University College London and from the European Union's Horizon 2020 research and innovation programme under the grant agreement No 776247 EWC. The PAU data center is hosted by the Port d'Informaci\'o Cient\'ifica (PIC), maintained through a collaboration of CIEMAT and IFAE, with additional support from Universitat Aut\`onoma de Barcelona and ERDF. We acknowledge the PIC services department team for their support and fruitful discussions.

\bibliography{bib/bibpaper2022_ML}

\newcommand{\noop}[1]{}
\begin{thebibliography}{}
\makeatletter
\relax
\def\mn@urlcharsother{\let\do\@makeother \do\$\do\&\do\#\do\^\do\_\do\%\do\~}
\def\mn@doi{\begingroup\mn@urlcharsother \@ifnextchar [ {\mn@doi@}
  {\mn@doi@[]}}
\def\mn@doi@[#1]#2{\def\@tempa{#1}\ifx\@tempa\@empty \href
  {http://dx.doi.org/#2} {doi:#2}\else \href {http://dx.doi.org/#2} {#1}\fi
  \endgroup}
\def\mn@eprint#1#2{\mn@eprint@#1:#2::\@nil}
\def\mn@eprint@arXiv#1{\href {http://arxiv.org/abs/#1} {{\tt arXiv:#1}}}
\def\mn@eprint@dblp#1{\href {http://dblp.uni-trier.de/rec/bibtex/#1.xml}
  {dblp:#1}}
\def\mn@eprint@#1:#2:#3:#4\@nil{\def\@tempa {#1}\def\@tempb {#2}\def\@tempc
  {#3}\ifx \@tempc \@empty \let \@tempc \@tempb \let \@tempb \@tempa \fi \ifx
  \@tempb \@empty \def\@tempb {arXiv}\fi \@ifundefined
  {mn@eprint@\@tempb}{\@tempb:\@tempc}{\expandafter \expandafter \csname
  mn@eprint@\@tempb\endcsname \expandafter{\@tempc}}}

\bibitem[\protect\citeauthoryear{{Abazajian} et~al.,}{{Abazajian}
  et~al.}{2009}]{Abazajian2009}
{Abazajian} K.~N.,  et~al., 2009, \mn@doi [\apjs]
  {10.1088/0067-0049/182/2/543}, \href
  {https://ui.adsabs.harvard.edu/abs/2009ApJS..182..543A} {182, 543}

\bibitem[\protect\citeauthoryear{{Aguerri}, {Huertas-Company}, {S{\'a}nchez
  Almeida}  \& {Mu{\~n}oz-Tu{\~n}{\'o}n}}{{Aguerri} et~al.}{2012}]{Aguerri2012}
{Aguerri} J.~A.~L.,  {Huertas-Company} M.,  {S{\'a}nchez Almeida} J.,
  {Mu{\~n}oz-Tu{\~n}{\'o}n} C.,  2012, \mn@doi [\aap]
  {10.1051/0004-6361/201117632}, \href
  {https://ui.adsabs.harvard.edu/abs/2012A&A...540A.136A} {540, A136}

\bibitem[\protect\citeauthoryear{{Alarcon} et~al.,}{{Alarcon}
  et~al.}{2021}]{Alarcon2021}
{Alarcon} A.,  et~al., 2021, \mn@doi [\mnras] {10.1093/mnras/staa3659}, \href
  {https://ui.adsabs.harvard.edu/abs/2021MNRAS.501.6103A} {501, 6103}

\bibitem[\protect\citeauthoryear{{Amor{\'\i}n}, {P{\'e}rez-Montero},
  {V{\'\i}lchez}  \& {Papaderos}}{{Amor{\'\i}n} et~al.}{2012}]{Amorin2012}
{Amor{\'\i}n} R.,  {P{\'e}rez-Montero} E.,  {V{\'\i}lchez} J.~M.,   {Papaderos}
  P.,  2012, \mn@doi [\apj] {10.1088/0004-637X/749/2/185}, \href
  {https://ui.adsabs.harvard.edu/abs/2012ApJ...749..185A} {749, 185}

\bibitem[\protect\citeauthoryear{{Amor{\'\i}n} et~al.,}{{Amor{\'\i}n}
  et~al.}{2017}]{Amorin2017}
{Amor{\'\i}n} R.,  et~al., 2017, \mn@doi [Nature Astronomy]
  {10.1038/s41550-017-0052}, \href
  {https://ui.adsabs.harvard.edu/abs/2017NatAs...1E..52A} {1, 0052}

\bibitem[\protect\citeauthoryear{{Arrabal Haro} et~al.,}{{Arrabal Haro}
  et~al.}{2018}]{ArrabalHaro2018}
{Arrabal Haro} P.,  et~al., 2018, \mn@doi [\mnras] {10.1093/mnras/sty1106},
  \href {https://ui.adsabs.harvard.edu/abs/2018MNRAS.478.3740A} {478, 3740}

\bibitem[\protect\citeauthoryear{{Arrabal Haro} et~al.,}{{Arrabal Haro}
  et~al.}{2020}]{Arrabal-Haro2020}
{Arrabal Haro} P.,  et~al., 2020, \mn@doi [\mnras] {10.1093/mnras/staa1196},
  \href {https://ui.adsabs.harvard.edu/abs/2020MNRAS.495.1807A} {495, 1807}

\bibitem[\protect\citeauthoryear{{Baron} \& {Poznanski}}{{Baron} \&
  {Poznanski}}{2017}]{Baron2017}
{Baron} D.,  {Poznanski} D.,  2017, \mn@doi [\mnras] {10.1093/mnras/stw3021},
  \href {https://ui.adsabs.harvard.edu/abs/2017MNRAS.465.4530B} {465, 4530}

\bibitem[\protect\citeauthoryear{{Barro} et~al.,}{{Barro}
  et~al.}{2019}]{Barro2019}
{Barro} G.,  et~al., 2019, \mn@doi [\apjs] {10.3847/1538-4365/ab23f2}, \href
  {https://ui.adsabs.harvard.edu/abs/2019ApJS..243...22B} {243, 22}

\bibitem[\protect\citeauthoryear{{Belfiore} et~al.,}{{Belfiore}
  et~al.}{2018}]{Belfiore2018}
{Belfiore} F.,  et~al., 2018, \mn@doi [\mnras] {10.1093/mnras/sty768}, \href
  {https://ui.adsabs.harvard.edu/abs/2018MNRAS.477.3014B} {477, 3014}

\bibitem[\protect\citeauthoryear{{Belli}, {Newman}  \& {Ellis}}{{Belli}
  et~al.}{2019}]{Belli2019}
{Belli} S.,  {Newman} A.~B.,   {Ellis} R.~S.,  2019, \mn@doi [\apj]
  {10.3847/1538-4357/ab07af}, \href
  {https://ui.adsabs.harvard.edu/abs/2019ApJ...874...17B} {874, 17}

\bibitem[\protect\citeauthoryear{{Boquien}, {Burgarella}, {Roehlly}, {Buat},
  {Ciesla}, {Corre}, {Inoue}  \& {Salas}}{{Boquien} et~al.}{2019}]{Boquien2019}
{Boquien} M.,  {Burgarella} D.,  {Roehlly} Y.,  {Buat} V.,  {Ciesla} L.,
  {Corre} D.,  {Inoue} A.~K.,   {Salas} H.,  2019, \mn@doi [\aap]
  {10.1051/0004-6361/201834156}, \href
  {https://ui.adsabs.harvard.edu/abs/2019A&A...622A.103B} {622, A103}

\bibitem[\protect\citeauthoryear{{Bordalo} \& {Telles}}{{Bordalo} \&
  {Telles}}{2011}]{Bordalo2011}
{Bordalo} V.,  {Telles} E.,  2011, \mn@doi [\apj] {10.1088/0004-637X/735/1/52},
  \href {https://ui.adsabs.harvard.edu/abs/2011ApJ...735...52B} {735, 52}

\bibitem[\protect\citeauthoryear{{Bruzual} \& {Charlot}}{{Bruzual} \&
  {Charlot}}{2003}]{Bruzual-Charlot2003}
{Bruzual} G.,  {Charlot} S.,  2003, \mn@doi [\mnras]
  {10.1046/j.1365-8711.2003.06897.x}, \href
  {https://ui.adsabs.harvard.edu/abs/2003MNRAS.344.1000B} {344, 1000}

\bibitem[\protect\citeauthoryear{{Buta}, {Mitra}, {de Vaucouleurs}  \&
  {Corwin}}{{Buta} et~al.}{1994}]{Buta1994}
{Buta} R.,  {Mitra} S.,  {de Vaucouleurs} G.,   {Corwin} H.~G. J.,  1994,
  \mn@doi [\aj] {10.1086/116838}, \href
  {https://ui.adsabs.harvard.edu/abs/1994AJ....107..118B} {107, 118}

\bibitem[\protect\citeauthoryear{{Cabayol-Garcia} et~al.,}{{Cabayol-Garcia}
  et~al.}{2020}]{CabayolGarcia2020}
{Cabayol-Garcia} L.,  et~al., 2020, \mn@doi [\mnras] {10.1093/mnras/stz3274},
  \href {https://ui.adsabs.harvard.edu/abs/2020MNRAS.491.5392C} {491, 5392}

\bibitem[\protect\citeauthoryear{Cabayol et~al.,}{Cabayol
  et~al.}{2021}]{Cabayol2021}
Cabayol L.,  et~al., 2021, \mn@doi [\mnras] {10.1093/mnras/stab1909}, 506, 4048

\bibitem[\protect\citeauthoryear{{Casas} et~al.,}{{Casas}
  et~al.}{2016}]{Casas2016}
{Casas} R.,  et~al., 2016, in {Evans} C.~J.,  {Simard} L.,   {Takami} H.,  eds,
   Society of Photo-Optical Instrumentation Engineers (SPIE) Conference Series
  Vol. 9908, Ground-based and Airborne Instrumentation for Astronomy VI. p.
  99084K, \mn@doi{10.1117/12.2232422}

\bibitem[\protect\citeauthoryear{{Cava} et~al.,}{{Cava}
  et~al.}{2015}]{Cava2015}
{Cava} A.,  et~al., 2015, \mn@doi [\apj] {10.1088/0004-637X/812/2/155}, \href
  {https://ui.adsabs.harvard.edu/abs/2015ApJ...812..155C} {812, 155}

\bibitem[\protect\citeauthoryear{{Charlot} \& {Fall}}{{Charlot} \&
  {Fall}}{2000}]{Charlot_Fall2000}
{Charlot} S.,  {Fall} S.~M.,  2000, \mn@doi [\apj] {10.1086/309250}, \href
  {https://ui.adsabs.harvard.edu/abs/2000ApJ...539..718C} {539, 718}

\bibitem[\protect\citeauthoryear{{Ch{\'a}vez}, {Terlevich}, {Terlevich},
  {Bresolin}, {Melnick}, {Plionis}  \& {Basilakos}}{{Ch{\'a}vez}
  et~al.}{2014}]{Chavez2014}
{Ch{\'a}vez} R.,  {Terlevich} R.,  {Terlevich} E.,  {Bresolin} F.,  {Melnick}
  J.,  {Plionis} M.,   {Basilakos} S.,  2014, \mn@doi [\mnras]
  {10.1093/mnras/stu987}, \href
  {https://ui.adsabs.harvard.edu/abs/2014MNRAS.442.3565C} {442, 3565}

\bibitem[\protect\citeauthoryear{{Ch{\'a}vez}, {Plionis}, {Basilakos},
  {Terlevich}, {Terlevich}, {Melnick}, {Bresolin}  \&
  {Gonz{\'a}lez-Mor{\'a}n}}{{Ch{\'a}vez} et~al.}{2016}]{Chavez2016}
{Ch{\'a}vez} R.,  {Plionis} M.,  {Basilakos} S.,  {Terlevich} R.,  {Terlevich}
  E.,  {Melnick} J.,  {Bresolin} F.,   {Gonz{\'a}lez-Mor{\'a}n} A.~L.,  2016,
  \mn@doi [\mnras] {10.1093/mnras/stw1813}, \href
  {https://ui.adsabs.harvard.edu/abs/2016MNRAS.462.2431C} {462, 2431}

\bibitem[\protect\citeauthoryear{{D'Abrusco}, {Longo}  \& {Walton}}{{D'Abrusco}
  et~al.}{2009}]{D'Abrusco2009}
{D'Abrusco} R.,  {Longo} G.,   {Walton} N.~A.,  2009, \mn@doi [\mnras]
  {10.1111/j.1365-2966.2009.14754.x}, \href
  {https://ui.adsabs.harvard.edu/abs/2009MNRAS.396..223D} {396, 223}

\bibitem[\protect\citeauthoryear{{D'Abrusco}, {Fabbiano}, {Djorgovski},
  {Donalek}, {Laurino}  \& {Longo}}{{D'Abrusco} et~al.}{2012}]{D'Abrusco2012}
{D'Abrusco} R.,  {Fabbiano} G.,  {Djorgovski} G.,  {Donalek} C.,  {Laurino} O.,
    {Longo} G.,  2012, \mn@doi [\apj] {10.1088/0004-637X/755/2/92}, \href
  {https://ui.adsabs.harvard.edu/abs/2012ApJ...755...92D} {755, 92}

\bibitem[\protect\citeauthoryear{{Dale}, {Helou}, {Magdis}, {Armus},
  {D{\'\i}az-Santos}  \& {Shi}}{{Dale} et~al.}{2014}]{Dale2014}
{Dale} D.~A.,  {Helou} G.,  {Magdis} G.~E.,  {Armus} L.,  {D{\'\i}az-Santos}
  T.,   {Shi} Y.,  2014, \mn@doi [\apj] {10.1088/0004-637X/784/1/83}, \href
  {https://ui.adsabs.harvard.edu/abs/2014ApJ...784...83D} {784, 83}

\bibitem[\protect\citeauthoryear{Dempster, Laird  \& Rubin}{Dempster
  et~al.}{1977}]{Dempster1977}
Dempster A.~P.,  Laird N.~M.,   Rubin D.~B.,  1977, Journal of the Royal
  Statistical Society. Series B (Methodological), 39, 1

\bibitem[\protect\citeauthoryear{{Doran} et~al.,}{{Doran}
  et~al.}{2013}]{Doran2013}
{Doran} E.~I.,  et~al., 2013, \mn@doi [\aap] {10.1051/0004-6361/201321824},
  \href {https://ui.adsabs.harvard.edu/abs/2013A&A...558A.134D} {558, A134}

\bibitem[\protect\citeauthoryear{{Duarte Puertas}, {Vilchez},
  {Iglesias-P{\'a}ramo}, {Kehrig}, {P{\'e}rez-Montero}  \&
  {Rosales-Ortega}}{{Duarte Puertas} et~al.}{2017}]{Duarte-Puertas2017}
{Duarte Puertas} S.,  {Vilchez} J.~M.,  {Iglesias-P{\'a}ramo} J.,  {Kehrig} C.,
   {P{\'e}rez-Montero} E.,   {Rosales-Ortega} F.~F.,  2017, \mn@doi [\aap]
  {10.1051/0004-6361/201629044}, \href
  {https://ui.adsabs.harvard.edu/abs/2017A&A...599A..71D} {599, A71}

\bibitem[\protect\citeauthoryear{{Dubois}, {Fraix-Burnet}, {Moultaka}, {Sharma}
   \& {Burgarella}}{{Dubois} et~al.}{2022}]{Dubois2022}
{Dubois} J.,  {Fraix-Burnet} D.,  {Moultaka} J.,  {Sharma} P.,   {Burgarella}
  D.,  2022, \mn@doi [\aap] {10.1051/0004-6361/202141729}, \href
  {https://ui.adsabs.harvard.edu/abs/2022A&A...663A..21D} {663, A21}

\bibitem[\protect\citeauthoryear{{Duda} \& {Hart}}{{Duda} \&
  {Hart}}{1973}]{Duda1973}
{Duda} R.~O.,  {Hart} P.~E.,  1973, {Pattern classification and scene
  analysis}.
Wiley New York

\bibitem[\protect\citeauthoryear{{Eriksen} et~al.,}{{Eriksen}
  et~al.}{2019}]{Eriksen2019}
{Eriksen} M.,  et~al., 2019, \mn@doi [\mnras] {10.1093/mnras/stz204}, \href
  {https://ui.adsabs.harvard.edu/abs/2019MNRAS.484.4200E} {484, 4200}

\bibitem[\protect\citeauthoryear{{Eriksen} et~al.,}{{Eriksen}
  et~al.}{2020}]{Eriksen2020}
{Eriksen} M.,  et~al., 2020, \mn@doi [\mnras] {10.1093/mnras/staa2265}, \href
  {https://ui.adsabs.harvard.edu/abs/2020MNRAS.497.4565E} {497, 4565}

\bibitem[\protect\citeauthoryear{{Fang} et~al.,}{{Fang}
  et~al.}{2018}]{Fang2018}
{Fang} J.~J.,  et~al., 2018, \mn@doi [\apj] {10.3847/1538-4357/aabcba}, \href
  {https://ui.adsabs.harvard.edu/abs/2018ApJ...858..100F} {858, 100}

\bibitem[\protect\citeauthoryear{{Ferreras}, {Cayon}, {Martinez-Gonzalez}  \&
  {Benitez}}{{Ferreras} et~al.}{1999}]{Ferreras1999}
{Ferreras} I.,  {Cayon} L.,  {Martinez-Gonzalez} E.,   {Benitez} N.,  1999,
  \mn@doi [\mnras] {10.1046/j.1365-8711.1999.02308.x}, \href
  {https://ui.adsabs.harvard.edu/abs/1999MNRAS.304..319F} {304, 319}

\bibitem[\protect\citeauthoryear{{Fioc} \& {Rocca-Volmerange}}{{Fioc} \&
  {Rocca-Volmerange}}{1999}]{Fioc_Rocca-Volmerange1999}
{Fioc} M.,  {Rocca-Volmerange} B.,  1999, \aap, \href
  {https://ui.adsabs.harvard.edu/abs/1999A&A...351..869F} {351, 869}

\bibitem[\protect\citeauthoryear{{Gil de Paz}, {Madore}  \& {Pevunova}}{{Gil de
  Paz} et~al.}{2003}]{Gil_de_Paz2003}
{Gil de Paz} A.,  {Madore} B.~F.,   {Pevunova} O.,  2003, \mn@doi [\apjs]
  {10.1086/374737}, \href
  {https://ui.adsabs.harvard.edu/abs/2003ApJS..147...29G} {147, 29}

\bibitem[\protect\citeauthoryear{{Gonz{\'a}lez-Mor{\'a}n}
  et~al.,}{{Gonz{\'a}lez-Mor{\'a}n} et~al.}{2019}]{Gonzalez-Moran2019}
{Gonz{\'a}lez-Mor{\'a}n} A.~L.,  et~al., 2019, \mn@doi [\mnras]
  {10.1093/mnras/stz1577}, \href
  {https://ui.adsabs.harvard.edu/abs/2019MNRAS.487.4669G} {487, 4669}

\bibitem[\protect\citeauthoryear{{Gonz{\'a}lez-Mor{\'a}n}
  et~al.,}{{Gonz{\'a}lez-Mor{\'a}n} et~al.}{2021}]{Gonzalez-Moran2021}
{Gonz{\'a}lez-Mor{\'a}n} A.~L.,  et~al., 2021, \mn@doi [\mnras]
  {10.1093/mnras/stab1385}, \href
  {https://ui.adsabs.harvard.edu/abs/2021MNRAS.505.1441G} {505, 1441}

\bibitem[\protect\citeauthoryear{{Grazian} et~al.,}{{Grazian}
  et~al.}{2015}]{Grazian2015}
{Grazian} A.,  et~al., 2015, \mn@doi [\aap] {10.1051/0004-6361/201424750},
  \href {https://ui.adsabs.harvard.edu/abs/2015A&A...575A..96G} {575, A96}

\bibitem[\protect\citeauthoryear{{Guo}, {Zheng}, {Wang}  \& {Fu}}{{Guo}
  et~al.}{2015}]{Guo2015}
{Guo} K.,  {Zheng} X.~Z.,  {Wang} T.,   {Fu} H.,  2015, \mn@doi [\apjl]
  {10.1088/2041-8205/808/2/L49}, \href
  {https://ui.adsabs.harvard.edu/abs/2015ApJ...808L..49G} {808, L49}

\bibitem[\protect\citeauthoryear{{Hern{\'a}n-Caballero}
  et~al.,}{{Hern{\'a}n-Caballero} et~al.}{2017}]{HernanCaballero2017}
{Hern{\'a}n-Caballero} A.,  et~al., 2017, \mn@doi [\apj]
  {10.3847/1538-4357/aa917f}, \href
  {https://ui.adsabs.harvard.edu/abs/2017ApJ...849...82H} {849, 82}

\bibitem[\protect\citeauthoryear{{Hinojosa-Go{\~n}i}, {Mu{\~n}oz-Tu{\~n}{\'o}n}
   \& {M{\'e}ndez-Abreu}}{{Hinojosa-Go{\~n}i} et~al.}{2016}]{Hinojosa2016}
{Hinojosa-Go{\~n}i} R.,  {Mu{\~n}oz-Tu{\~n}{\'o}n} C.,   {M{\'e}ndez-Abreu} J.,
   2016, \mn@doi [\aap] {10.1051/0004-6361/201527066}, \href
  {https://ui.adsabs.harvard.edu/abs/2016A&A...592A.122H} {592, A122}

\bibitem[\protect\citeauthoryear{{Hubble}}{{Hubble}}{1926}]{Hubble1926}
{Hubble} E.~P.,  1926, \mn@doi [\apj] {10.1086/143018}, \href
  {https://ui.adsabs.harvard.edu/abs/1926ApJ....64..321H} {64, 321}

\bibitem[\protect\citeauthoryear{{Hubble}}{{Hubble}}{1936}]{Hubble1936}
{Hubble} E.~P.,  1936, {Realm of the Nebulae}.
New Haven: Yale University Press

\bibitem[\protect\citeauthoryear{{Ilbert} et~al.,}{{Ilbert}
  et~al.}{2009}]{Ilbert2009}
{Ilbert} O.,  et~al., 2009, \mn@doi [\apj] {10.1088/0004-637X/690/2/1236},
  \href {https://ui.adsabs.harvard.edu/abs/2009ApJ...690.1236I} {690, 1236}

\bibitem[\protect\citeauthoryear{{Izotov}, {Thuan}  \& {Guseva}}{{Izotov}
  et~al.}{2012}]{Izotov2012}
{Izotov} Y.~I.,  {Thuan} T.~X.,   {Guseva} N.~G.,  2012, \mn@doi [\aap]
  {10.1051/0004-6361/201219733}, \href
  {https://ui.adsabs.harvard.edu/abs/2012A&A...546A.122I} {546, A122}

\bibitem[\protect\citeauthoryear{{Jeans}}{{Jeans}}{1928}]{Jeans1928}
{Jeans} J.~H.,  1928, {Astronomy and cosmogony}.
Cambridge [Eng.] The University press

\bibitem[\protect\citeauthoryear{{Johnston} et~al.,}{{Johnston}
  et~al.}{2021}]{Johnston2021}
{Johnston} H.,  et~al., 2021, \mn@doi [\aap] {10.1051/0004-6361/202039682},
  \href {https://ui.adsabs.harvard.edu/abs/2021A&A...646A.147J} {646, A147}

\bibitem[\protect\citeauthoryear{{Kehrig} et~al.,}{{Kehrig}
  et~al.}{2016}]{Kehrig2016}
{Kehrig} C.,  et~al., 2016, \mn@doi [\mnras] {10.1093/mnras/stw806}, \href
  {https://ui.adsabs.harvard.edu/abs/2016MNRAS.459.2992K} {459, 2992}

\bibitem[\protect\citeauthoryear{{Kehrig}, {V{\'\i}lchez}, {Guerrero},
  {Iglesias-P{\'a}ramo}, {Hunt}, {Duarte-Puertas}  \& {Ramos-Larios}}{{Kehrig}
  et~al.}{2018}]{Kehrig2018}
{Kehrig} C.,  {V{\'\i}lchez} J.~M.,  {Guerrero} M.~A.,  {Iglesias-P{\'a}ramo}
  J.,  {Hunt} L.~K.,  {Duarte-Puertas} S.,   {Ramos-Larios} G.,  2018, \mn@doi
  [\mnras] {10.1093/mnras/sty1920}, \href
  {https://ui.adsabs.harvard.edu/abs/2018MNRAS.480.1081K} {480, 1081}

\bibitem[\protect\citeauthoryear{{Kennicutt}}{{Kennicutt}}{1992}]{Kennicutt1992}
{Kennicutt} Robert~C. J.,  1992, \mn@doi [\apjs] {10.1086/191653}, \href
  {https://ui.adsabs.harvard.edu/abs/1992ApJS...79..255K} {79, 255}

\bibitem[\protect\citeauthoryear{{Kewley}, {Nicholls}  \&
  {Sutherland}}{{Kewley} et~al.}{2019}]{Kewley2019}
{Kewley} L.~J.,  {Nicholls} D.~C.,   {Sutherland} R.~S.,  2019, \mn@doi [\araa]
  {10.1146/annurev-astro-081817-051832}, \href
  {https://ui.adsabs.harvard.edu/abs/2019ARA&A..57..511K} {57, 511}

\bibitem[\protect\citeauthoryear{{Kunth} \& {{\"O}stlin}}{{Kunth} \&
  {{\"O}stlin}}{2000}]{Kunth2000}
{Kunth} D.,  {{\"O}stlin} G.,  2000, \mn@doi [\aapr] {10.1007/s001590000005},
  \href {https://ui.adsabs.harvard.edu/abs/2000A&ARv..10....1K} {10, 1}

\bibitem[\protect\citeauthoryear{{Labb{\'e}} et~al.,}{{Labb{\'e}}
  et~al.}{2005}]{Labbe2005}
{Labb{\'e}} I.,  et~al., 2005, \mn@doi [\apjl] {10.1086/430700}, \href
  {https://ui.adsabs.harvard.edu/abs/2005ApJ...624L..81L} {624, L81}

\bibitem[\protect\citeauthoryear{{Lilly} et~al.,}{{Lilly}
  et~al.}{2009}]{Lilly2009}
{Lilly} S.~J.,  et~al., 2009, \mn@doi [\apjs] {10.1088/0067-0049/184/2/218},
  \href {https://ui.adsabs.harvard.edu/abs/2009ApJS..184..218L} {184, 218}

\bibitem[\protect\citeauthoryear{{Lumbreras-Calle} et~al.,}{{Lumbreras-Calle}
  et~al.}{2019}]{Lumbreras-Calle2019}
{Lumbreras-Calle} A.,  et~al., 2019, \mn@doi [\aap]
  {10.1051/0004-6361/201731670}, \href
  {https://ui.adsabs.harvard.edu/abs/2019A&A...621A..52L} {621, A52}

\bibitem[\protect\citeauthoryear{{Madau}, {Ferguson}, {Dickinson},
  {Giavalisco}, {Steidel}  \& {Fruchter}}{{Madau} et~al.}{1996}]{Madau1996}
{Madau} P.,  {Ferguson} H.~C.,  {Dickinson} M.~E.,  {Giavalisco} M.,  {Steidel}
  C.~C.,   {Fruchter} A.,  1996, \mn@doi [\mnras] {10.1093/mnras/283.4.1388},
  \href {https://ui.adsabs.harvard.edu/abs/1996MNRAS.283.1388M} {283, 1388}

\bibitem[\protect\citeauthoryear{{Maraston}, {Str{\"o}mb{\"a}ck}, {Thomas},
  {Wake}  \& {Nichol}}{{Maraston} et~al.}{2009}]{Maraston2009}
{Maraston} C.,  {Str{\"o}mb{\"a}ck} G.,  {Thomas} D.,  {Wake} D.~A.,   {Nichol}
  R.~C.,  2009, \mn@doi [\mnras] {10.1111/j.1745-3933.2009.00621.x}, \href
  {https://ui.adsabs.harvard.edu/abs/2009MNRAS.394L.107M} {394, L107}

\bibitem[\protect\citeauthoryear{{Mehrabi} et~al.,}{{Mehrabi}
  et~al.}{2022}]{Mehrabi2022}
{Mehrabi} A.,  et~al., 2022, \mn@doi [\mnras] {10.1093/mnras/stab2915}, \href
  {https://ui.adsabs.harvard.edu/abs/2022MNRAS.509..224M} {509, 224}

\bibitem[\protect\citeauthoryear{{Melnick}, {Terlevich}  \& {Moles}}{{Melnick}
  et~al.}{1988}]{Melnick1988}
{Melnick} J.,  {Terlevich} R.,   {Moles} M.,  1988, \mn@doi [\mnras]
  {10.1093/mnras/235.1.297}, \href
  {https://ui.adsabs.harvard.edu/abs/1988MNRAS.235..297M} {235, 297}

\bibitem[\protect\citeauthoryear{{Melnick}, {Terlevich}  \&
  {Terlevich}}{{Melnick} et~al.}{2000}]{Melnick2000}
{Melnick} J.,  {Terlevich} R.,   {Terlevich} E.,  2000, \mn@doi [\mnras]
  {10.1046/j.1365-8711.2000.03112.x}, \href
  {https://ui.adsabs.harvard.edu/abs/2000MNRAS.311..629M} {311, 629}

\bibitem[\protect\citeauthoryear{{Moutard} et~al.,}{{Moutard}
  et~al.}{2016}]{Moutard2016a}
{Moutard} T.,  et~al., 2016, \mn@doi [\aap] {10.1051/0004-6361/201527945},
  \href {https://ui.adsabs.harvard.edu/abs/2016A&A...590A.102M} {590, A102}

\bibitem[\protect\citeauthoryear{{Noll}, {Burgarella}, {Giovannoli}, {Buat},
  {Marcillac}  \& {Mu{\~n}oz-Mateos}}{{Noll} et~al.}{2009}]{Noll2009}
{Noll} S.,  {Burgarella} D.,  {Giovannoli} E.,  {Buat} V.,  {Marcillac} D.,
  {Mu{\~n}oz-Mateos} J.~C.,  2009, \mn@doi [\aap]
  {10.1051/0004-6361/200912497}, \href
  {https://ui.adsabs.harvard.edu/abs/2009A&A...507.1793N} {507, 1793}

\bibitem[\protect\citeauthoryear{{Padilla} et~al.,}{{Padilla}
  et~al.}{2019}]{Padilla2019}
{Padilla} C.,  et~al., 2019, \mn@doi [\aj] {10.3847/1538-3881/ab0412}, \href
  {https://ui.adsabs.harvard.edu/abs/2019AJ....157..246P} {157, 246}

\bibitem[\protect\citeauthoryear{{Papovich}, {Dickinson}  \&
  {Ferguson}}{{Papovich} et~al.}{2001}]{Papovich2001}
{Papovich} C.,  {Dickinson} M.,   {Ferguson} H.~C.,  2001, \mn@doi [\apj]
  {10.1086/322412}, \href
  {https://ui.adsabs.harvard.edu/abs/2001ApJ...559..620P} {559, 620}

\bibitem[\protect\citeauthoryear{{P{\'e}rez-Gonz{\'a}lez}, {Gil de Paz},
  {Zamorano}, {Gallego}, {Alonso-Herrero}  \&
  {Arag{\'o}n-Salamanca}}{{P{\'e}rez-Gonz{\'a}lez}
  et~al.}{2003}]{Perez-Gonzalez2003}
{P{\'e}rez-Gonz{\'a}lez} P.~G.,  {Gil de Paz} A.,  {Zamorano} J.,  {Gallego}
  J.,  {Alonso-Herrero} A.,   {Arag{\'o}n-Salamanca} A.,  2003, \mn@doi
  [\mnras] {10.1046/j.1365-8711.2003.06078.x}, \href
  {https://ui.adsabs.harvard.edu/abs/2003MNRAS.338..525P} {338, 525}

\bibitem[\protect\citeauthoryear{{P{\'e}rez-Gonz{\'a}lez}
  et~al.,}{{P{\'e}rez-Gonz{\'a}lez} et~al.}{2013}]{PerezGonzalez2013}
{P{\'e}rez-Gonz{\'a}lez} P.~G.,  et~al., 2013, \mn@doi [\apj]
  {10.1088/0004-637X/762/1/46}, \href
  {https://ui.adsabs.harvard.edu/abs/2013ApJ...762...46P} {762, 46}

\bibitem[\protect\citeauthoryear{{Plionis}, {Terlevich}, {Basilakos},
  {Bresolin}, {Terlevich}, {Melnick}  \& {Chavez}}{{Plionis}
  et~al.}{2011}]{Plionis2011}
{Plionis} M.,  {Terlevich} R.,  {Basilakos} S.,  {Bresolin} F.,  {Terlevich}
  E.,  {Melnick} J.,   {Chavez} R.,  2011, \mn@doi [\mnras]
  {10.1111/j.1365-2966.2011.19247.x}, \href
  {https://ui.adsabs.harvard.edu/abs/2011MNRAS.416.2981P} {416, 2981}

\bibitem[\protect\citeauthoryear{{Renard} et~al.,}{{Renard}
  et~al.}{2022}]{Renard2022}
{Renard} P.,  et~al., 2022, \mn@doi [\mnras] {10.1093/mnras/stac1730}, \href
  {https://ui.adsabs.harvard.edu/abs/2022MNRAS.515..146R} {515, 146}

\bibitem[\protect\citeauthoryear{{Renzini} \& {Peng}}{{Renzini} \&
  {Peng}}{2015}]{Renzini2015}
{Renzini} A.,  {Peng} Y.-j.,  2015, \mn@doi [\apjl]
  {10.1088/2041-8205/801/2/L29}, \href
  {https://ui.adsabs.harvard.edu/abs/2015ApJ...801L..29R} {801, L29}

\bibitem[\protect\citeauthoryear{{Roberts} \& {Haynes}}{{Roberts} \&
  {Haynes}}{1994}]{Roberts_Haynes1994}
{Roberts} M.~S.,  {Haynes} M.~P.,  1994, \mn@doi [\araa]
  {10.1146/annurev.aa.32.090194.000555}, \href
  {https://ui.adsabs.harvard.edu/abs/1994ARA&A..32..115R} {32, 115}

\bibitem[\protect\citeauthoryear{{Rodriguez Espinosa} et~al.,}{{Rodriguez
  Espinosa} et~al.}{2014}]{Rodriguez-Espinosa2014}
{Rodriguez Espinosa} J.~M.,  et~al., 2014, \mn@doi [\mnras]
  {10.1093/mnrasl/slu099}, \href
  {https://ui.adsabs.harvard.edu/abs/2014MNRAS.444L..68R} {444, L68}

\bibitem[\protect\citeauthoryear{Rousseeuw}{Rousseeuw}{1987}]{Rousseeuw1987}
Rousseeuw P.~J.,  1987, \mn@doi [Journal of Computational and Applied
  Mathematics] {https://doi.org/10.1016/0377-0427(87)90125-7}, 20, 53

\bibitem[\protect\citeauthoryear{{Ruan}, {Melia}, {Chen}  \& {Zhang}}{{Ruan}
  et~al.}{2019}]{Ruan2019}
{Ruan} C.-Z.,  {Melia} F.,  {Chen} Y.,   {Zhang} T.-J.,  2019, \mn@doi [\apj]
  {10.3847/1538-4357/ab2ed0}, \href
  {https://ui.adsabs.harvard.edu/abs/2019ApJ...881..137R} {881, 137}

\bibitem[\protect\citeauthoryear{{Salpeter}}{{Salpeter}}{1955}]{Salpeter1955}
{Salpeter} E.~E.,  1955, \mn@doi [\apj] {10.1086/145971}, \href
  {http://adsabs.harvard.edu/abs/1955ApJ...121..161S} {121, 161}

\bibitem[\protect\citeauthoryear{{S{\'a}nchez Almeida} \& {Allende
  Prieto}}{{S{\'a}nchez Almeida} \& {Allende
  Prieto}}{2013}]{SanchezAlmeida2013}
{S{\'a}nchez Almeida} J.,  {Allende Prieto} C.,  2013, \mn@doi [\apj]
  {10.1088/0004-637X/763/1/50}, \href
  {https://ui.adsabs.harvard.edu/abs/2013ApJ...763...50S} {763, 50}

\bibitem[\protect\citeauthoryear{{S{\'a}nchez Almeida}, {Aguerri},
  {Mu{\~n}oz-Tu{\~n}{\'o}n}  \& {de Vicente}}{{S{\'a}nchez Almeida}
  et~al.}{2010}]{Sanchez-Almeida2010}
{S{\'a}nchez Almeida} J.,  {Aguerri} J.~A.~L.,  {Mu{\~n}oz-Tu{\~n}{\'o}n} C.,
  {de Vicente} A.,  2010, \mn@doi [\apj] {10.1088/0004-637X/714/1/487}, \href
  {https://ui.adsabs.harvard.edu/abs/2010ApJ...714..487S} {714, 487}

\bibitem[\protect\citeauthoryear{{S{\'a}nchez Almeida}, {Aguerri},
  {Mu{\~n}oz-Tu{\~n}{\'o}n}  \& {Huertas-Company}}{{S{\'a}nchez Almeida}
  et~al.}{2011}]{Sanchez-Almeida2011}
{S{\'a}nchez Almeida} J.,  {Aguerri} J.~A.~L.,  {Mu{\~n}oz-Tu{\~n}{\'o}n} C.,
  {Huertas-Company} M.,  2011, \mn@doi [\apj] {10.1088/0004-637X/735/2/125},
  \href {https://ui.adsabs.harvard.edu/abs/2011ApJ...735..125S} {735, 125}

\bibitem[\protect\citeauthoryear{{S{\'a}nchez} et~al.,}{{S{\'a}nchez}
  et~al.}{2019}]{Sanchez2019}
{S{\'a}nchez} S.~F.,  et~al., 2019, \mn@doi [\mnras] {10.1093/mnras/sty2730},
  \href {https://ui.adsabs.harvard.edu/abs/2019MNRAS.482.1557S} {482, 1557}

\bibitem[\protect\citeauthoryear{{Sandage}}{{Sandage}}{1961}]{Sandage1961}
{Sandage} A.,  1961, {The Hubble Atlas of Galaxies}.
Washington: Carnegie Institution

\bibitem[\protect\citeauthoryear{{Schwarz}}{{Schwarz}}{1978}]{Schwarz1978}
{Schwarz} G.~E.,  1978, \mn@doi [Ann. Stat.] {10.1007/s11222-013-9416-2}, 6,
  461

\bibitem[\protect\citeauthoryear{{Scoville} et~al.,}{{Scoville}
  et~al.}{2007}]{Scoville2007}
{Scoville} N.,  et~al., 2007, \mn@doi [\apjs] {10.1086/516585}, \href
  {https://ui.adsabs.harvard.edu/abs/2007ApJS..172....1S} {172, 1}

\bibitem[\protect\citeauthoryear{{Serrano} et~al.,}{{Serrano}
  et~al.}{2022}]{2022arXiv220614022S}
{Serrano} S.,  et~al., 2022, arXiv e-prints, \href
  {https://ui.adsabs.harvard.edu/abs/2022arXiv220614022S} {p. arXiv:2206.14022}

\bibitem[\protect\citeauthoryear{{Shin}, {Ly}, {Malkan}, {Malhotra}, {de los
  Reyes}  \& {Rhoads}}{{Shin} et~al.}{2021}]{Shin2021}
{Shin} K.,  {Ly} C.,  {Malkan} M.~A.,  {Malhotra} S.,  {de los Reyes} M.,
  {Rhoads} J.~E.,  2021, \mn@doi [\mnras] {10.1093/mnras/staa3307}, \href
  {https://ui.adsabs.harvard.edu/abs/2021MNRAS.501.2231S} {501, 2231}

\bibitem[\protect\citeauthoryear{{Siudek} et~al.,}{{Siudek}
  et~al.}{2018}]{Siudek2018}
{Siudek} M.,  et~al., 2018, \mn@doi [\aap] {10.1051/0004-6361/201832784}, \href
  {https://ui.adsabs.harvard.edu/abs/2018A&A...617A..70S} {617, A70}

\bibitem[\protect\citeauthoryear{{Sobral}, {Santos}, {Matthee},
  {Paulino-Afonso}, {Ribeiro}, {Calhau}  \& {Khostovan}}{{Sobral}
  et~al.}{2018}]{Sobral2018}
{Sobral} D.,  {Santos} S.,  {Matthee} J.,  {Paulino-Afonso} A.,  {Ribeiro} B.,
  {Calhau} J.,   {Khostovan} A.~A.,  2018, \mn@doi [\mnras]
  {10.1093/mnras/sty378}, \href
  {https://ui.adsabs.harvard.edu/abs/2018MNRAS.476.4725S} {476, 4725}

\bibitem[\protect\citeauthoryear{{Soo} et~al.,}{{Soo} et~al.}{2021}]{Soo2021}
{Soo} J. Y.~H.,  et~al., 2021, \mn@doi [\mnras] {10.1093/mnras/stab711}, \href
  {https://ui.adsabs.harvard.edu/abs/2021MNRAS.503.4118S} {503, 4118}

\bibitem[\protect\citeauthoryear{{Strateva} et~al.,}{{Strateva}
  et~al.}{2001}]{Strateva2001}
{Strateva} I.,  et~al., 2001, \mn@doi [\aj] {10.1086/323301}, \href
  {https://ui.adsabs.harvard.edu/abs/2001AJ....122.1861S} {122, 1861}

\bibitem[\protect\citeauthoryear{{Taylor} et~al.,}{{Taylor}
  et~al.}{2009}]{Taylor2009}
{Taylor} E.~N.,  et~al., 2009, \mn@doi [\apj] {10.1088/0004-637X/694/2/1171},
  \href {https://ui.adsabs.harvard.edu/abs/2009ApJ...694.1171T} {694, 1171}

\bibitem[\protect\citeauthoryear{{Teimoorinia}, {Archinuk}, {Woo}, {Shishehchi}
   \& {Bluck}}{{Teimoorinia} et~al.}{2022}]{Teimoorinia2022}
{Teimoorinia} H.,  {Archinuk} F.,  {Woo} J.,  {Shishehchi} S.,   {Bluck} A.
  F.~L.,  2022, \mn@doi [\aj] {10.3847/1538-3881/ac4039}, \href
  {https://ui.adsabs.harvard.edu/abs/2022AJ....163...71T} {163, 71}

\bibitem[\protect\citeauthoryear{{Telles} \& {Melnick}}{{Telles} \&
  {Melnick}}{2018}]{Telles2018}
{Telles} E.,  {Melnick} J.,  2018, \mn@doi [\aap]
  {10.1051/0004-6361/201732275}, \href
  {https://ui.adsabs.harvard.edu/abs/2018A&A...615A..55T} {615, A55}

\bibitem[\protect\citeauthoryear{{Terlevich} \& {Melnick}}{{Terlevich} \&
  {Melnick}}{1981}]{Terlevich1981}
{Terlevich} R.,  {Melnick} J.,  1981, \mnras, \href
  {http://adsabs.harvard.edu/abs/1981MNRAS.195..839T} {195, 839}

\bibitem[\protect\citeauthoryear{{Terlevich}, {Terlevich}, {Melnick},
  {Ch{\'a}vez}, {Plionis}, {Bresolin}  \& {Basilakos}}{{Terlevich}
  et~al.}{2015}]{Terlevich2015}
{Terlevich} R.,  {Terlevich} E.,  {Melnick} J.,  {Ch{\'a}vez} R.,  {Plionis}
  M.,  {Bresolin} F.,   {Basilakos} S.,  2015, \mn@doi [\mnras]
  {10.1093/mnras/stv1128}, \href
  {http://adsabs.harvard.edu/abs/2015MNRAS.451.3001T} {451, 3001}

\bibitem[\protect\citeauthoryear{{Tortorelli} et~al.,}{{Tortorelli}
  et~al.}{2021}]{Tortorelli2021}
{Tortorelli} L.,  et~al., 2021, \mn@doi [\jcap]
  {10.1088/1475-7516/2021/12/013}, \href
  {https://ui.adsabs.harvard.edu/abs/2021JCAP...12..013T} {2021, 013}

\bibitem[\protect\citeauthoryear{{Tsiapi} et~al.,}{{Tsiapi}
  et~al.}{2021}]{Tsiapi2021}
{Tsiapi} P.,  et~al., 2021, \mn@doi [\mnras] {10.1093/mnras/stab1933}, \href
  {https://ui.adsabs.harvard.edu/abs/2021MNRAS.506.5039T} {506, 5039}

\bibitem[\protect\citeauthoryear{{Turner} et~al.,}{{Turner}
  et~al.}{2021}]{Turner2021}
{Turner} S.,  et~al., 2021, \mn@doi [\mnras] {10.1093/mnras/stab653}, \href
  {https://ui.adsabs.harvard.edu/abs/2021MNRAS.503.3010T} {503, 3010}

\bibitem[\protect\citeauthoryear{{Vilella-Rojo} et~al.,}{{Vilella-Rojo}
  et~al.}{2021}]{Vilella-Rojo2021}
{Vilella-Rojo} G.,  et~al., 2021, \mn@doi [\aap] {10.1051/0004-6361/202039156},
  \href {https://ui.adsabs.harvard.edu/abs/2021A&A...650A..68V} {650, A68}

\bibitem[\protect\citeauthoryear{{Whitaker} et~al.,}{{Whitaker}
  et~al.}{2011}]{Whitaker2011}
{Whitaker} K.~E.,  et~al., 2011, \mn@doi [\apj] {10.1088/0004-637X/735/2/86},
  \href {https://ui.adsabs.harvard.edu/abs/2011ApJ...735...86W} {735, 86}

\bibitem[\protect\citeauthoryear{{Whitaker}, {van Dokkum}, {Brammer}  \&
  {Franx}}{{Whitaker} et~al.}{2012}]{Whitaker2012}
{Whitaker} K.~E.,  {van Dokkum} P.~G.,  {Brammer} G.,   {Franx} M.,  2012,
  \mn@doi [\apjl] {10.1088/2041-8205/754/2/L29}, \href
  {https://ui.adsabs.harvard.edu/abs/2012ApJ...754L..29W} {754, L29}

\bibitem[\protect\citeauthoryear{{Williams}, {Quadri}, {Franx}, {van Dokkum}
  \& {Labb{\'e}}}{{Williams} et~al.}{2009}]{Williams2009}
{Williams} R.~J.,  {Quadri} R.~F.,  {Franx} M.,  {van Dokkum} P.,   {Labb{\'e}}
  I.,  2009, \mn@doi [\apj] {10.1088/0004-637X/691/2/1879}, \href
  {https://ui.adsabs.harvard.edu/abs/2009ApJ...691.1879W} {691, 1879}

\bibitem[\protect\citeauthoryear{{Wofford}, {Vidal-Garc{\'\i}a}, {Feltre},
  {Chevallard}, {Charlot}, {Stark}, {Herenz}  \& {Hayes}}{{Wofford}
  et~al.}{2021}]{Wofford2021}
{Wofford} A.,  {Vidal-Garc{\'\i}a} A.,  {Feltre} A.,  {Chevallard} J.,
  {Charlot} S.,  {Stark} D.~P.,  {Herenz} E.~C.,   {Hayes} M.,  2021, \mn@doi
  [\mnras] {10.1093/mnras/staa3365}, \href
  {https://ui.adsabs.harvard.edu/abs/2021MNRAS.500.2908W} {500, 2908}

\bibitem[\protect\citeauthoryear{{Wu}, {Cao}, {Zhang}, {Liu}, {Liu}, {Geng}  \&
  {Lian}}{{Wu} et~al.}{2020}]{Wu2020}
{Wu} Y.,  {Cao} S.,  {Zhang} J.,  {Liu} T.,  {Liu} Y.,  {Geng} S.,   {Lian} Y.,
   2020, \mn@doi [\apj] {10.3847/1538-4357/ab5b94}, \href
  {https://ui.adsabs.harvard.edu/abs/2020ApJ...888..113W} {888, 113}

\bibitem[\protect\citeauthoryear{{Wuyts} et~al.,}{{Wuyts}
  et~al.}{2007}]{Wuyts2007}
{Wuyts} S.,  et~al., 2007, \mn@doi [\apj] {10.1086/509708}, \href
  {https://ui.adsabs.harvard.edu/abs/2007ApJ...655...51W} {655, 51}

\bibitem[\protect\citeauthoryear{{Yennapureddy} \& {Melia}}{{Yennapureddy} \&
  {Melia}}{2017}]{Yennapureddy2017}
{Yennapureddy} M.~K.,  {Melia} F.,  2017, \mn@doi [\jcap]
  {10.1088/1475-7516/2017/11/029}, \href
  {https://ui.adsabs.harvard.edu/abs/2017JCAP...11..029Y} {2017, 029}

\bibitem[\protect\citeauthoryear{{de Vaucouleurs}}{{de
  Vaucouleurs}}{1959}]{deVaucouleurs1959}
{de Vaucouleurs} G.,  1959, \mn@doi [Handbuch der Physik]
  {10.1007/978-3-642-45932-0_7}, \href
  {https://ui.adsabs.harvard.edu/abs/1959HDP....53..275D} {53, 275}

\bibitem[\protect\citeauthoryear{{de Vaucouleurs}}{{de
  Vaucouleurs}}{1961}]{de_Vaucouleurs1961}
{de Vaucouleurs} G.,  1961, \mn@doi [\apjs] {10.1086/190056}, \href
  {https://ui.adsabs.harvard.edu/abs/1961ApJS....5..233D} {5, 233}

\makeatother
\end{thebibliography}
\label{lastpage}

\newpage

\appendix


\clearpage

\end{document}